%% file: 00_main.tex
\documentclass[
reprint,
superscriptaddress,
nofootinbib,
amsmath,amssymb,
aps,
prx,
showkeys
]{revtex4-2}

\usepackage{graphicx}
\usepackage{bm}
\usepackage{xspace}
\usepackage[table]{xcolor}
\usepackage{float}  
\usepackage[caption=false]{subfig}
\usepackage[
  colorlinks=true,
  linkcolor=blue,
  citecolor=blue,
  urlcolor=blue,
  breaklinks=true
]{hyperref}  
\newcommand{\rc}[0]{RC\xspace}
\newcommand{\etal}[0]{\textit{et al.}\xspace}
\newcommand{\eg}[0]{e.g.\xspace}
\newcommand{\ie}[0]{i.e.\xspace}
\newcommand{\figref}[2][]{Fig.\ \ref{#2}#1\xspace}  
\newcommand{\eqqref}[1]{Eq.\ \ref{#1}\xspace}
\newcommand{\sectref}[1]{Sect.\ \ref{#1}\xspace}
\newcommand{\axref}[1]{Appendix\ \ref{#1}\xspace}
\newcommand{\figrefs}[2]{Figs.\ \ref{#1} - \ref{#2}\xspace}
\newcommand{\refref}[1]{Ref.\ #1\xspace}
\newcommand{\tabref}[1]{Tab.\ \ref{#1}\xspace}
\newcommand{\vidref}[2]{Supplementary Video #1/#2\xspace}

\newcommand{\vidTableIdColumnWidth}{.5cm}
\newcommand{\vidTableSnapshotColumnWidth}{2.6cm}
\newcommand{\vidTableDescriptionColumnWidth}{8cm}

\newcommand{\new}[1]{#1}

\newcommand{\affilsimtech}{Stuttgart Center for Simulation Science,
Cluster of Excellence EXC 2075,\\ University of Stuttgart,
Universit\"atsstra{\ss}e 32, 70569 Stuttgart, Germany}
\newcommand{\affilwin}{WIN-Kolleg of the Young Academy $\vert$ Heidelberg Academy of Science and the Humanities, Karlstraße 4, 69117 Heidelberg, Germany}
\begin{document}

\title{Robustly optimal dynamics for active matter reservoir computing}

\author{Mario U. Gaimann}
\affiliation{\affilsimtech}%

\author{Miriam Klopotek}
\email{miriam.klopotek@simtech.uni-stuttgart.de}
\affiliation{\affilsimtech}%
\affiliation{\affilwin}

\date{\today}

\include{abstract}


\maketitle
\newpage

\begingroup
\let\clearpage\relax
\include{intro}
\endgroup

\begingroup
\let\clearpage\relax
\section{\label{sec:materials-methods}Methods}
\include{methods_model}
\include{methods_observables}
\include{methods_simulation_details}
\endgroup

\begingroup
\let\clearpage\relax
\section{\label{sec:results}Results} 

\include{results_speed_controller}
\include{results_phenomenology_overdamped}
\include{results_fluctuations}
\include{results_benchmarks}
\include{results_few_particles}
\include{results_alignment}
\include{results_homing}

\endgroup

\begingroup
\let\clearpage\relax
\include{discussion}
\include{contributions}
\include{acknowledgements}
\endgroup

\bibliography{export, bib_2}

\clearpage

\appendix
\begingroup
\let\clearpage\relax
\include{supp_benchmarks_details}

\include{supp_noise}
\include{supp_memory_capacity}
\include{supp_esn_comparison}

\include{supp_data}
\endgroup

\begin{widetext}
\raggedbottom
\include{supp_figs}
\include{supp_vids}
\end{widetext}

\end{document}

%% file: abstract.tex

\begin{abstract}
Information processing abilities of active matter are studied in the reservoir computing (RC) paradigm to infer the future state of a chaotic signal. 
We uncover an exceptional regime of agent dynamics that has been overlooked previously. It appears robustly optimal for performance under many conditions, thus providing valuable insights into collective and physical computation more generally. The key to forming effective mechanisms for information processing appears in the system's intrinsic relaxation abilities. 
These are probed without actually enforcing a specific inference goal. The dynamical regime that achieves optimal computation is located just below a critical damping threshold, involving a relaxation with multiple stages, and is readable at the single-particle level. At the many-body level, it yields substrates robustly optimal for RC across varying physical parameters and inference tasks. A system in this regime exhibits a strong diversity of dynamic mechanisms under highly fluctuating driving forces.
Correlations of agent dynamics can express a tight relationship between the responding system and the fluctuating forces driving it. 
As this model is interpretable in physical terms, it facilitates re-framing inquiries regarding learning and computing with a fresh rationale for many-body physics far from equilibrium.
\end{abstract}

%% file: intro.tex

\section{\label{sec:introduction}Introduction}

Computation is not limited to the digital domain: 
\emph{In materio} computation is about analog physical computing directly with matter, on a continuous domain of space and time \cite{Horsman2014-when_phys_compute,Dale2017-rc_in-materio,Kaspar2021-intelligent_matter}. Unlike Turing machines, a computation will not require sequential transitions between well-defined discrete internal states \cite{Maass2002-real-time_computing,Horsman2014-when_phys_compute,maclennan2004natural}. This is part of a wider push towards unconventional computing: using the intrinsic dynamics of physical systems for computation can lessen the computational overhead and resulting energy expenditure arising from traditional electronic and digital means \cite{Dale2017-rc_in-materio,Stepney2008-materialcompute,Konkoli2018-rc_comput_matter,stepney2012}. The idea is to take inspiration from how nature computes \cite{Zenil2013-book_computable_universe}, for example the brain \cite{Jaeger2021-unconventional_general,Hopfield1982}, import basic physical concepts and systems \cite{Horsman2014-when_phys_compute,Markovic2020-phys_for_neuromoph,jaeger2023toward}, and depart from the von Neumann computing paradigm, where information processing and memory storage are executed in separate compartments \cite{beyondVonNeumann2020}.

Can soft condensed matter or biological systems present a complementary class of systems adept for computation and inference? \new{(How) does computation rely on collectivity, inherent to many systems}? \new{Swarm intelligence hinges on the cooperation of many parts acting together} -- which is gaining rapid interest for artificial intelligence algorithms \cite{Liu2025-preprint,Marl2024-book} as well as robotics \cite{elShowk2025}. From a physical point of view, active matter systems \cite{teVrugt2025} are interesting candidates because they are inherently out of equilibrium \cite{OByrne2022}, rendering rich collective behavior at the macroscopic scale like swarming and flocking \cite{Yadav2018,Bowick2022-review_active_mat}.
Also, soft materials offer a way to distribute forces and integrate responses important for modern robotic sensing and actuation \cite{Rothemund2018}.
How can we \emph{generally} assess the suitability of bio-inspired systems for both analog computation and learning algorithms, which represent significantly distinct contexts? 

The reservoir computing (\rc) paradigm is ideal for exploring these basic questions. It directly exploits dynamical systems for information processing and inference in real-time \cite{Jaeger2010,Atiya2000, Maass2002-real-time_computing}.
The power of \rc for scientific inquiry into physical computation is its generic and straightforward setup: Information of a low-dimensional input (\eg a time series) is injected into the reservoir computing substrate, which is a high-dimensional nonlinear dynamical system that enters different dynamical states depending on the input signal and its history \cite{Jaeger2001-short_term_memory}. 
In the \rc paradigm, the reservoir substrate is optimized based on heuristic global properties (such as the spectral radius in Echo State Networks). In the canonical case of utilizing a digital neural network, substrates are simply initialized in a random way. The readout layer \cite{Lukosevicius2009} typically consists of a \emph{linear} model trained using Ridge regression. 
\new{This separation between an `untrained' nonlinear substrate and a trainable-but-simple readout layer constitutes} the original discovery of \rc as a method \cite{Jaeger2010, Atiya2000, Maass2002-real-time_computing}, avoiding costly backpropagation through time \cite{Lukosevicius2009}.

Currently, \rc{} \new{enables} state-of-the-art performance for time-series forecasting \cite{Pathak2018_model_free,Liu2010,Gilpin2023,Shahi2022}, in particular, chaotic time series prediction \cite{Liu2010}.
Despite the plethora of other machine learning techniques available for time-series forecasting (nonlinear vector autoregression (``next-generation \rc'') \cite{Gauthier2021}, NBEATS \cite{Oreshkin2019}, NHiTS \cite{Challu2023}, transformer models \cite{Vaswani2017} or LSTMs \cite{Hochreiter1997}), the highly generic method of \rc using \eg Echo State Networks has often remained an architecture of choice. It requires far less training data \cite{Gilpin2023, Shahi2022} and allows for predictions \new{ahead of} several characteristic times \new{(Lyapunov times) for some chaotic systems -- when in a closed-loop autoregressive setup} \cite{Gilpin2023}. It is considered ``model-free'' in the sense that it can estimate properties of the input without the need for an explicit model thereof \cite{Roehm2021-rc_attractors}. \new{It can be implemented using generic substrates; photonics-based machine learning is one promising generic direction that can harness RC design paradigms} \cite{Vandoorne2014,Tanaka2019,Hlser2022,Nakajima2021b,Duport2016}.

In \rc, only a few general properties of the substrate are believed to play a role for maximum performance: the degree of nonlinearity, reproducibility (where similar inputs yield similar outputs), separability (where the reservoir differentiates distinct inputs into different outputs), and the echo state property (which refers to fading memory, meaning the current reservoir states depend only on recent history and not on initial conditions) \cite{Dale2019-theory_general_rc,Maass2002-real-time_computing, Schrauwen2007a, Verstraeten2009, Lukosevicius2009,Yildiz2012,Jaeger2001-short_term_memory}. 
However, many issues remain open: For example, the optimal mix of nonlinearity and linearity in responses \cite{Inubushi2017}; \new{how exactly} memory capacity and nonlinearity present trade-offs \cite{Dambre2012,Inubushi2017}; \new{the basic role of} time delay in response \cite{Appeltant2011,Appeltant2012,Hlser2022,Tanaka2019}; and how information should be injected \new{into and }read out from \cite{Ohkubo2024} the reservoir.

Yet, for reservoir computation \emph{in materio} or using biological-like swarms, deep physical concepts remain hidden beneath the surface. For example, how do nonequilibrium phase diagrams and dissipative as well as correlative dynamics relate to computational properties? 
Understanding these links would be valuable for recognizing the true feasibility of neuromorphic computing systems \cite{Lee2023, Jaeger2021,Markovic2020-phys_for_neuromoph}, and connecting different realms of matter-based and digital computing, both conceptually and practically \cite{Kendon2015}.
Generally, interpretable \rc frameworks should lead to better integration of theory with practice \cite{Yan2024-opportunitiesRC,Gauthier2021}. 
Artificial neural networks have been employed as the standard substrate for \rc; they are essentially black boxes \cite{wetzel2025interpretable}.
A \emph{different paradigm for thinking about computation and learning} is required when dealing with physical substrates \cite{Horsman2014-when_phys_compute, jaeger2023toward}. 
Notably, hard condensed matter, \eg, magnetic systems, has already  \new{enabled RC implementations for } prediction and pattern recognition by exploiting topologically protected phases (skyrmions) \cite{Prychynenko2018,Everschor2024-review-magnetic_RC,Love2023,Yokouchi2022,Lee2023}.

Regarding living systems, it may not be clear whether the power of collective information processing is attributable to physical dynamics alone, or whether other factors may play a role \cite{Couzin2002,Couzin2007,Couzin2009,Flack2017}.  
A basic physical model system for biologically-inspired \rc can be enormously clarifying in these respects. Moreover, there is a rapidly growing interest in the use of biological matter for artificial intelligence, particularly within the \rc paradigm, using self-organized brain organoids and objects alike \cite{Smirnova2023,Cai2023,Kobayashi2025}.

\rc has hence been demonstrated for soft matter, though extremely rarely, and developments remain in an early stage: Reported were cases of gel materials, soft macroscopic bodies, and other examples of cell cultures \cite{Strong2022-rc_polymer_gels,Nakajima2015,Obien2015}. 
In one exotic case, water waves in a bucket were used as a proof-of-principle \cite{Fernando2003}.
The usage of active matter models for \rc is in its infancy \cite{Jeggle2025}. Lymburn \etal simulated biologically inspired swarms that avoid an externally controlled predator as a reservoir computer \cite{Lymburn2021}, which motivated our work. Note that the converse idea of using a reservoir computer to control a swarm has also been investigated \cite{Algar2019}, and there is a growing interest in controlling active particles using artificial intelligence \cite{Loewen2025, Cichos2020, Tovey2024b}. Wang \etal presented an experimental realization of \rc with an array of active gold nano-particles driven by a laser \cite{Wang2024}. Strong \etal designed \rc around active hydrogen ions residing in an electroactive polymer gel, which can be excited by applying an electric field \cite{Strong2022}. Two more experimental setups have been conceptually proposed by Jeggle and Wittkowski that leverage driving the active matter agents via spatiotemporally modulated fields \cite{Jeggle2025}: micro-particles in a medium driven by ultrasonic pressure waves, and micro-particles with a symmetry-broken refractive index profile driven by light fields.

Physical reservoir computing involves continuously probing matter and using the information gathered by observing responses at various spatial locations, scales, and times. The nature of these responses is the essence of computation, which occurs out of thermodynamic equilibrium \cite{Szilard1929,Landauer1961,Bennett1982}.
When (driven) out of thermodynamic equilibrium, complex systems composed of molecules, colloidal, or active particles exhibit collective and correlative features not found in counterparts near equilibrium \cite{Solon2015,Rahmani2012,Nicolis1971,Klopotek2017,Cross1993,Horowitz2019,kubo2012statistical,Keim2019}.  The fluctuations internal to these systems are generally broader, and nonlinear coupling mechanisms play a role \cite{Das2004}, rendering heterogeneous dynamics. These can stimulate the formation of new structures \cite{Cross1993}. Notably, intelligent systems in nature or elsewhere share these and related key physical markers \cite{England2015,Flack2013,Rabinovich2008,Brush2018}.

In this paper, we use an active matter model externally driven far from equilibrium by a chaotically varying signal. This active system serves as the reservoir computing substrate to predict the future state of the external signal. We employ particle-based simulations. The physical information is coarse-grained using Gaussian kernels in an observation layer, and a readout layer is trained using Ridge regression (\sectref{sec:materials-methods}), following the design of Lymburn \etal \cite{Lymburn2021}. We conduct extensive physical parameter scans (\sectref{sec:results}).

These scans reveal a novel nonequilibrium dynamical regime that robustly establishes optimal performance in the RC (\sectref{subsec:slow-regime}). It is situated near a critical damping point in the intrinsic relaxation dynamics. These individual-particle, non-collective properties give rise to abilities essential for reliable computation. Favorable collective qualities emerge from them.
Contrary to previous beliefs, optimal computing is not observed near a gas-droplet phase transition regime of a dynamically rich ``critical'' or ``boiling'' swarm, as suggested by Lymburn \etal \cite{Lymburn2021}. Instead, our discovered regime achieves approximately 20\,\% better predictive performance (using the same metric) than previously reported best values.

As we find, the success of computing capabilities is powerfully associated with the intrinsic relaxation behavior of the system, which is a direct effect of the microscopic dynamics (\sectref{subsec:damping_and_vac}). Under driving, a complex interplay is revealed in dynamical autocorrelation functions, including an adaptation to the chaotically changing environment. Intriguing effective mechanisms at mesoscopic scales appear (\sectref{subsec:rationale_RC_critically damped}). 
The intensity and reach of velocity fluctuations (\sectref{subsec:correlated_current_fluctuations}) serve as key physical proxies for computational quality. 

We vary prediction tasks for our reservoir computer and experiment with different dynamic classes of chaotic driving (\sectref{subsec:other-benchmarks}). Moreover, we test \rc with only one- or two-particle substrates (\sectref{subsec:few-particle}), eliminating collective effects, \ie, the theoretical minimum of active matter \rc in this model.

We also vary agent alignment forces, \ie, non-Hamiltonian interactions between particles (\sectref{subsec:active-crystal}) and static external forces (\sectref{subsec:homing}) to test our hypothesis of generically optimal computation in the same regime.

\new{In this work, our primary objective is to elucidate how and under which conditions active matter can serve as a computational medium, rather than to rigorously optimize or benchmark its performance against conventional echo state networks or other physical reservoirs. Accordingly, we focus on standard tasks (open-loop Lorenz-63 n-steps-ahead prediction) and basic performance metrics (Pearson correlation coefficient of the x-coordinate) established in prior work \cite{Lymburn2021}, and defer a systematic performance benchmarking and task-specific optimization to future studies.}
The roots of optimal computation thus embody generic features in the microscopic dynamical mechanisms involved. These are non-specific to the particular chaotic input signal and suggest a deeper reason for the characteristic robustness of certain physical substrates, which perform well across various scenarios. We synthesize results at the end of the paper (\sectref{sec:discussion}).

%% file: methods_model.tex

\subsection{\label{sec:mm-interactions}Steady-state active matter simulations}

\begin{figure*}[htbp]
    \centering
    \includegraphics{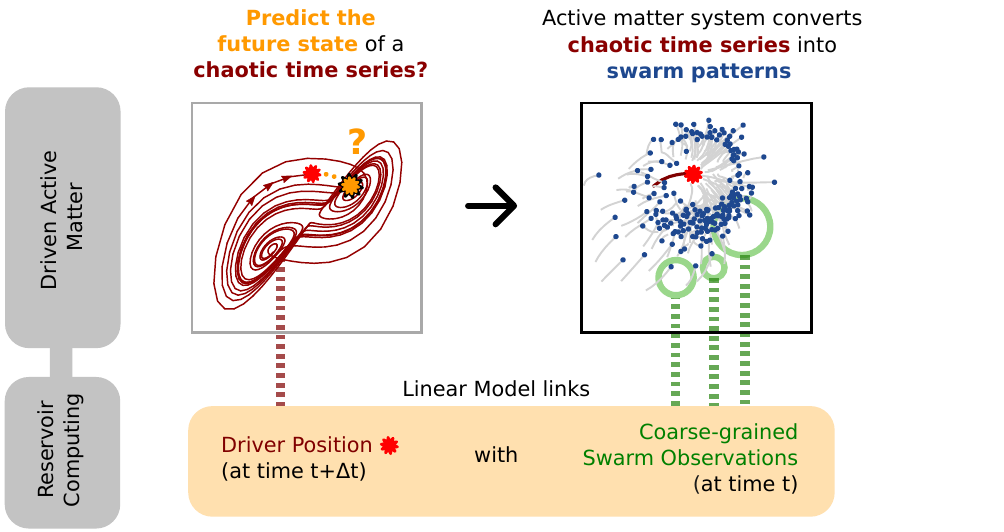}%
    \caption{\textbf{Reservoir computing with active matter concept.}
    The problem to solve is predicting the future state of a chaotic time series $\bm{Y}_{\text{target}}^{\text{train}}$ (the Lorenz-63 attractor, red trajectory) using reservoir computing. To this end, the time series is continuously injected into a swarm of active particles (blue) as a perturbing signal called the driver (red spiked ball, driver position $\bm{x}_d(t) \equiv \bm{Y}_{\text{target}}^{\text{train}}(t)$). This perturbation is modeled as a repulsive force $\bm{F}_d$ between the driver and the agents. A set of intrinsic interactions governs the swarm agents: They align their direction of motion ($\bm{F}_a$), avoid collisions ($\bm{F}_r$), seek a constant speed ($\bm{F}_{sc})$, return to the center of the simulation box ($\bm{F}_h$), and are limited in their overall response (force clamp) (see \sectref{sec:mm-interactions}). The non-linear, non-equilibrium response of the swarm is then measured in a coarse-grained fashion (green observation kernels, see also \figref{fig:gaussian_kernels_placement}) and stored in a matrix $\mathrm{X}^{\text{train}}$ for each time step. Together with the original driver trajectory, this enables training a linear readout layer with weights $\bm{W}_{\text{out}}$. Using these weights, one can predict the future state $t + \Delta t$ at each time step $t$ of a new time series $\bm{Y}_{\text{target}}^{\text{test}}$. This new time series must follow the same dynamics as the one used for training, but may have different initial conditions. In this paper, we perform reservoir optimization: we aim to find the optimal set of agent interactions and understand the resulting agent dynamics in terms of underlying physics. 
    }
    \label{fig:concept}
\end{figure*}

Our simulation design follows the basic active matter reservoir computing set-up described by Lymburn \etal \cite{Lymburn2021} (see \figref{fig:concept}).
The active matter agents constitute the reservoir, and the high-dimensional reservoir state can be constructed from the system state properties (positions, velocities, forces).

A particle $i$ experiences from neighboring particles $j=1,\dots,N_r$ a repulsive force
\begin{equation} \label{eq:repulsion_force}
    \bm{F}_{r,i}=\sum_{j=1}^{N_r} \frac{\bm{x}_i-\bm{x}_j}{\left\|\bm{x}_i-\bm{x}_j\right\|^2}
\end{equation}
where $N_r$ is the number of neighbors within a radial neighborhood $r_r$ around particle $i$. This force component drives agents apart from each other. \\
The particles are attracted to a defined home location $\bm{x}_{h}$, which is chosen as the center of the simulation box, by a linear force
\begin{equation} \label{eq:homing_force}
    \bm{F}_{h,i}= \bm{x}_{h}-\bm{x}_i\,.
\end{equation}
We introduce two non-Hamiltonian components that render our system an active matter system. All particles locally align their direction of motion with their $N_a$ neighbors in a radial neighborhood $r_a$, given by the force 
\begin{equation} \label{eq:alignment_force}
    \bm{F}_{a,i}=\sum_{j=1}^{N_a} \bm{v}_j-\bm{v}_i\,.
\end{equation}
All particles experience a force that regulates their speed towards a stationary particle speed $s$. This friction-like force, called speed-controller, is given as
\begin{equation} \label{eq:speed-controller}
    \bm{F}_{{sc},i}=-\bm{v}_i \frac{\left(\left\|\bm{v}_i\right\|-s\right)}{s}\,,
\end{equation}
where $\bm{v}_i$ is the velocity of particle $i$. It accelerates or decelerates each particle based on its current speed and on its relative deviation from the target agent speed $s$.\\ 

The total force acting on a particle is then given as the weighted sum of all force contributions
\begin{equation} \label{eq:total_force}
    \bm{F}_i(t)=K_a \bm{F}_{a,i}+K_r \bm{F}_{r,i}+K_{sc} \bm{F}_{sc,i}+K_h\bm{F}_{h,i}\,,
\end{equation}
where the weights $K$ indicate the respective force strengths.
The resulting action that the active matter agent undertakes based on the total force it experiences is restricted using the sigmoid wrapper function
\begin{equation} \label{eq:force_wrapper}
    \bm{F}_i(t) \mapsto \alpha \tanh \left(\beta \bm{F}_i(t)\right)\,,
\end{equation}
where $\alpha$ describes the scale of the particle response (the asymptotic force limit) and $\beta$ tunes the slope of the $\tanh$ function. This additional non-linearity is motivated by the disparity between a biologically inspired agent's intended motion and its physical limitations \cite{Lymburn2021} and increases numerical stability.

\subsection{\label{sec:mm-interactions-driven} Injecting information of a chaotic time-series into the system via external driving}

So far, we have described an active matter model that does not receive any external information. To perform reservoir computing and inject time series information into the swarm, we introduce an additional particle called \textit{driver} (red spiked ball in \figref{fig:concept}). It is called the driver because it drives the swarm out of its steady state. We then model a repulsive force between this driver particle and the swarm. 
The repulsive force 
\begin{equation} \label{eq:driver_repulsion}
    \bm{F}_{d,i}=\theta\left(r_d-\left\|\bm{x}_i-\bm{x}_d\right\|\right) \frac{\bm{x}_i-\bm{x}_d}{\left\|\bm{x}_i-\bm{x}_d\right\|^2}
\end{equation}
is induced by the driver and experienced by an agent $i$. A cut-off radius $r_d$ is in place modeled by the Heaviside step function ($\theta ( r_d- \left\|\bm{x}_i-\bm{x}_d\right\|) = 1$ if $\left\|\bm{x}_i-\bm{x}_d\right\| \leq r_d$, and $0$ otherwise). This term is added to the total force sum in \eqqref{eq:total_force} with a term $K_d \bm{F}_{d,i}$ that enters the $\tanh$ function in \eqqref{eq:force_wrapper}. The driver particle itself does not experience a force from the agents; its trajectory is fully determined by an external input time series.
This time series is to be processed by the swarm reservoir computer to predict its future state.

Chaotic attractors have been shown to serve as useful benchmarking problems for prediction \cite{Gilpin2021, Gilpin2023, dysts, Wringe2024}. They are challenging in prediction tasks because small changes in the initial conditions lead to exponentially different states due to their deterministically chaotic nature. 
We choose the famous chaotic Lorenz-63 attractor \cite{Lorenz1963} as the main benchmark time series. In addition, chaotic attractors from different dynamical classes (as classified in \refref{\cite{Gilpin2023}}) are chosen to assess the performance of our active matter reservoir computer in different scenarios: the attractors Hénon–Heiles \cite{Henon1964} (twists), Rössler \cite{Rossler1976} (rotations), Chua \cite{Chua1969} (lobes), and \new{the four-dimensional} Lorenz-96 \new{system} \cite{Lorenz1995} (ripples) \new{(see also \axref{subsec:prediction_tasks} for details).}

We report on the resulting phenomenological differences of active matter \rc for these different attractors in \sectref{subsec:other-benchmarks}. 

To compute the driver trajectory $\textbf{x}_d (t)$, we project the original \new{higher dimensional} time series on a 2D simulation canvas by omitting excess dimensions, if applicable. For example, for the Lorenz-63 trajectory, the third ($z$) dimension is omitted, \new{and for the four-dimensional Lorenz-96 system, the third and fourth dimensions ($x_3, x_4$) are omitted. Hence, the first two dynamical variables are used as input. }
We scale the trajectory to cover a specified fraction of the simulation box size and center it in the simulation box. Details regarding chaotic time series generation and pre-processing can be found in \axref{subsec:prediction_tasks}.
Taking all model components together, the driven swarm is evolved by integrating the forces from \eqqref{eq:total_force} and \eqqref{eq:force_wrapper} with an Euler-forward scheme \cite{Griffiths2010}, followed by moving the driver to its next time-series position \new{(a new input is fed into the system for each time step)}.

\subsection{\label{sec:mm-observations} Extracting a set of coarse-grained reservoir state variables with Gaussian observation kernels}

After injecting the target time series via a driver particle into the swarm reservoir, the next step in the reservoir computing method is to extract information from the driven, out-of-equilibrium swarm. 
To extract a high-dimensional swarm response, the immediate positions and velocities of the particles cannot be used for reservoir computing due to the permutation symmetry of swarm agents (see \refref{\cite{Lymburn2021}} for a detailed discussion).
To overcome this problem, we introduce Gaussian observation kernels to capture local densities. The $m$th kernel records three quantities $r_{\left\{ 1,2,3\right\} ,m}$ in each time step $t$: the Gaussian-integrated (effectively local) densities for 1) particle count, 2) total velocity in $x$ direction, and 3) total velocity in $y$ direction. The observations of each kernel are described by the equations
\begin{equation}
    \begin{aligned}
    & r_{1, m}(t)=\sum_{i=1}^{N} \psi_m\left(\bm{x}_i(t)\right)\,, \\
    & r_{2, m}(t)=\sum_{i=1}^{N} \psi_m\left(\bm{x}_i(t)\right) v_{x i}(t)\,, \text{and} \\
    & r_{3, m}(t)=\sum_{i=1}^{N} \psi_m\left(\bm{x}_i(t)\right) v_{y i}(t)\,,
    \end{aligned}
\end{equation}
where $v_{x i}$ and $v_{y i}$ are the $x$ and $y$ velocity components of an agent $i$, and $N$ is the total number of agents. The factors $\psi_m(\bm{x}_i)$ describe the ``weight'' that each particle $i$ contributes to the sums $\sum_{i=1}^{N}$, based on the proximity of each particle to the center position $\bm{c}_m$ of the $m$th kernel and kernel width $w_m$. They are given as
\begin{equation}
    \psi_m(\bm{x}_i) =e^{-\frac{\left(\bm{x}_i-\bm{c}_m\right)^2}{2 w_m}}\,.
    \label{eq:gaussian-kernel}
\end{equation}
The closer agents are to a kernel's center point, the higher their relative contribution to the kernel sum.
At the beginning of a simulation, we place $M$ Gaussian observation kernels randomly on the square simulation canvas with lengths $l_{\text{box}}$. This is done by drawing $\bm{c}_m$ randomly from a uniform distribution $[0; l_{\text{box}})$, and $w_m$ from a Gaussian distribution with mean $\mu = 0.0$ and standard deviation $\sigma = 1/\sqrt{2}$.
We note that in \refref{\cite{Lymburn2021}} a kernel placement protocol was employed that takes into account agent locations, but our simpler choice leads to similar results as verified in \figref{fig:reproduction_lymburn_alignment_repulsion_performance}. 
For each time step $t$ of $T$ total simulation time steps, we collect $3M$ observations, where $M$ is the number of kernels placed on the canvas, and save them in the observation matrix $\bm{X} \in \mathbb{R}^{3M \times T}$.

\subsection{\label{sec:mm-training} Readout layer training and prediction}
To make a prediction using the reservoir computing approach, we train a linear readout layer $\bm{W}_{\mathrm{out}} \in \mathbb{R}^{2 \times \new{(3M+1)}}$. It connects via weights the $3M$ observations of the swarm at time $t$ \new{(and a bias)} with future states of the two-dimensional input time series (the driver trajectory $\bm{x}_d$) at a time $t + \Delta t_{\mathrm{pred}}$. \new{$\Delta t_{\mathrm{pred}}$ can also be written as multiples of the integration time step, $n_{\text{pred}} \Delta t$, allowing for an  $n_{\text{pred}}$-step-ahead prediction at each time step (open-loop prediction)}. To train the readout layer, we write the input (target) time series
\begin{equation}
    \bm{Y}_{\mathrm{target}} = (\bm{x}_d(t_1), ..., \bm{x}_d(t_T)) \in \mathbb{R}^{2 \times T}\,
\end{equation}
as a sequence of driver positions.
\new{To train the readout layer for an $n_{\text{pred}}$-step-ahead prediction, we remove the first and last $n_{\text{pred}}$ time steps from the target time series $\bm{Y}_{\mathrm{target}}^{\mathrm{train}}$ and the observations $\bm{X}^\mathrm{train}$, respectively.
Concretely, we define the aligned (reduced) matrices
\begin{align}
        \tilde{\bm{X}}_{\text{train}} &= \bm{X}[:,\, : - n_{\text{pred}}]
        \in \mathbb{R}^{3M \times (T - n_{\text{pred}})}\,,\\
    \tilde{\bm{Y}}_{\text{target}} &= \bm{Y}[:,\,  n_{\text{pred}}:]
        \in \mathbb{R}^{2 \times (T - n_{\text{pred}})}\,.
\end{align}
Thus, each column of $\tilde{\bm{Y}}_{\text{target}}$ contains the driver position $n_{\text{pred}}$ steps ahead of the observations in the corresponding column in $\tilde{\bm{X}}_{\text{train}}$.}
We compute $\bm{W}_{\mathrm{out}}$ in the overdetermined linear system
\begin{equation} \label{eq:rc-prediction}
    \bm{W}_{\mathrm{out}} \tilde{\bm{X}}^\mathrm{train} = \tilde{\bm{Y}}_{\mathrm{target}}^{\mathrm{train}}
\end{equation}
in a standard way using Ridge regression (also known as Tikhonov regularization) \new{by minimizing the L2-regularized least squares objective for each dimension independently}.
With the calculated readout weights $\bm{W}_{\mathrm{out}}$ one can predict the future trajectory of a chaotic time series $\bm{Y}_{\mathrm{target}}^{\mathrm{test}}$ (the testing time series) with the same dynamics but different initial conditions (and hence a different trajectory). This can be achieved by injecting $\bm{Y}_{\mathrm{target}}^{\mathrm{test}}$ into the swarm and simulating the driven active matter system for $T$ time steps, collecting new coarse-grained observations $\bm{X}^{\mathrm{test}}$ in each time step, and then computing 
\begin{equation}
    \bm{Y}_{\mathrm{pred}}^{\mathrm{test}} = \bm{W}_{\mathrm{out}} \bm{X}^\mathrm{test}\,. 
\end{equation}
\new{Finally, to assess the predictive performance $P$ of our reservoir and to compare it to previous work \cite{Lymburn2021}, we first obtain the reduced predictions by cutting the last $n_{\text{pred}}$ steps that have no target time series counterpart:
\begin{equation}
    \tilde{\bm{Y}}_{\text{pred}}^{\text{test}} = \bm{Y}_{\text{pred}}^{\text{test}}[:,\, :-n_{\text{pred}}]
        \in \mathbb{R}^{2 \times (T - n_{\text{pred}})}\,.
\end{equation}
We then center both time series around zero by subtracting half of the simulation box size, and compute the previously employed correlation coefficient}
\new{
\begin{equation}
    P
    \;=\;
    \frac{
        \displaystyle \sum_{t=1}^{T}
        \bigl( y_{\mathrm{pred},t,x}^{\mathrm{test}} - \bar{y}_{\mathrm{pred},x} \bigr)
        \bigl( y_{\mathrm{target},t,x}^{\mathrm{test}} - \bar{y}_{\mathrm{target},x} \bigr)
    }{
        \sqrt{\displaystyle \sum_{t=1}^{T}
        \bigl( y_{\mathrm{pred},t,x}^{\mathrm{test}} - \bar{y}_{\mathrm{pred},x} \bigr)^{2}}
        \;\sqrt{\displaystyle \sum_{t=1}^{T}
        \bigl( y_{\mathrm{target},t,x}^{\mathrm{test}} - \bar{y}_{\mathrm{target},x} \bigr)^{2}}
    },
    \label{eq:correlation-coefficient}
\end{equation}
between the first dimension (the $x$ coordinate) of the target time series and the predicted time series, both centered around zero by subtracting half of the simulation box size,
where $
    y_{\mathrm{pred},t,d}^{\mathrm{test}},\;
    y_{\mathrm{target},t,d}^{\mathrm{test}}$
    denote the entries of the (length matched, i.e.\ reduced by $\Delta t_{\text{pred}}$) predicted and the actual time series at a time point $t = 1,\dots,T$, and dimension $d \in \{x,y\}$,
\begin{equation}
        \quad
    \bar{y}_{\mathrm{pred},x} = \frac{1}{T} \sum_{t=1}^{T} y_{\mathrm{pred},t,x}^{\mathrm{test}},
    \quad
    \bar{y}_{\mathrm{target},x} = \frac{1}{T} \sum_{t=1}^{T} y_{\mathrm{target},t,x}^{\mathrm{test}}.
\end{equation}
denote the means over the predicted and target time series, respectively.
A perfect active matter reservoir with a trained readout layer would feature a predictive performance of 1.0, while a prediction that does not correlate with the target time series shows performances around 0.0. Values below but close to 0.0 correspond to small anti-correlations.\\
To further characterize RC performance, we also compute the Pearson correlation coefficient for the $y$ dimension (the second dimension of the chaotic attractors) using the previous equations.
We also compute three other common metrics using the first two dimensions of the chaotic attractors:
the Normalized Root Mean Squared Error by standard deviation (NRMSE),
\begin{align}
    \mathrm{NRMSE}_{d}
    \;&=\;
    \frac{
        \sqrt{\displaystyle \frac{1}{T} \sum_{t=1}^{T}
        \bigl( y_{\mathrm{pred},t,d}^{\mathrm{test}} - y_{\mathrm{target},t,d}^{\mathrm{test}} \bigr)^{2}}
    }{
        \sqrt{\displaystyle \frac{1}{T} \sum_{t=1}^{T}
        \bigl( y_{\mathrm{target},t,d}^{\mathrm{test}} - \mu_{d} \bigr)^{2}}
    },\\
    \mu_{d} &= \frac{1}{T} \sum_{t=1}^{T} y_{\mathrm{target},t,d}^{\mathrm{test}},\\
    \mathrm{NRMSE}
    \;&=\;
    \frac{1}{2}\,\bigl(\mathrm{NRMSE}_{x} + \mathrm{NRMSE}_{y}\bigr),
    \label{eq:nrmse}
\end{align}
the Normalized Mean Squared Error (NMSE) by variance,
\begin{align}
    \mathrm{NMSE}_{d}
    \;&=\;
    \frac{
        \displaystyle \frac{1}{T} \sum_{t=1}^{T}
        \bigl( y_{\mathrm{pred},t,d}^{\mathrm{test}} - y_{\mathrm{target},t,d}^{\mathrm{test}} \bigr)^{2}
    }{
        \displaystyle \frac{1}{T} \sum_{t=1}^{T}
        \bigl( y_{\mathrm{target},t,d}^{\mathrm{test}} - \mu_{d} \bigr)^{2}
    },\\
    \mathrm{NMSE}
    \;&=\;
    \frac{1}{2}\,\bigl(\mathrm{NMSE}_{x} + \mathrm{NMSE}_{y}\bigr),
    \label{eq:nmse}
\end{align}
and the Symmetric Mean Absolute Percentage Error (sMAPE)
\begin{align}
    \mathrm{sMAPE}_{d}
    \;&=\;
    \frac{100\%}{T}
    \sum_{t=1}^{T}
    \frac{
        \bigl| y_{\mathrm{pred},t,d}^{\mathrm{test}}
             - y_{\mathrm{target},t,d}^{\mathrm{test}} \bigr|
    }{
        \bigl| y_{\mathrm{pred},t,d}^{\mathrm{test}} \bigr|
        +
        \bigl| y_{\mathrm{target},t,d}^{\mathrm{test}} \bigr|
    },\\
    \mathrm{sMAPE}
    \;&=\;
    \frac{1}{2}\,\bigl(\mathrm{sMAPE}_{x} + \mathrm{sMAPE}_{y}\bigr)\,.
    \label{eq:smape}
\end{align}
For NRMSE, a value of 0.0 corresponds to a perfect prediction, while a value of 1.0 signifies an error magnitude comparable to the standard deviation of the target. For NMSE, a value of 0.0 corresponds to a perfect prediction, while a value over 1.0 signifies a prediction worse than using the mean of the target as a trivial predictor. For sMAPE, a score of 0\,\% is a perfect prediction, while a score of 100\,\% means that time-averaged the absolute error is as large as the average magnitude of the two values. We use the \texttt{sktime} library \cite{Kiraly2025, Loening2019} to compute these measures.
}

%% file: methods_observables.tex

\subsection{\label{sec:mm-observables}Observables for driven nonequilibrium soft matter systems}

Canonical active matter order parameters comprise polarity $\Phi_P$ and rotation $\Phi_R$ \cite{Lymburn2021, Vicsek1995},
\begin{align}
    \Phi_P & = \frac{1}{N}\left|\sum_i \frac{\bm{v}_i}{\left|\bm{v}_i\right|}\right|\, \text{and} \\
    \Phi_R & =\frac{1}{N}\left|\sum_i \frac{\bm{x}_{C_i} \times \bm{v}_i}{\left|\bm{x}_{C_i} \times \bm{v}_i\right|}\right|\,,
\end{align}
where $\bm{x}_{C_i} = \bm{x}_i - \bm{x}_{CoM}$ is the position of agent $i$ in a center of mass frame $\bm{x}_{CoM}$. They characterize the polar alignment and rotational order of agent velocity vectors, respectively.

Under non-equilibrium conditions, polar or rotational order alone may not suffice to distinguish different phases of matter because it is not a definitive indicator for collective behavior, in general \cite{Attanasi2014a-collective_biology, Attanasi2014b-finite-size-scaling_swarms}.
A more generic observable that detects interactions between particles in the absence of order is based on the most basic form of collectivity: correlations between velocity fluctuations of particles \cite{Attanasi2014_collective_behaviour}.

To introduce a more generic observable for the collective behavior of our driven system, we define the velocity fluctuation of an agent $i$ as
\begin{equation}
    \delta \bm{v}_i=\bm{v}_i - \bm{V} \,,
    \label{eq:velocity-fluctuation}
\end{equation}
which is the difference between the velocity of agent $i$ and the mean velocity 
\begin{equation}
    \bm{V}=\frac{1}{N} \sum_{i=1}^N \bm{v}_i 
\end{equation}
of all $N$ agents.
The velocity fluctuation $\delta \bm{v}_i$ can also be viewed as agent velocity in a center of momentum frame (if all particles have unit mass).
We note that for simplicity we take only translational modes into account, as originally introduced in \refref{\cite{Cavagna2010}}, but one could also incorporate rotational and dilatational (expansive/contractive) modes in $\bm{V}$ \cite{Attanasi2014a-collective_biology}.
To compare velocity fluctuations independently of swarm scaling, we normalize the velocity fluctuations by the standard deviation of the velocity fluctuations as
\begin{equation}
    \delta \bm{\varphi}_i=\frac{\delta \bm{v}_i}{\sqrt{\frac{1}{N} \sum_k \delta \bm{v}_k \cdot \delta \bm{v}_k}} \,.
\end{equation}

To spatially resolve how the change of behavior of an agent $i$ correlates with the change of behavior of another agent $j$ we introduce the connected correlation function
\begin{equation}
    \mathrm{CVC}(r)=\frac{\sum_{i \neq j}^N \delta \bm{\varphi}_i \cdot \delta \bm{\varphi}_j \ \delta\left(r-r_{i j}\right)}{\sum_{i \neq j}^N \delta\left(r-r_{i j}\right)} \,,
    \label{eq:connected_velocity_correlation}
\end{equation}
where $r$ is the radial distance, $r_{ij}$ is the distance between particle $i$ and particle $j$, and $\delta(r-r_{ij})$ is the Dirac delta function ($\delta\left(r-r_{i j}\right) = 1 \text { if } r < r_{i j} < r + dr, \text{ and } 0 \text { otherwise }$).
$\mathrm{CVC}(r)$ is a radially binned function, \ie all particles $j$ are considered that reside within a radial distance $r < r_{ij} < r + dr$.
In natural swarms, particles align their direction of motion with their close neighbors, hence the connected correlation is positive for small distances. For intermediate distances, anti-alignment and negative correlations are observed. For large distances, the connected correlation is close to zero, since there is no correlation between particles that are far apart.
The first root of the connected correlation function is called correlation length $r_0$ and is a measure of the range of interactions between particles.
To compute the spatial range of correlations in a system we consider a quantity related to the space integral of $\mathrm{CVC}(r)$, the cumulative velocity correlation function
\begin{equation}
    \text{CumVC}(r)=\frac{1}{N} \sum_{i \neq j}^N \delta \bm{\varphi}_i \cdot \delta \bm{\varphi}_j \theta\left(r-r_{i j}\right)\,,
\end{equation}
where $\theta$ is the Heaviside step function ($\theta\left(r-r_{i j}\right)= 1$ if $r_{i j} \leq r$, and $0$ otherwise).

$\mathrm{CumVC}(r)$ has its maximum at the radial distance where positive correlations turn into anti-correlations, \ie at the correlation length $r_0$ where $\text{CVC}(r_0) = 0$. This maximum value of $\mathrm{CumVC}(r)$,
\begin{equation}
    \chi \equiv \mathrm{CumVC}\left(r_0\right)\,,
    \label{eq:dynsus}
\end{equation}
is called the dynamical susceptibility $\chi$. For a stationary system, this quantity captures the response of the system to uniform external perturbations \cite{Attanasi2014a-collective_biology}. It can also be interpreted as the total amount of correlation in the swarm \cite{Attanasi2014a-collective_biology} because it captures the intensity of correlations together with its spatial reach. To quantify anti-correlations we measure the anti-correlation strength at the first local minimum of the connected correlation function, $\text{CVC}(r_{min})$. \\

In addition to correlations between particles, we also assess correlations of single-agent dynamics. 
We compute the absolute velocity auto-correlation function
\begin{equation}
    \left| \mathrm{VAC}(\tau) \right| = \left| \frac{1}{N N_{\mathcal{T}}} \sum_{i=1}^N \sum_{j=1}^{N_{\mathcal{T}}} \bm{v}_i(t_{0,j}) \cdot \bm{v}_i(t_{0,j}+\tau) \right|
\end{equation}
averaged over all agents $N$ and all $N_{\mathcal{T}}$ time windows $\mathcal{T}_{t_{0,j}} = [t_{0,j}, t_{0,j} + \tau_\text{max})$ within a total number of simulation steps $T$, for a given time lag $\tau \in \mathcal{T}$. Windows $\mathcal{T}_{t_{0,j}}$ do not overlap and are separated by a random waiting time drawn uniformly from $\mathcal{T}$ each time.\\
We use the mean squared displacement (MSD) of agents after a characteristic time $\tau_c$ as a simple proxy to quantify the strength and duration of intrinsic reservoir dynamics of the soft matter response to the external driving. Agents that are strongly excited and enter trajectories that prevent them from returning to their initial positions will show high MSDs. Stable swarm excitations that enable a fast return to the original configuration show low MSDs.
We define the MSD per agent $i$ after a lag $\tau$ (in steps) as
\begin{widetext}
    \begin{equation}
        \text{MSD}_i(\tau) = \frac{1}{T-\tau} \sum_{t=0}^{T-\tau-1}\left(\bm{x}_i(t+\tau)-\bm{x}_i(t)\right)^2\,,
    \end{equation}
\end{widetext}
where we average over all windows with length $\tau$ within a total number of simulation steps $T$. We use the default MSD implementation in the \texttt{freud} Python library \cite{Ramasubramani2020}.\\
To quantify the intrinsic agent relaxation dynamics -- in the absence of external driving and agent-agent interactions -- we measure both a structural and dynamical quantity. For structural relaxation, we measure the structural excitation 
\begin{equation}
    \Delta r_c (t) = \left\langle \frac{||\bm{x}(t) - \bm{x}_{\text{center}}||}{||\bm{x}(t_0)||} \right\rangle_{N}
    \label{eq:structual_relaxation_observable}
\end{equation}
as radial distance of agents to the center of the simulation box, normalized by the initial agent positions $\bm{x}(t_0)$ and averaged over all $N$ agents.\\
To investigate the dynamical relaxation, we measure the dynamical excitation
\begin{equation}
    \Delta v_s (t) = \left\langle \left( \frac{||\bm{v}(t)|| - s}{s} \right)^2  \right\rangle_{N}\,,
    \label{eq:dynamical_relaxation_observable}
\end{equation}
which is given as the deviation of current agent speeds $||\bm{v}(t)||$ relative to the target agent speed $s$, averaged over all $N$ agents.\\

%% file: methods_simulation_details.tex

\subsection{\label{sec:mm-simulation-details}Simulation and reservoir computing set-up}

Our reservoir computer is based on active matter simulations with $N=200$ particles in a square simulation box with length $l_{\text{box}} = 16.0$ and periodic boundary conditions. The chaotic input trajectory is scaled to fit in a square box of size $l_{\text{box}}^{\text{driver}} = 8.0$ centered in the simulation box. Refer to \axref{subsec:prediction_tasks} for details regarding input trajectory generation and pre-processing using the \texttt{dysts} library \cite{Gilpin2021, Gilpin2023, dysts}. At the beginning of each run, particles are uniformly at random placed on the simulation canvas, with velocity vectors with a length of $\left|\mathbf{v}_i\right| = 1.0$ and uniformly at random directions.

Agent-agent and driver-agent dynamics are governed by the forces described in \sectref{sec:mm-interactions} and \sectref{sec:mm-interactions-driven}. The default simulation parameters correspond to the parameters presented in Fig.\ 7B of \refref{\cite{Lymburn2021}} for the ``critical'' regime: a local alignment force with $r_a = 1.0$ and $K_a = 0.01$; a local repulsion force with $r_r = 1.0$ and $K_r = 2.0$; a global homing force with $K_h = 2.0$; a speed-controller with $s = 10.0$ and $K_{sc} = 2.0$; a local driver-agent repulsion with $r_d = 2.0$ and $K_d = 100.0$; and a sigmoid force clamp with $\alpha = 200.0$ and $\beta = 0.1$.
We note that in \refref{\cite{Lymburn2021}} the speed-controller strength is given as $K_{sc} = 20.0$, but the system in the ``critical'' regime could only be reproduced with $K_{sc} = 2.0$ (see also \vidref{\tabref{tab:supplementary_videos_lymburn_reproductions}}{2}).

We vary the speed-controller parameters $K_{sc}, s \in [10^{-5}; 10^{2}]$ in \sectref{subsec:slow-regime}, \ref{subsec:correlated_current_fluctuations}, \ref{subsec:other-benchmarks} and \ref{subsec:few-particle}; the alignment force parameters $K_{a} \in [10^{-3}; 10^{1}]$ and $r_{a} \in [10^{-1}; 10^{1}]$ for two speed-controller settings $(K_{sc}, s) = (2.0, 10.0)$ and $(0.02069, 0.0483)$ in \sectref{subsec:active-crystal}; the homing force strength $K_{h} \in [10^{-3}; 10^{3}]$ and the speed-controller strength $K_{sc} \in [10^{-5}; 10^{2}]$ for a fixed target agent speed $s = 0.04833$ in \sectref{subsec:homing}. 
These parameter scans are performed with 20 values spaced logarithmically for each parameter and plotted in contour plots using linear interpolation (Gouraud shading). All figures show data using testing run initial conditions if not mentioned otherwise; figures showing undriven steady-state systems show data using training run initial conditions. All other parameters are fixed throughout this work unless stated otherwise. 

Observables are recorded in each time step. Radially binned observables, such as connected and cumulative correlation functions, are recorded using a bin width of $\Delta r_{\text{bin}} = 1.0$. To compute mean squared displacements, we consider windows within the first $T = 1,000$ time steps of a trajectory. 

To extract coarse-grained swarm information from the reservoir, we place $M=200$ Gaussian observation kernels uniformly at random on the simulation canvas. We set the Ridge parameter for Ridge regression to $\lambda_{\text{Ridge}} = 1.0$.

We integrate the equations of motion using a time step of $\Delta t = 0.02$ with an Euler-forward scheme to obtain particle velocities and positions from forces in each time step; every step is recorded. The large step size of 0.02 is numerically feasible because the force clamp (\eqqref{eq:force_wrapper}) ensures that forces cannot become excessively large.

To perform reservoir computing, we simulate our active matter system in two runs: training and testing. For each run, we choose different initial conditions for the chaotic input time series, leading to exponentially different trajectories due to their deterministically chaotic nature. Each run consists of an initial burn-in (equilibration) time of 1,000 steps that are neither recorded nor considered for the readout layer training and 50,000 recorded steps for the main simulation.
After training the readout layer using the Gaussian kernel observations as described in \sectref{sec:mm-observations} and \sectref{sec:mm-training}, we predict $\Delta t_{\text{pred}} = 25$ fixed integration time steps ahead by default at each time point in a non-autoregressive fashion, which corresponds to $\approx 0.45283$ Lyapunov times for the Lorenz attractor. \new{The forecast horizon of the other chaotic attractors matches that of the Lorenz-63 system in terms of Lyapunov times.}

%% file: results_speed_controller.tex

\subsection{\label{subsec:slow-regime} Around the critical damping regime: optimal performance}

\begin{figure*}[htbp]
    \centering
    \includegraphics{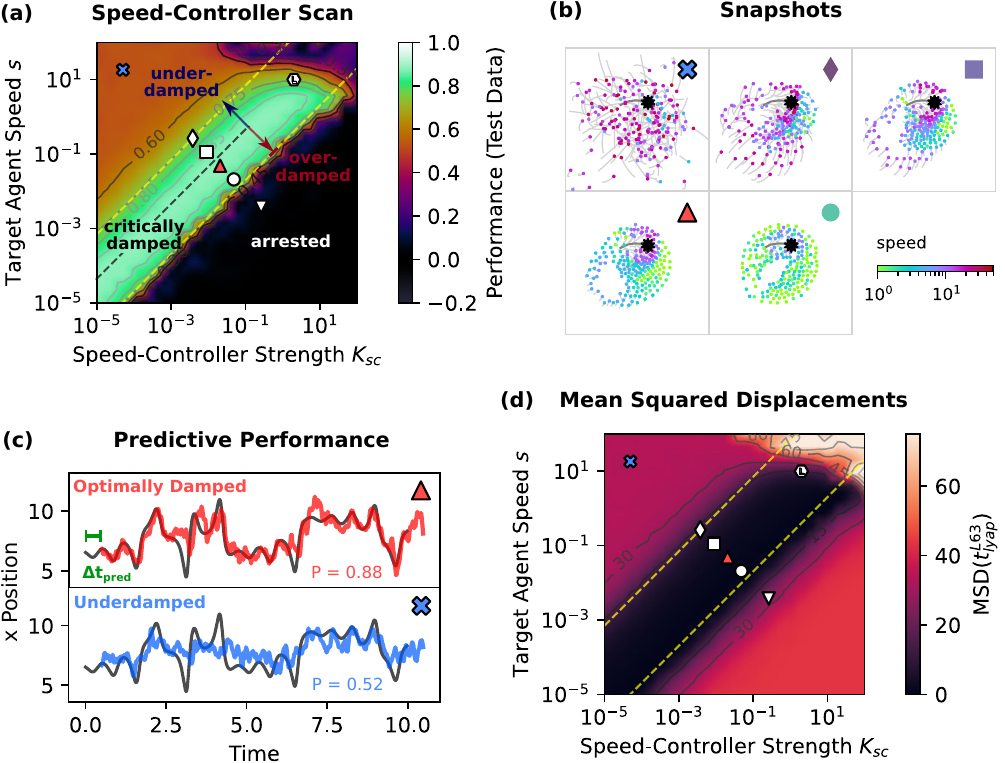}
    
    \caption{
        \textbf{A critically damped dynamical regime shows optimal predictive performance for active matter reservoir computing.} 
        (a) Varying the parameters of the speed-controlling force reveals regimes with stark differences in predictive performance, corresponding to different agent dynamics. The heatmap shows varied target agent speeds $s \in [10^{-5}; 10^{2}]$ that a particle will accelerate or decelerate towards (see \eqqref{eq:speed-controller}), and the strength coefficient $K_{sc} \in [10^{-5}; 10^{2}]$ of this force. 
        In the center diagonal enclosed by the dashed yellow lines, there is a critically damped regime with suppressed agent excitations and a spatially well-confined collective swarm (circle, pyramid, square symbols).  Moving towards lower speed-controller strengths and higher target agent speeds yields a response that appears much more disordered than coherent-collective (diamond, star symbols). Higher speed-controller strengths and lower target agent speeds lead to arrested particle motion (nabla symbol), which becomes overdamped when choosing a smaller integration time step (see \figref{fig:Lymburn_critical_scans_speed_controller_smaller_integration_time_step}). The hexagon/``L'' denotes the optimal dynamical regime presented in \refref{\cite{Lymburn2021}} (see \vidref{\tabref{tab:supplementary_videos_lymburn_reproductions}}{2} for a snapshot and visualization).
        (b) Snapshots of swarms in different dynamical regimes marked as symbols in sub-figure (a); color indicates agent speed. Refer to \tabref{tab:supplementary_videos_speed_controller} for corresponding videos.
    (c) Predictive performances $P$ are higher in the critically damped regime (pyramid symbol, $P = 0.88$, 
    \new{$P_y = 0.79$, 
    $\mathrm{NRMSE}= 0.54$, 
    $\mathrm{NMSE}=0.29$, 
    $\mathrm{sMAPE}= 6.4\,\%$}
    ) than in the underdamped regime (cross symbol, $P = 0.52$, 
    \new{$P_y = 0.40$, 
    $\mathrm{NRMSE}= 0.88$, 
    $\mathrm{NMSE}=0.78$, 
    $\mathrm{sMAPE}= 12\,\%$
    }). \new{$P \equiv P_x$} is defined as the correlation coefficient specified in \eqqref{eq:correlation-coefficient} between the \new{first dimension of the} actual and the predicted time series. The gray curve shows the \new{first dimension of the} actual time series, while the colored curves show predictions $\Delta t = 0.5$ time units ahead made with the active matter reservoir at each time point.
        (d) Agent-averaged mean squared displacements (MSDs) after the Lyapunov time of the Lorenz-63 attractor $t^{\text{L63}}_{\text{lyap}} \approx 1.1$ \cite{Viswanath1998} for different speed-controller parameter combinations. Low MSDs separate the underdamped from the near-critical (and arrested) regime, and correlate with high predictive performances. 
        They are thus reliable proxies for consistent, controlled agent responses to changing external stimuli. Parameter values $(K_{sc}, s)$ for symbols: cross: (0.00005, 18.32981); diamond: (0.00379, 0.26367); square: (0.00886, 0.11288); pyramid: (0.02069, 0.04833); circle: (0.04833, 0.02069); nabla: (0.26367, 0.00379).
    }
    \label{fig:Lymburn_critical_speed_controller_scan_larger}
\end{figure*}

So far, it has been assumed that active matter reservoirs show optimal performance in a ``critical'' regime, around a fluid-to-gas-like phase transition \cite{Lymburn2021}, supporting the hypothesis of information processing being optimal around a phase transition \cite{Langton1990}.
Here we present a novel dynamical regime for improved reservoir computing with active matter: the critically damped regime.
We uncover this regime by varying the parameters of the speed-controlling force $\bm{F}_{sc}$: the target agent speed $s$ and the strength $K_{sc}$ (see \eqqref{eq:speed-controller}). 
\figref[a]{fig:Lymburn_critical_speed_controller_scan_larger} shows how well different active matter systems at different $(K_{sc}, s)$ parameter combinations perform at predicting a chaotic input time series in a reservoir computing framework. We observe a broad region with high predictive performances which we define with yellow dashed lines ($s = r (K_{sc})^{m}$ with power $m = 1$ and coefficients $r_{AB} \approx 0.2$ and $r_{BC} \approx 70$). The optimal ratio $r$ that delivers performances around $P = 0.884$ is located at $K_{sc} = 10^{-5}$ and $s = 5.0 \cdot 10^{-5}$.
This is a significant increase in performance from the previously reported optimal regime \cite{Lymburn2021} (reproduced as $P = 0.719$) indicated by the hexagon/``L'' symbol in \figref[a]{fig:Lymburn_critical_speed_controller_scan_larger}. Videos of the driven active matter systems (see \axref{ax:supplementary_videos}, \tabref{tab:supplementary_videos_speed_controller}) reveal that, upon driver excitation, particles do not embark on long trajectories or oscillate across the simulation box before returning to their steady-state positions. Instead, the speed-controller setting ensures that agents swiftly return to the box center. This is why we call this regime ``critically damped''. The quick return and the higher order in this regime can also be observed from the snapshots in \figref[b]{fig:Lymburn_critical_speed_controller_scan_larger} (square, pyramid, circle symbols). The swarm maintains a coherent droplet shape with low mean speeds (see Supplementary \figref{fig:Lymburn_critical_scans_speed_controller_observable_heatmaps_mean_speed}), even if perturbed by the driver.

For $s \gg K_{sc}$, in the upper-left corner of \figref[a]{fig:Lymburn_critical_speed_controller_scan_larger}, agent speeds are only weakly regulated by the speed-controlling force. The overall response is much more disordered at the local particle level, which can be seen from the snapshot in \figref[a]{fig:Lymburn_critical_speed_controller_scan_larger} (cross symbol). Because agent momenta do not relax quickly, we call this regime underdamped. This notion becomes even more evident in the corresponding supplementary video. The predictive performance is on the order of $P = 0.60$ and lower.
\figref[c]{fig:Lymburn_critical_speed_controller_scan_larger} compares the strong differences in predictive performance between the actual and predicted time series in the near-critically damped ($P = 0.88$) and the underdamped ($P = 0.52$) regime.

The critically damped regime breaks down for high target agent speeds $s \gtrapprox 5.0$ and strengths $K_{sc} \gtrapprox 3.0$ (\figref[a,d]{fig:Lymburn_critical_speed_controller_scan_larger}, upper right-hand corner). Mean agent speeds and MSDs increase in this regime and agents are spatially less confined (see also \vidref{\tabref{tab:supplementary_videos_lymburn_reproductions}}{2}). The previously reported ``critical'' regime (hexagon/``L'' symbol, referring to the fluid-droplet-to-gas phase transition in \refref{\cite{Lymburn2021}}), is situated here.

Below the lower yellow dashed line, agent dynamics begin to be dominated by the speed-controlling force. Here, the agents almost rest apart from local vibrational motion. 
For higher $K_{sc}$ and lower $s$ values (nabla symbol), agents perform alternating steps, where one direction (forward) is slightly stronger than the other (backwards), yielding an overall forward movement. Because agents do not interact strongly with each other or the driver, and instead oscillate alone, we call this regime ``arrested''. The performance in \figref[a]{fig:Lymburn_critical_speed_controller_scan_larger} shows a sharp drop here, originating from the nearly complete absence of agent-driver interaction. Integrating the system with a smaller time step $\Delta t = 0.002$ leads to smoother dynamics (see \vidref{\tabref{tab:supplementary_videos_speed_controller}}{7}) and removes the sharp performance drop for a local strip around the lower yellow boundary (see Supplementary \figref{fig:Lymburn_critical_scans_speed_controller_smaller_integration_time_step}). Thus, the numerical integration time step plays a key role in this regime and its existence.
Without a driver, both the underdamped and the critically damped regimes feature highly ordered steady states (see \tabref{tab:supplementary_videos_no_driver}).\\

\new{Using other performance measures in \figref{fig:speed-controller-more-metrics} (correlation coefficient for the $y$ dimension $P_y$, $\mathrm{NRMSE}$, $\mathrm{NMSE}$, $\mathrm{sMAPE}$) we confirm that the overall performance landscapes are similar, with the optimal performance remaining in the near-critically damped regime (pyramid symbol). These complementary measures capture accuracies of magnitude better than the correlation coefficient and allow comparisons across reservoir computing methods and datasets. They indicate that in the optimal dynamical regime, there are still discrepancies between the features of the predicted time series and the target time series, as shown in \figref[c]{fig:Lymburn_critical_speed_controller_scan_larger}, leaving potential for further improvements.}
\new{We compare these performances to a standard Echo State Network (ESN) using a comparable number of $N=600$ neurons in \axref{subsec:esn_comparison}. We find that our near-critically damped active matter reservoir has a similar (5.5\,\% higher) predictive performance compared to the (hyperparameter optimized) ESN, suggesting that optimized active matter reservoir computing could be an interesting candidate for \textit{in materio} computing.}

One simple, microscopic physical indicator of the difference between regimes is the mean squared displacement (MSD) after a delay time $\tau$. We choose as a suitable delay timescale the characteristic Lyapunov time of the driver, the Lorenz-63 system. Individual agent MSDs vary strongly even after multiple Lyapunov times in the underdamped regime, while MSDs remain low in the critically damped regime (see Supplementary \figref[a]{fig:msds}). The agent-averaged MSDs presented in \figref[d]{fig:Lymburn_critical_speed_controller_scan_larger} are consistently higher in the underdamped regime compared with the critically damped regime. This low MSD region correlates with the high-performance prediction region in \figref[a]{fig:Lymburn_critical_speed_controller_scan_larger}.
We also observe a stronger consistency of the types of responses visible in the critically damped regime in \vidref{\tabref{tab:supplementary_videos_speed_controller}}{1-3}, while this consistency is not clear in the underdamped regime in \vidref{\tabref{tab:supplementary_videos_speed_controller}}{5}.
We hypothesize that underdamped system responses tend to generate more ``different'' spatio-temporal swarm patterns for similar driver trajectory patterns. This possibly impairs a key reservoir quality, the \new{\textit{consistency property} (the reliability of the input to reservoir state mapping).}

\subsection{Signatures from microscopic dynamics with and without driving}
\label{subsec:damping_and_vac}

\begin{figure*}[htbp]
    \centering
    \includegraphics{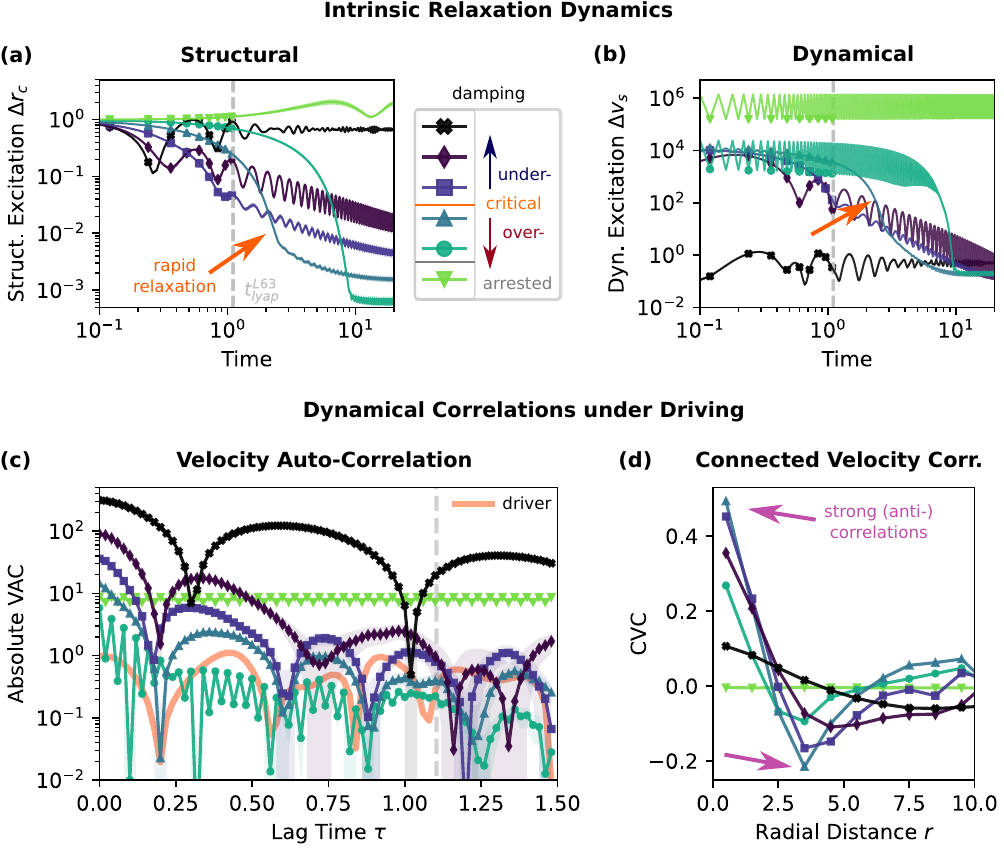}
    \caption{\textbf{Characterizing intrinsic microscopic relaxation dynamics (a,b) as well as dynamical correlations under driving (c,d).} Symbols correspond to parameter combinations changing speed-controller settings $(K_{sc}, s)$ (and resulting damping dynamics) as shown in \figref[]{fig:Lymburn_critical_speed_controller_scan_larger}. (a,b) Intrinsic relaxation dynamics for a system without external driving in the absence of any collective effects (without agent-agent interactions, only subject to the homing force, the speed controller, and the sigmoid force wrapper described in \sectref{sec:mm-interactions}), the time step used is $\Delta t = 0.02$. The system relaxes from an initial, uniformly random distribution to a steady state, analogous to a damped harmonic oscillator. (a)  Structural relaxation of a population of non-interacting agents measured as the mean radial distance to the center of the simulation box, normalized by the initial radial distance at time $t_0 = 0$. Sustained oscillations around the center of the simulation box (underdamped regime, cross symbol) become suppressed when moving towards higher speed-controller strengths $K_{sc}$ and lower target agent speeds $s$ (diamond symbol). Critical damping is situated between the square and pyramid symbols, at the threshold where a single oscillation disappears during the initial phase of the relaxation. (b) Corresponding dynamical relaxation: the relative deviation of agent speed from the set target agent speed $s$. Error bands indicate the standard error of the mean over $N=200$ agent trajectories. For reference, the Lyapunov timescale of the Lorenz-63 driver is shown in the time plots (a)-(c), \ie,  $t^{\text{L63}}_{\text{lyap}} \approx 1.1$ \cite{Viswanath1998} as a gray dashed line. 
    (c,d) Dynamical correlations under driving for driven agents subject to the full active matter model described in \sectref{sec:mm-interactions}. (c) The absolute value of the velocity auto-correlations for selected damping dynamics. The optimal dynamics (around pyramid) entail unique features described in the main text. (d) The connected velocity correlation function (CVC) for different parameter combinations. The optimally damped regime shows the strongest connected velocity fluctuation correlations and anti-correlations, measured spatially via a radial distance around each particle.
    }
    \label{fig:damping}
\end{figure*}

\subsubsection{Intrinsic damping dynamics: No driving}

The underdamped, critically damped, and arrested dynamical regimes of the active matter substrates render visibly distinct RC performances. These terms are inspired by control theory \cite{Flower2002} and by observing the damping behavior of individual agents (see \figref{fig:damping}). Intriguingly, one can identify these different regimes without the influence of an external driver, \ie, without an explicit injection of information. 
Even though we exclude the initial transient (``burn-in'') phase from training and inference in \rc, it is highly informative about dynamics and associated computational properties: In \figref[a,b]{fig:damping} the relaxation behavior of a system without external driving is studied and categorized -- in the absence of any collective effects (without agent-agent interactions, only subject to the homing force, the speed controller, and the sigmoid force wrapper described in \sectref{sec:mm-interactions}) (refer to \tabref{tab:supplementary_videos_damping_non_interacting} for the corresponding videos). These results show subtle features that are more visible than in the case of including interactions, displayed in the  Supplementary \figref{fig:damping_variations}. An ensemble of active agents relaxes from an initial uniformly random distribution to a steady-state configuration, similar to a damped harmonic oscillator. In sub-figure (a), structural relaxation is made visible via the mean radial distance to the center of the simulation box, normalized by the initial radial distance at $t_0 = 0.0$.  Sustained oscillations around the center of the simulation box (underdamped regime, cross symbol) become suppressed when moving towards higher speed-controller strengths $K_{sc}$ and lower target agent speeds $s$, highlighted by the diamond symbol.

\figref[b]{fig:damping} shows the intrinsic microscopic relaxation dynamics as the relative deviation of agent speed from the set target agent speed from the initial condition. Seen in both (a) and (b), critical damping is situated between the square and pyramid symbols, at the threshold where a single oscillation disappears during the initial phase of the relaxation, which becomes exponential for \emph{both} the structure and dynamics at the square symbol. Further, the microscopic relaxation dynamics around the diamond symbol has an appearance of a power-law regime with superimposed oscillations. The oscillations decay and the power-law duration drastically shortens when introducing particle interactions \figref[d]{fig:damping_variations}). In (b), the power-law(-like) regime is retained for the square and pyramid symbols (nearly critically damped), and disappears for the circle symbol (overdamped). Small oscillations persist at later phases at the steady state. We refer to the crossover case near critical damping (pyramid symbol) as the ``optimal  damping'' regime for active matter \rc \figref{fig:Lymburn_critical_speed_controller_scan_larger}; it is the first overdamped parameter combination below critical damping.

For the optimal damping case (pyramid symbol), there is a rapid relaxation to a low value in the structure (a), while in the dynamics, there is an intermediate stage exhibited that appears power-law-like. Intriguingly, for both cases of pyramid and circle (more overdamped), the steady state values in dynamics are reached at the same time and with the same value.  Yet, this is an effect of the lack of particle interactions: In the case of interacting particles, shown in the Supplementary \figref[d,f]{fig:damping_variations}, the optimally damped system (pyramid) indeed converges earlier and more efficiently in dynamics than the case of the circle symbol. The combination of intrinsic exponential decay and a power-law-like regime, hence, seems to have some utility for relaxing the dynamics (velocities) more effectively. It could be a dynamical feature enabling favorable properties of the system, discussed later in \sectref{sec:discussion}. Refer to \tabref{tab:supplementary_videos_damping_interacting} for the corresponding videos for the interacting agents case.

For the default time step of $\Delta t = 0.02$ used here, agents at the nabla symbol parameter combination show arrested behavior. The mean normalized radial distances in (a) go above 1.0 because of periodic boundary conditions and the dynamical excitation in (b) attains a high value, an effect that disappears with a smaller timestep of $\Delta t = 0.002$ (refer to \figref[a,b]{fig:damping_variations}). In the latter, the onset of overdamping is shifted, which could explain how the optimal computing regime is effectively widened in \figref{fig:Lymburn_critical_scans_speed_controller_smaller_integration_time_step}.

\subsubsection{Correlative dynamics under driving}

We further characterize the system dynamics -- now under the same external chaotic driving as in the RC setup and with full interactions described in \figref[c,d]{fig:damping} -- in terms of spatial and temporal correlations of agent velocities, the velocity autocorrelation function (VAC), and the connected velocity correlation function (CVC).
We take direct inspiration from the physics of metastable fluids and glassy systems, where one way of understanding dynamics is the time-autocorrelation functions of particle mobilities \cite{Williams2006,vanMegen2006,Williams2006b,Nuevo1998}. Velocities or related current autocorrelations render information on general transport and response properties of fluids \cite{Palmer1994,Zwanzig1970}.

The velocity autocorrelation (VAC) tells how a system's microscopic dynamics, on average, relaxes; it has information on the second time derivative of basic structural scattering functions, \ie, the self-intermediate scattering function or the mean square displacements \cite{HansenMcDonald-book}. It is generally suitable for studying systems out of equilibrium. It requires that long-time stationarity or quasi-repeatability of the dynamics persists, as it accumulates lag-time information over the absolute course of the forward evolution. Here, all data is collected after an initial ``burn-in'' period where the swarm system gets ``warmed up'' under driving. Generally, the VAC can reveal signs of the heterogeneous dynamics visible in these systems, multiple processes responsible for relaxation, as well as signatures of coherent and incoherent motion, and provide a measure of the \new{(physical)} memory of a disturbance in a fluid  \cite{Williams2006,vanMegen2006,Williams2006b,Nuevo1998}. 

In \figref[c]{fig:damping}, the absolute value of the velocity auto-correlations is plotted for selected variations of the speed-controller parameters (symbols) as shown in \figref[a]{fig:Lymburn_critical_speed_controller_scan_larger}. (Here, oscillations are turned into positive peaks with steep zero-crossings on the long-scale; the first peak always corresponds to a negative correlation, and positive peaks alternate thereafter.)
The optimal dynamics (around the pyramid symbol) entail initial short-time viscous or overdamped motion (an initial exponential decay). This is followed by harmonic-like signals with rounded peaks, exhibiting a shorter frequency closely matching that of the external driving force (the time of the first zero crossing of the VAC). Intriguingly, a near-coincidence of the driver's Lyapunov timescale, $t^{\text{L63}}_{\text{lyap}} \approx 1.1$ \cite{Viswanath1998}, with the minimum of the second (negative) peak. 

The strongly underdamped system (black cross) represents one (extreme) type of dynamics we encountered in \figref[a]{fig:Lymburn_critical_speed_controller_scan_larger}.
What first strikes the eye is the slow decay of its envelope in the absolute VAC function (cross symbol), in \figref[a]{fig:speed_controller_fluctuations} -- a first agreement with our classification that the motion is underdamped, too, under chaotic driving. Moreover, it showcases harmonic-like motion with very long characteristic time scales: The successive peaks display a rounded profile, and the approximate wavelength (the distances between the sharp minima of the absolute VAC) is comparatively large, fitting more than half of a Lyapunov timescale indicated by the gray dotted line.
As the first peak corresponds to a minimum in the VAC, partially coherent anti-correlated motion persists during the bulk of the characteristic chaotic timescale of the Lorenz-63 driver.
There is a delay in the first zero crossing of the VAC -- the first onset of anticorrelated velocities, which occurs later than for the other cases with more damping.
The initial slope near $\tau=0$ is nearly flat and lacks a clear decay, indicating nearly ballistic or inertial-like motion at the shortest timescales--which can be visually confirmed as shown in
\vidref{\tabref{tab:supplementary_videos_speed_controller}}{5}. 
The driven system in the underdamped regime exhibits slow expansive and contractive pulsations, which might generate a major part of these oscillations in the VAC (see the corresponding video).
Further, the VAC for underdamped dynamics lacks a clear signature of separated dissipative (incoherent) dynamics: There is no sign of an exponential decay, \ie, viscous motion at short time-scales, which we do find for the overdamped dynamical regime.
All these features together effectively can point to noticeable memory effects in the system, which would \new{likely} impair a key reservoir computing quality, the \textit{fading memory property}. The system state should only depend on recent input signals and not on past inputs.

\new{The physical memory traceable in time delayed quantities like the VAC is very difficult to compare directly to the short-term memory capacity (see \axref{subsec:memcap}), a standard measure for RC. Reasons are likely complex, but the kind of driving patterns (randomness versus chaotic) may be a key point. In any case, the measured memory capacity seems to be sensitive to the dynamical regimes -- i.e., is maximal in the transition regime from underdamped to overdamped dynamics (diamond symbol), and remains high for the near-critically damped regime (pyramid symbol). Our memory capacity analysis suggests that the near-critically damped regime possesses a reasonable amount of memory -- albeit not the highest found among different speed-controller settings -- and likely possesses other favourable properties such as non-linearity or consistency to obtain optimal performance in this regime.}

Moving towards higher damping strengths, we observe much shorter timescales between extrema or zero crossings (i.e., shorter oscillations in the VAC). The optimally damped regime (pyramid symbol), for example, displays an early first minimum and overall three minima in the absolute VAC within the Lyapunov time of the chaotic driver, compared to only two minima in the underdamped regime. Moreover, the peaks are nearly paraboloid-shaped for the former. 
Overall, around optimal damping, a diversity of discernible types of dynamics appears (\ie, short-time exponential relaxation and harmonic-like oscillations). 
Note here that at the crossover (green circle) from the critically damped to the arrested regime, there are additional high-frequency oscillations in the VAC due to its vicinity to the arrested regime (nabla symbol, light green), where particles undergo arrested, back-and-forth oscillatory motion.
 
Just above critical damping (at purple square), the absolute VAC begins to have overlapping features with that of the driver's VAC (orange line), in particular, having the same first zero-crossing and mimicking the approximate frequency of the following two oscillations, to just before the Lyapunov time (there, the statistics are sensitive to sampling errors and a clear statement cannot be made). The physical system may be synchronizing some of its mechanisms to the driver  \cite{Pecora1991}.

Another important quantity that expresses spatial correlations of dynamical fluctuations within the agent system is the CVC (\eqqref{eq:connected_velocity_correlation}) \cite{Attanasi2014a-collective_biology}. It measures how strongly a change in the velocity of one agent correlates with a change in the velocity of neighboring agents at a relative radial distance $r$.
\figref[d]{fig:damping} shows that the critically damped regime (between square and pyramid symbols) is characterized by the strongest velocity correlations with next neighbors and the strongest anti-correlations at a radial distance of about $ r \approx 3.5$. We also note the short correlation length $r_0$ (the first zero crossing of the CVC). A stronger initial decay observed for optimal damping means that fluctuations in velocity rapidly turn from positive into negative correlations. 

%% file: results_phenomenology_overdamped.tex

\subsection{\label{subsec:rationale_RC_critically damped}Emergent mechanisms in the optimal regime}

\begin{figure*}[htbp]
    \centering
    \includegraphics{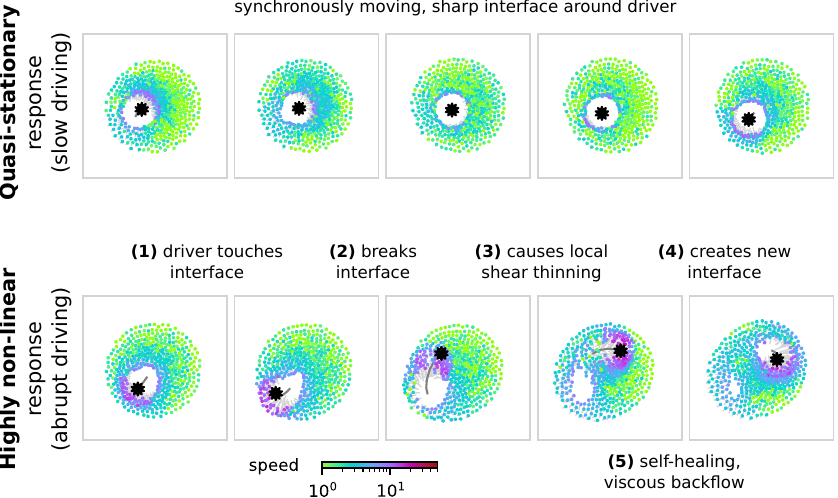}
    \caption{
        \textbf{Phenomenology of optimally damped soft matter systems under driving.} Displayed are the response dynamics of $N=500$ agents under quasi-stationary and abrupt driving, which are both part of Lorenz-63 dynamics. 
        Upper panel: Under adiabatic, quasi-stationary driving, agents form a stable interface around the driver, established by the global attractive force to the center of the simulation box (homing force) and the local repulsive force from the driver.
        Lower panel: Abrupt driving causes the agent-driver interface to break. The driver dashes through the swarm, causing local shear thinning. A new interface forms once the driver slows down. Self-healing through viscous backflow converts the former interface to bulk.
        The driver tail length corresponds to five integration time steps ($5 \Delta t = 0.1$), which is also the time between each snapshot. The snapshots were taken at times $t_i = \{1.0, \dots, 1.4\}$ (quasi-stationary response) and $t_i = \{3.0, \dots, 3.4\}$ (highly non-linear response); the Lorenz-63 driving protocol is the same as shown in \figref[c]{fig:Lymburn_critical_speed_controller_scan_larger}.
        Simulation parameters correspond to those marked by the pyramid symbol in \figref[a]{fig:Lymburn_critical_speed_controller_scan_larger} ($K_{sc} = 0.02069$, $s = 0.04833$).
        Refer to \vidref{\tabref{tab:supplementary_video_phenomenology}}{1} for the corresponding video.
    }
    \label{fig:phenomenology}
\end{figure*}

Here, we visually investigate the information processing mechanisms behind \rc in the optimal (nearly critically damped) regime. 
For the optimal regime under Lorenz-63 driving, two alternating dynamical states are consistently observed: In \figref{fig:phenomenology}, we observe a quasi-stationary response state. The system forms a moving interface synchronous with the slow driver, enabled largely by the overdamped dynamics at short time scales, and marginally showing subtle long-wavelength oscillations in structure and velocities. Upon rapid changes in the driver's dynamics, which occur around the transitions between the two attractors, synchronization breaks and a multi-step feedback cycle are triggered. This cycle amplifies signals of the rupture, converts them into different visible mechanisms, and eventually restores the quasi-stationary, driver-synchronized state. Note that we used a larger system ($N=500$) for visual purposes, which showed very similar phenomenology qualitatively and quantitatively to the smaller default system ($N=200$), even though the local density is slightly higher. 
 
\textit{Quasi-stationary response:}
When the driver moves slowly (in \figref{fig:phenomenology}, upper panel), an interface with marked surface tension is formed. This interface creates an exclusion zone (a vacuum) around the driver. The interface moves synchronously with the driver and remains stable as long as the driver moves at slow enough speeds. This ability of the system to synchronize an interface to a driver even as the latter moves at intermediate speeds might be enabled by the active propulsion. Notably, long-range wave motion can be seen in the videos colored by speed (\vidref{\tabref{tab:supplementary_video_phenomenology}}{1}), or in the videos of velocity fluctuations (\vidref{\tabref{tab:supplementary_videos_speed_controller_velocity_fluctuations}}{2}). We hypothesize that the localization of the driver position with an interface could constitute a main beneficial collective mechanism for the high reservoir computing performance. 

\textit{Highly nonlinear response:}
When there is abrupt driver motion (\figref{fig:phenomenology}, lower panel), the driver first touches the interface, whereby agents experience a strong local repulsion force. The whole system experiences a sudden perturbation, showing oscillatory behavior. As the interface breaks, the driver moves faster than agents may coherently respond, causing local shear thinning. A corridor briefly forms through the bulk of the agents, followed by self-healing of the old interface via viscous flow.
Once the driver slows down at a spatially separated position, the local shear thinning caused by the driver in the bulk allows for the efficient formation of the next quasi-stationary interface. The end of this cycle effectively erases the memory effects of the most recent rupture.

%% file: results_fluctuations.tex

\subsection{\label{subsec:correlated_current_fluctuations}Correlative spatiotemporal velocity fluctuations indicate high performance regimes}

\begin{figure*}[htbp]
    \centering
    \includegraphics{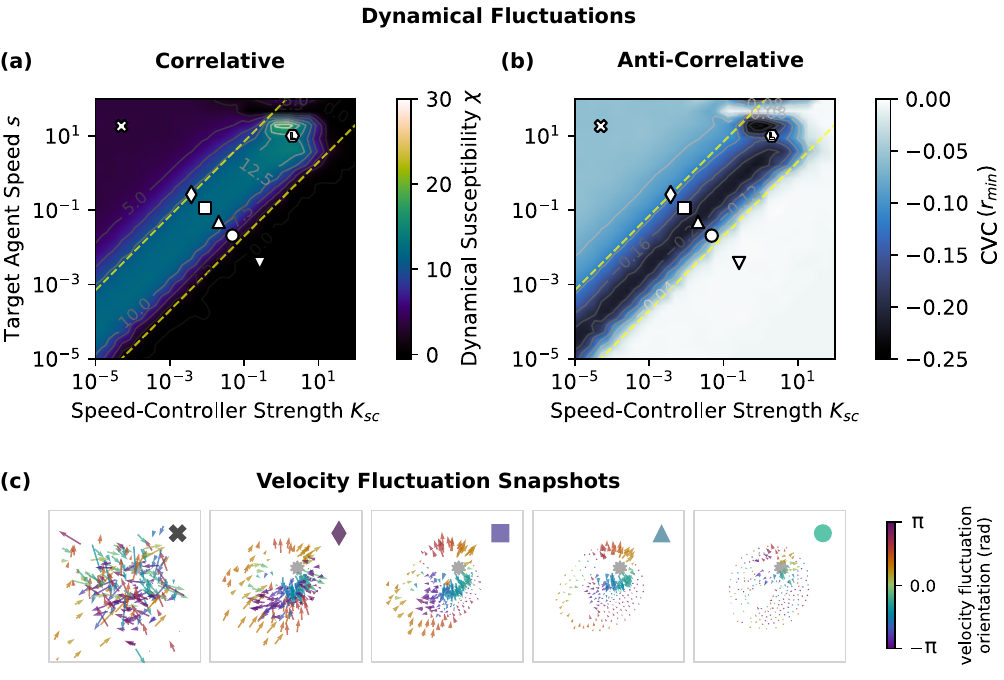}
    \caption{
        \textbf{The near-critically damped regime shows strong correlations and anti-correlations of agent velocity fluctuations.}.
        (a) The dynamical susceptibility $\chi$ (\eqqref{eq:dynsus}) captures the intensity and spatial reach between velocity fluctuation correlations (\eqqref{eq:velocity-fluctuation}, deviations of individual agent velocities from the system's center of mass velocity). The high predictive performance regime found in \figref[a]{fig:Lymburn_critical_speed_controller_scan_larger} correlates with the $\chi > 5.0$ region.
        (b) The anti-correlative dynamical response $\text{CVC}(r_{\text{min}})$, the minimum of the connected velocity correlation function (\eqqref{eq:connected_velocity_correlation}) shown in \figref[d]{fig:damping}), quantifies the maximum extent of anti-correlations between agent velocity fluctuations and correlates with the predictive performance \figref[a]{fig:Lymburn_critical_speed_controller_scan_larger}. The strongest signal is observed just below critical damping at the pyramid symbol. Extrema around the ``L'' symbol could stem from strong rotational modes that were not considered for velocity fluctuation computations. (c) Visualizations of velocity fluctuations for different parameter combinations (symbols) in the speed-controller parameter scans of (a,b). Arrow length indicates strength, and color indicates orientation of velocity fluctuations.  
        Refer to \tabref{tab:supplementary_videos_speed_controller_velocity_fluctuations} for corresponding videos.
    }
    \label{fig:speed_controller_fluctuations}
\end{figure*}

The total amount of correlation in the system up to the correlation length $r_0$ (at the first root of the CVC, until anti-correlation begins to dominate) is called the dynamical susceptibility $\chi$. It can be viewed as the susceptibility of our active matter system to the external driving signal \cite{Attanasi2014a-collective_biology}. In \figref[c]{fig:speed_controller_fluctuations} we find that $\chi$ is particularly high in the critically damped regime. The dynamical susceptibility remains high within the yellow dashed boundaries that demarcate the high predictive performance regime. Intuitively, a high susceptibility to the external driving signal enables the soft matter system to better process this signal and convert it into spatio-temporal patterns.

A different viewpoint is considering the strength of anti-correlations at the first minimum (at $r_{min}$) of the CVC in \figref[b]{fig:speed_controller_fluctuations}. We observe that at the optimal parameter combination (pyramid symbol), the greatest degree of anti-correlative strength is reached. It offers a complementary indicator for high predictive performance compared to our calculated dynamical susceptibility, particularly at the crossover region between underdamped/ overdamped and optimally damped regimes (the yellow dashed diagonals), respectively.
At the upper-right-hand corner of the diagram, we observe extrema in $\chi$ and $\text{CVC}(r_{\text{min}})$, around the ``L''/hexagon symbol. Strong rotational modes (see Supplementary \figref[b]{fig:Lymburn_critical_speed_controller_scan_heatmap_canoncial_order_parameters}) that occur in these regimes and that were not accounted for when computing velocity fluctuations could explain these extrema.

Examining the velocity fluctuation snapshots in \figref[c]{fig:speed_controller_fluctuations} shows that in the underdamped regime (cross symbol), the amount of correlation among velocity fluctuations -- in their spatial distribution -- is low.
As the speed-controlling force increases (diamond symbol), the velocity fluctuations become more correlated, indicated by a smoother color gradient of the arrows -- a more gradual change of orientation of neighboring velocity fluctuations. 
Only near the optimal and overdamped cases, represented by the pyramid and circle symbols, do we visually recognize the localization of prominent anti-correlated velocity fluctuations. Here, those agents that are close to the source of the disturbance, the driver, experience strong velocity fluctuations. The remainder of the swarm features velocity fluctuations on a much smaller scale, creating a neutral background for a sharp contrast of velocity fluctuations localized near the driver. In contrast, in the underdamped regime (cross symbol), driver-induced velocity fluctuations cannot be distinguished as easily from intrinsic velocity fluctuations of the system. This highlights the favorability of the following duality for active matter reservoir computing: a well-defined swarm ``ground state'' and local, driver-induced agent excitations. This finding solidifies Lymburn \etal's observation that there must be a functional dependency of the active matter response on the input signal \cite{Lymburn2021} -- which is here stronger in the critically damped regime compared to the underdamped regime.

In general, correlated velocity fluctuations could form in several ways, such as the propagation of a wavefront or pulse. These are tied to basic viscous and elastic, or mechanical properties of the fluid. The CVC calculates correlations at the same time-points just like the dynamic susceptibilities do. The anti-correlative behavior arising from wavefronts or pulses forming around the driver's source is probably the strongest contributor to the full signal, as seen also in the visuals of the velocity fluctuations (\figref[c]{fig:speed_controller_fluctuations}). In the absolute VAC's negative-valued peaks in \figref[c]{fig:damping}, incoherent contributions manifesting at diverse timescales can broaden or distort the shape \figref[c]{fig:damping}. Depending on the types of interactions allowed, these correlations can also arise differently, though \cite{Attanasi2014_collective_behaviour}.

%% file: results_benchmarks.tex

\subsection{\label{subsec:other-benchmarks}Robust optimality across various classes of chaotic driving}

\begin{figure*}[htbp]
     \centering
     \includegraphics{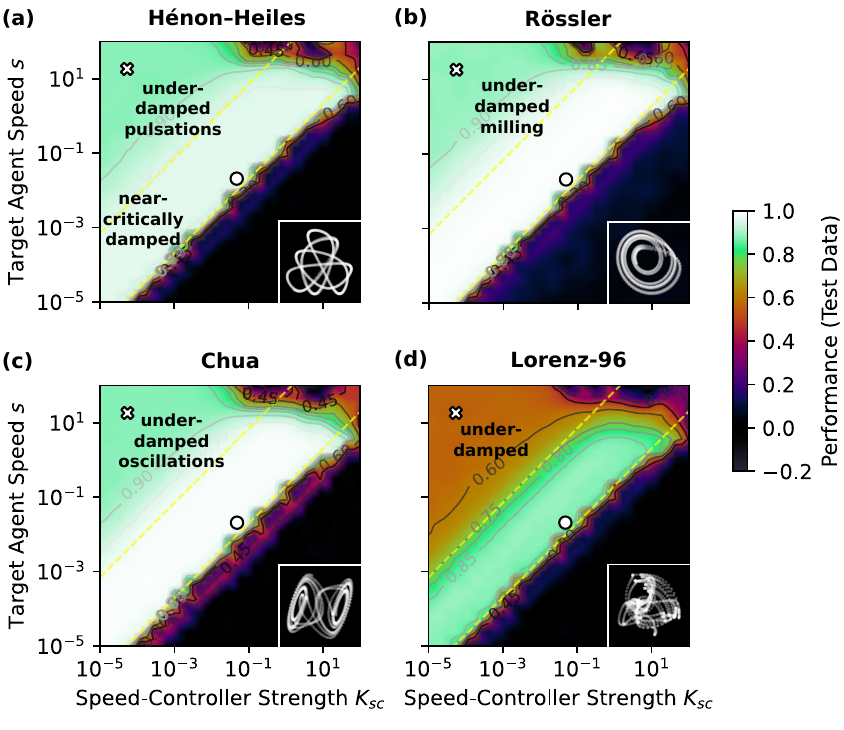}
     \caption{\label{fig:benchmarks}\textbf{The near-critically damped regime yields the highest predictive performance for attractors from different dynamical classes.} Predictive performance for speed-controller parameter scans using different chaotic attractor driving protocols: (a) Hénon-Heiles, (b) Rössler, (c) Chua, and (d) Lorenz-96. The mean speed of the drivers and the Lyapunov time predicted ahead are adjusted to match those of the Lorenz-63 driver used in \figref{fig:Lymburn_critical_speed_controller_scan_larger} ($\approx 0.45283\ t_{\text{lyap}}^{\text{attractor}}$). We again find the near-critically damped and underdamped regimes for the drivers displayed here, while some attractors exhibit characteristic swarm motion in the underdamped regime (pulsations, milling, oscillations). For all prediction tasks, the highest performances are obtained in the near-critically damped regime. Insets show the first 1,000 data points of the chaotic attractor input trajectories. Refer to \tabref{tab:supplementary_videos_benchmarks} for swarm snapshots and videos at parameter combinations indicated in the figure (cross and circle symbols).
    }
\end{figure*}

In the previous sections, we found that a near-critically damped active matter reservoir could optimally predict the future state of our chaotic Lorenz-63 input trajectory. 
This raises the question of whether critically damped active matter systems would also enable optimal predictive performance on different input trajectories with distinctive dynamical properties. 

To investigate this inquiry, we empirically choose four additional time-series benchmarks 
belonging to different dynamic classes \cite{Gilpin2023}:
the chaotic attractors Hénon-Heiles \cite{Henon1964} (twists),
Rössler \cite{Rossler1976} (rotations), 
Chua \cite{Chua1969} (lobes), 
and Lorenz-96 \cite{Lorenz1995} (ripples).
While these attractors belong to different dynamical classes and feature different Lyapunov exponents, we normalize their trajectories such that the drivers' mean speeds match that of the previously studied Lorenz attractor (see \axref{subsec:prediction_tasks} for details). 
The Lyapunov times of the chaotic trajectories indicate the varying difficulties of the prediction problems.  
Hence, to compare \rc predictive performance across chaotic attractors, we predict the same Lyapunov time ahead as for the Lorenz-63 system studied earlier ($\approx 0.45283\ t_{\text{lyap}}$, different real times) and shown in \figref{fig:Lymburn_critical_speed_controller_scan_larger}.

The predictive performances of active matter \rc systems are presented in \figref{fig:benchmarks} for the chosen attractors with different speed-controller settings. 
Predicting the Rössler, Chua, and Hénon–Heiles attractors consistently yields higher performances in the underdamped and critically damped regimes than the Lorenz-63 and Lorenz-96 attractors.
Yet, some technical challenges are making a fair comparison difficult.
Firstly, the repository used to generate these attractor trajectories sometimes reports greatly different values for the maximum estimated Lyapunov exponent determined using different methods. For example, for Chua $\approx$ 1.23 and 0.46 were reported \cite{dysts}, while we computed $\approx$ 0.30 using the standard QR method \cite{Wiesel1993}.
Secondly, we consider the Lyapunov exponents computed for the full multidimensional attractors while using only their 2D projections in this paper.

At the circle symbols, the Rössler attractor can be predicted with a performance of up to 0.9862, Chua with up to 0.9828, Hénon-Heiles with up to 0.9560, and Lorenz-96 with up to 0.8737.
When comparing the underdamped and critically damped regimes, we observe marked differences for attractors of varying difficulty.
The underdamped regime still yields quite useful reservoir dynamics for less ``difficult'' chaotic attractors (Rössler, Hénon–Heiles, Chua). Compared to the critically damped regime, the performance differences are around 0.10. For the chaotic attractors Lorenz-63 and Lorenz-96, we observe higher differences of 0.30 between the two regimes. While the numerical difference between 0.90 and 0.990 might seem small, there is a significant difference in the smoothness of the prediction quality, as seen in the Supplementary \figref{fig:benchmarks-actual-vs-predicted-timeseries}. The predicted, less challenging time series of Rössler, Hénon–Heiles and Chua are resolved by our near-critically damped active matter reservoirs much better compared to their underdamped counterparts. The amplitudes of the fluctuations in the predicted time series are much smaller. 

The surprisingly high prediction scores for the underdamped regimes for less chaotic attractors could have another reason beyond the lower prediction difficulty. Each chaotic attractor naturally drives the soft matter system into out-of-steady-state configurations distinct from those of the Lorenz-63 (and -96) systems, which may be more or less favorable for making predictions.
For example, for the Rössler attractor, we measure high scalar rotation parameters in the underdamped regime (see Supplementary \figref[b]{fig:benchmarks-scalar-rotation}. This is because the Rössler driving protocol induces strong milling in the underdamped swarm (see \vidref{\tabref{tab:supplementary_videos_benchmarks}}{4}). These sustained rotations increase the structural and dynamical order of the swarm, which could improve predictive performances via a higher consistency of swarm responses in this particular driving context.
Similar considerations could hold for Chua (underdamped oscillations, see \vidref{\tabref{tab:supplementary_videos_benchmarks}}{6}) and possibly Hénon-Heiles (underdamped pulsations, see \vidref{\tabref{tab:supplementary_videos_benchmarks}}{2}).
This observation highlights the intricacies and variability of non-equilibrium dynamics in the underdamped regime, which vary more strongly for each prediction problem and depend on the driving context. 
Yet, even the critically damped regime displays non-equilibrium phenomenology different from that of the Lorenz-63 case: the distinction between quasi-stationary and highly non-linear response (\figref{fig:phenomenology}) is less visible in the corresponding videos. (Note, however, that we here display videos for slightly higher underdamping (circle symbol)). Moreover, the time-scales of particular emergent mechanisms like shear thinning are not transient, but occur over prolonged time-scales with a weaker degree of intensity.

We also provide predictive performances for a constant predicted real-time ahead (instead of constant Lyapunov time ahead) of $\Delta t_{\text{pred}} = 0.50$ for all attractors in Supplementary \figref{fig:benchmarks-same-real-time-predicted-ahead}.

A high dynamical susceptibility, in the way it was calculated, is still a reasonable indicator of the vicinity of the critically damped regime (see Supplementary \figref{fig:benchmarks-attanasi-susceptibility}). By accounting for the swarm rotations and dilatations as proposed in \refref{\cite{Attanasi2014b-finite-size-scaling_swarms}}, a properly corrected dynamical susceptibility could yield a more generic indicator for the high-performance region for benchmarks from different dynamical classes.
Refer to Supplementary \figrefs{fig:benchmarks-scalar-polarity}{fig:benchmarks-minimum-conn-vel-corr} for an overview of observables for the four studied benchmark systems. 

To sum up, we found that the critically damped regime is the optimal dynamical regime across various chaotic time series benchmark problems, with different inherent prediction difficulties and dynamical properties of the resulting driving protocol. The threshold to overdamping in the intrinsic dynamics of a soft matter system could be a useful, generic, dynamical property for physical reservoir computing.

%% file: results_few_particles.tex

\subsection{\label{subsec:few-particle}The optimal damping regime is recoverable through reservoir computing with a single agent}

\begin{figure*}[htbp]
    \centering
     \includegraphics{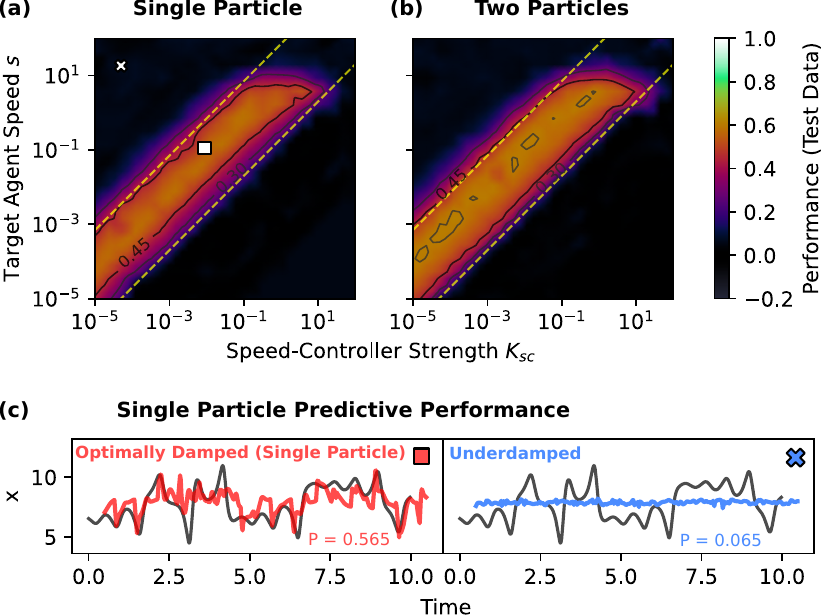}
    \caption{
        \textbf{The fundamental advantage of the critically damped regime is observed already for active matter reservoirs with few particles.} (a,b) Predictive performance for reservoir computing with few agents for varied speed-controller parameters analogous to the $N=200$ particle system in \figref[a]{fig:Lymburn_critical_speed_controller_scan_larger}. The critically damped high-performance region found in the $N=200$ agent systems appears already for the limiting cases of (a) $N=1$ and (b) $N=2$ agents. (c) Actual (gray) versus predicted (colored) time series obtained using reservoir computing with a single particle at different parameter combinations (cross, square symbols) shown in sub-figure (a). Refer to \tabref{tab:supplementary_videos_few_agents} for snapshots and videos. 
    }
    \label{fig:few_particles}
\end{figure*}

To investigate how elemental the superiority of the critically damped regime is for high-performance active matter reservoir computing, we study the edge case of $N=1$ agents in \figref[a]{fig:few_particles}.
The reservoir here comprises a single particle, again driven by the external time series. Studying a single agent allows us to focus on the pure single-agent dynamics, in the absence of any collective or statistical effects that play a role for higher $N$. 
We find that the performance is again optimal in the critically damped regime (\figref[a]{fig:few_particles}, square symbol), but the region with the best performance is here smaller compared to the $N=200$ agents case in \figref[a]{fig:Lymburn_critical_speed_controller_scan_larger} and confined to the center of the area enclosed by the yellow dashed lines. 
Surprisingly, the single-particle reservoir already delivers a relatively high predictive performance of around $P \approx 0.55$.
In contrast, the regime with underdamped speed-controller settings (\figref[a]{fig:few_particles}, cross symbol) provides no predictive capability here ($P \approx 0.05$). 
This stark difference is highlighted when comparing actual versus predicted time series in \figref[c]{fig:few_particles}. While in the critically damped regime, the single-particle reservoir produces a response that even captures some of the peaks of the driver signal. In the underdamped regime, it generates a constant, noisy output. One possible reason for these massive differences in prediction quality can be observed in the corresponding videos in \tabref{tab:supplementary_videos_few_agents}. In the critically damped regime, the particle quickly returns to the center of the simulation box. The damping causes the momenta acquired through driving to decay quickly. Because of this, it tracks the driver much more closely than in the underdamped case. Here, the single particle maintains its inertia and orbits the driver with only occasional repulsive interactions. This finding underlines the importance of ensuring a constant flow of information from the driving signal into the reservoir for optimal predictions.  
The surprisingly high performance in the critically damped regime could originate largely from the structural information in the readout layer, learning to process the radial displacement of the single particle systems from the box center, which contains basic information for the future state of the time series at a later time. 
For $N=2$ agents in \figref[b]{fig:few_particles}, the maximum performance increases slightly to $P \approx 0.60$ while the band of highest performances becomes extended towards the yellow dashed lines.

Together, these findings underline that the individual dynamic properties of agents, in the absence of any collective effects, play a crucial role in the predictive performance of active matter reservoirs. It allows insights into optimal spatiotemporal transfer of information from one particle (the driver) to another particle (the agent). 
Few-particle systems could serve as cheap indicators to find fundamental parameters of high-performance reservoirs, which could, in a second step, be optimized for enhanced collective properties. 
Further studies could link these few-particle systems with ideal excitation behavior and the ideal physical memory for different classes of dynamical systems.

%% file: results_alignment.tex

\subsection{\label{subsec:active-crystal}Revisiting the role of particle alignment}

\begin{figure*}[htbp]
    \centering
    \includegraphics{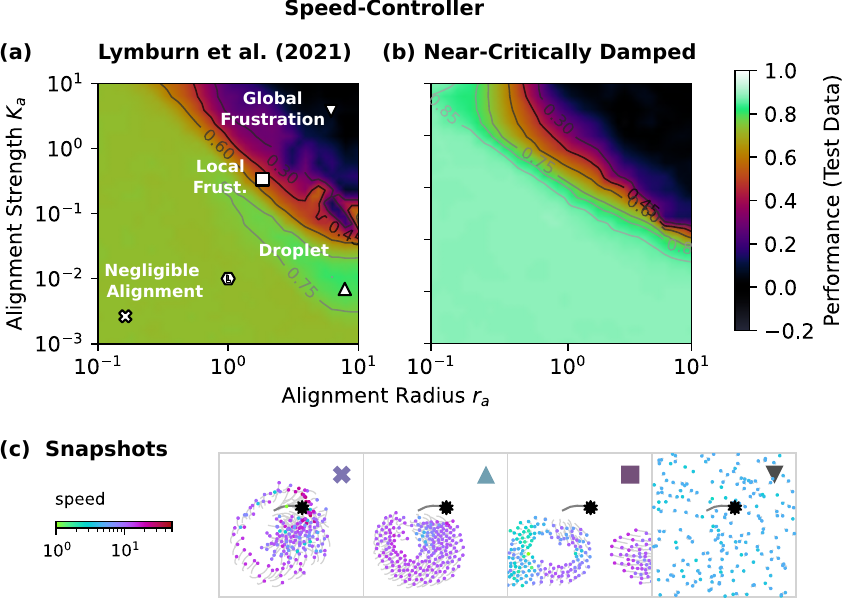}
    \caption{
        \textbf{Alignment forces can improve performance in the speed-controller regime of Lymburn \etal (2021), but not in the optimally damped regime.} (a,b) Scans over the parameters of the agent-agent alignment interaction -- the interaction (cut-off) radius $r_a \in [10^{-1}; 10^1]$ around each agent and the strength $K_a \in [10^{-3}; 10^1]$ of the interaction -- using (a) speed-controller settings of Lymburn \etal (2021) ($K_{sc} = 2.0$, $s = 10.0$) \cite{Lymburn2021} and (b) critically damped speed-controller settings ($K_{sc} = 0.02069$, $s = 0.04833$; pyramid symbol in \figref[a]{fig:Lymburn_critical_speed_controller_scan_larger}). 
        (a) Global, weak alignment leads to the formation of droplets, which increases the coherence of the active matter system and its predictive performance. High alignment strengths induce local (intermediate $r_a$) and global (high $K_a$) frustration in the system. The hexagon/``L'' symbol corresponds to the parameter combination used in \refref{\cite{Lymburn2021}} ($K_a = 0.01$, $r_a = 1.0$; see \vidref{\tabref{tab:supplementary_videos_lymburn_reproductions}}{2} for the corresponding snapshot and video).
        (b) In the critically damped regime, the presence of an alignment force with a strength $K_a \gtrapprox 0.1$ decreases predictive performance in dependence on the reach of the interaction. This demonstrates that alignment is detrimental to predictive performance in an optimally damped setting.     
        (c) Snapshots of driven swarms at different parameter combinations indicated by symbols in sub-figure (a).
        Refer to \tabref{tab:supplementary_videos_alignment} for corresponding videos.
    }
    \label{fig:alignment}
\end{figure*}

It has been hypothesized that a swarm operating around a liquid-to-gas-like phase transition (a ``critical'' point) has a superior performance, compared to a condensed droplet \cite{Lymburn2021}. At this ``critical'' point, the swarm would be able to display a broad variety of patterns in response to the external driving signal. In contrast, the condensed droplet would be less dynamically rich and come with a lower predictive performance.
Here we present evidence that a condensed droplet can deliver higher reservoir computing performances than a dynamically richer swarm. 

In \figref[a]{fig:alignment}, we tune the alignment force strength $K_a$ and the interaction radius $r_a$ in a swarm with the default speed-controller setting used in \refref{\cite{Lymburn2021}}.
For low $r_a$ and $K_a$, the alignment effect is negligible (\figref[c]{fig:alignment}: the performance $P = 0.726$ at this parameter combination is comparable to the one measured for the ``critical'' swarm at the liquid-to-gas-like phase transition swarm presented in \refref{\cite{Lymburn2021}} and discussed in \sectref{subsec:slow-regime} (hexagon/``L'' symbol; reproduced as $P = 0.719$, see \vidref{\tabref{tab:supplementary_videos_lymburn_reproductions}}{3}).

Higher interaction radii $r_a$ and alignment strengths $K_a$ then lead to the formation of an active crystal droplet phase (pyramid symbol). This phase yields performances of $P = 0.798$ and shows high polar order (see Supplementary \figref[a]{fig:alignment-observables}). From \vidref{\tabref{tab:supplementary_videos_alignment}}{1} we observe that the long-ranging weak alignment force creates a condensed swarm. Here, the alignment force effectively suppresses driver-induced agent inertia, and repelled agents quickly return to the flock. This leads to the formation of an effectively critically damped droplet with more consistent dynamics upon driving, which is a desirable property for reservoir computing with soft matter as analyzed in \sectref{subsec:slow-regime}. One reason why the performance could be lower here compared with the actual critically damped regime (shown in \figref[a]{fig:Lymburn_critical_speed_controller_scan_larger}) is that the droplet is not confined to the center of the simulation box. Due to the high mean agent speed, a result of the chosen speed controller settings, the droplet rotates around the box center as an active crystal. This could reduce the prediction quality because similar driving patterns could lead to different swarm states due to this inherent rotational swarm motion.
Furthermore, we note an interesting property of the flock: for slow driving, we observe a rotating active crystal that avoids the driver, but is only weakly perturbed by it. For abrupt driving, the active crystal flock aligns its direction of motion to that of the driver. This additional information processing mechanism could improve the predictive quality of the droplet because information about recent driving is conserved in the swarm. 

Further increasing radius and strength then leads to alignment force contributions that dominate over the other force terms. This creates a locally frustrated swarm motion defying the homing force (square symbol). The frustrated swarm splits when crossing the (periodic) box boundaries. We hypothesize that the waning influence of the driver due to frustrated motion, hence creating different patterns for similar driving histories, reduces the predictive performance in this regime.
For extremely high alignment strengths and radii, we observe a globally frustrated swarm state (nabla symbol) with performances close to zero. The freezing of the initial orientation is indicated by an absence of order, measured by polarity values close to zero (Supplementary \figref[a]{fig:alignment-observables}).
We note that for this speed-controller setting, the dynamical susceptibility has its maximum and the connected velocity correlation function has its first local minimum in the locally frustrated regime (square symbol), not in the optimal droplet regime. The mean squared displacement at the Lorenz-63 Lyapunov time is low in the optimal regime but features another minimum at the transition from locally to globally frustrated regimes. This hints that these indicators are not absolute and must be applied and interpreted with care.

To investigate the importance of the alignment force in the critically damped regime that we presented in \sectref{subsec:slow-regime}, we perform another parameter scan in \figref[b]{fig:alignment} using the previously found optimal (near-critically damped) speed-controller settings ($K_{sc} = 0.02069$, $s = 0.04833$; pyramid symbol in \figref[a]{fig:Lymburn_critical_speed_controller_scan_larger}). The performance heatmap reveals that weak to intermediate alignment does not improve predictive performance in the critically damped regime. We report a maximum performance of $P = 0.884$ here, which is equal to the highest performance measured in the speed-controller scan in \figref[a]{fig:Lymburn_critical_speed_controller_scan_larger}. For high alignment strengths and intermediate to high alignment radii, the performance decreases down to zero, as observed before. Overall, this supports the hypothesis that near-critically damped dynamics are optimal because alignment interactions cannot improve the quality of response dynamics and thus only deteriorate the predictive quality. Importantly, this suggests that we can remove alignment as an ingredient in the quest to find the highest-performance soft matter reservoirs.  

We conclude that for a swarm with high mean speed (enforced by Lymburn \etal (2021) speed-controller settings), weak and long-ranged alignment forms an active crystal droplet which improves predictive performance. Nevertheless, the high predictive performance in the critically damped regime in the absence of any alignment interaction suggests that even passive matter could be a suitable candidate for soft matter reservoir computing.

%% file: results_homing.tex

\subsection{\label{subsec:homing}The optimal regime is robust against variations in homing force strength} 

\begin{figure*}[htbp]
    \centering
    \includegraphics{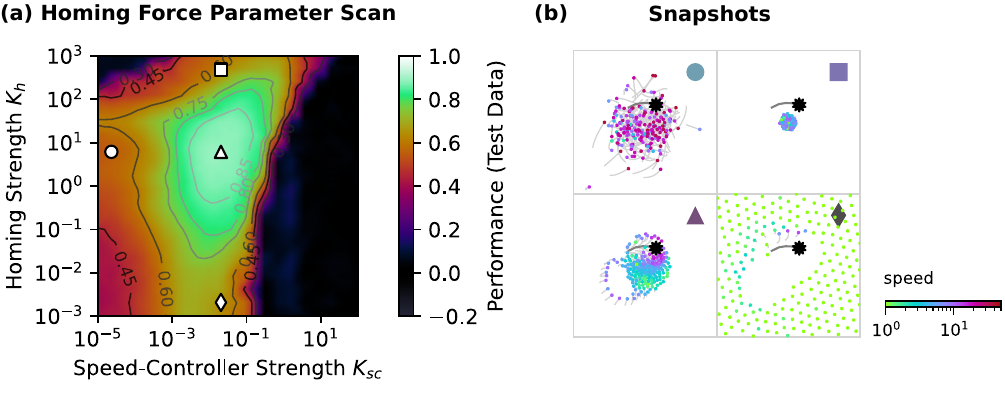}
    \caption{\textbf{Viscoelastic droplets forming in the optimal regime are remarkably robust against variations in homing force strength, under driven conditions.} (a)  Predictive performance for varied homing force strength $K_h \in [10^{-3}; 10^{3}]$ (global attraction to the center of the simulation box) and speed controller strength $K_{sc} \in [10^{-5}; 10^{2}]$, using a fixed target agent speed $s = 0.04833$. This target agent speed corresponds to \figref{fig:Lymburn_critical_speed_controller_scan_larger}, pyramid symbol. The optimal homing strength regime in the critically damped regime (pyramid symbol) yields viscoelastic soft matter droplets and spans about two orders of magnitude ($1.0 \lessapprox K_h \lessapprox 20.0$). (b) Snapshots corresponding to parameter combinations marked as symbols in sub-figure (a): underdamped (circle), highly condensed (square), viscoelastic droplet (pyramid), loosely bound crystal under near-zero homing force conditions (diamond). Refer to \tabref{tab:supplementary_videos_homing_strength_and_speed_controller_strength} for corresponding videos.}
    \label{fig:homing}
\end{figure*}

By investigating the optimally damped regime, we found that soft matter systems with agents that immediately return to the center of the simulation box, rather than maintaining inertia and entering longer orbits around the center, yield higher predictive performance. These resetting dynamics to the box center are crucially influenced by the homing force strength $K_h$. In this section, we explain how to achieve optimal predictive performance by tuning the homing force.

In \figref[a]{fig:homing} we vary the homing force strength $K_h$ for a fixed target agent speed $s$ of the speed-controller, and a varied speed-controller strength $K_{sc}$. The optimal performance $P = 0.889$ is reached at $(K_{sc}, K_h) = (0.02069, 6.158)$ (pyramid symbol). This performance is almost identical to those with a homing force strength $K_h = 2.0$, presented in \figref[a]{fig:Lymburn_critical_speed_controller_scan_larger} (pyramid symbol). The corresponding snapshot in \figref[b]{fig:homing} and the \vidref{\tabref{tab:supplementary_videos_homing_strength_and_speed_controller_strength}}{2} reveal that the soft matter system forms the previously studied condensed, viscoelastic droplet with an interface enclosing the driver. We note that high performances $P > 0.85$ are stable within about two orders of magnitude of varying the homing force strength, underlining its robustness.
Lowering the speed-controller strength (circle symbol) recovers the underdamped soft matter regime with a predictive performance $P < 0.60$ (see also \figref{fig:Lymburn_critical_speed_controller_scan_larger}, cross symbol).
Increasing the homing force strength in the critically damped regime (square symbol) leads to dominating homing forces, which cause a strong compression of the soft matter system. Here, agents barely react to the repulsive driver force, resulting in a decrease in performance. Despite this weak driver influence, a prediction score of $P \approx 0.60$ can still be achieved. 
\new{While the driver influence is weak here, it is not negligible. This can be seen when studying the canonical scalar agent observables polarity and rotation, which fluctuate strongly upon driving as seen in \figref{fig:strong-homing-actual-observables} -- much stronger than in the undriven case. This suggests that, despite the driver-agent interaction being barely visible to a human observer due to the extreme density of the agents in the video, the interaction is nonetheless significant. The active matter response to the driver -- while being noisy -- seems to be significant enough allow prediction of some features of the target time series. }

Decreasing the homing force strength significantly in the critically damped regime (represented by the diamond symbol) effectively eliminates the agent's pull toward the center of the simulation box. Because of this, agents distribute themselves across the simulation box. The agent-agent repulsive forces cause them to occupy the entire simulation box while organizing themselves into a highly ordered crystalline structure. The repulsive force of the driver cuts out a vacancy area in the crystal. This area is not immediately reoccupied by agents due to the relative weakness of agent-agent repulsions. Strong and abrupt driving causes traveling waves through the bulk, which can be nicely observed in the corresponding video \vidref{\tabref{tab:supplementary_videos_homing_strength_and_speed_controller_strength}}{4}. This phononic information transfer is yet another example of the rich phenomenology found in our driven soft matter systems.
The relatively low performance of $P = 0.642$ at this parameter combination may be due to the lack of a strong interface that ensures continued information transfer from the driver to the system under slow driving. Small, quasi-stationary driving changes the system state only minimally. On top of that, the overall heterogeneity of (collective) agent responses becomes limited compared to the critically damped droplet, which should lower the responsiveness and thus expressivity of the soft matter system for representing and processing relevant information.
A complementary parameter scan over homing force strength $K_h$ and target agent speed $s$ with a fixed speed-controller strength is shown in Supplementary \figref{fig:homing_target_agent_speed_scan}.

%% file: discussion.tex

\section{\label{sec:discussion}Discussion and Conclusion}

\subsection*{Summary}

We conducted an empirical simulation study of reservoir computing using a physical model system composed of active matter agents, driven chaotically via local repulsive forces. 
In comparison to earlier work \cite{Lymburn2021}, we discovered an exceptional dynamical regime of the reservoir that appears to be robustly optimal for computation.

We find that the parameters that regulate the microscopic particle speed directly, via non-Hamiltonian forces at a microscopic scale, are \new{highly} influential physical controls: They \new{could be considered} lowest-order dynamical causes for attaining high computational performance; they further enable the system to adapt and render high performances in various controlled contexts. \new{Collective dynamics (originating through e.g. particle repulsions, plus other factors) \new{have an amplification effect}, as a comparison between the one- and two-particle reservoirs versus the full $N=200$ systems shows: a boost of around 30\,\% performance. Surely, details of driver-agent interactions play a key role as well in how they generate collectivity.} The effects of velocity-dependent particle-particle alignment forces (as defined in this model) \new{seem to be} of higher order; in the optimal computing regime, these can be eliminated without harming performance.
\new{In \figref{fig:n-agents-scaling} we observe a saturation of predictive performance for a lower agent number of $N\approx 100$ in the overdamped regime, compared to a higher agent number $N > 2,000$ for the underdamped regime. Before the saturation, the performance scales roughly with $\log(N)$.}

The substrate's responsiveness and adaptability can be understood in the sense that it both detects changes in its environment (the driver's fluctuations) and efficiently self-restores itself from recent history. As these abilities appear to originate at the microscopic level, they are generic. Investigated were the intrinsic relaxation characteristics of noninteracting (or, likewise, interacting) agents that are free from external driving forces: In an ensemble starting from an initial state, the mean microscopic relaxation dynamics exhibit a rapid exponential decay, followed by a brief power-law decay, eventually transitioning to a final steady state characterized by weak oscillations. 

\new{Across the different chaotic input time series studied here, along with further variations such as different integration time steps, different driver force shapes \cite{Gaimann2025b}, and different mean driver speeds, the optimal performance consistently appeared around the critically damped regime, with only minor shifts in the precise optimum. This indicates that certain generic reservoir computing properties (e.g., balance of memory and nonlinearity, plus sufficient consistency) are particularly well realized near critical damping, whereas they degrade more clearly outside this regime. However, we cannot rule out that very different tasks or input schemes could qualitatively alter the error landscape, so further systematic studies would be needed to confirm the generality of this behavior.}

As previously noted in \refref{\cite{Lymburn2021}}, canonical (mean) order parameters (polarity, rotation) indicating the global order of a swarm, do not capture the essence of adapting to external changes and thus cannot reveal areas of high predictive performance in general. Our approach demonstrated that correlation functions on agent dynamics (\ie correlated velocity fluctuations and velocity autocorrelations in time) capture the responsiveness of an active matter reservoir substrate to its external perturbation and could serve as a more reliable indicator.

The notion of microscopic behavior leading to adaptivity is highlighted particularly in the setup where only one- or two-particle substrates are used for \rc, which actually work as minimal models of computing. All dynamics arise solely from interactions with the surrounding (hidden) environment in the form of non-Hamiltonian forces controlling the speed and the repulsive forces from the driver. 

When we altered the specific class of chaotic prediction tasks (the external driving protocol), we consistently found that the near-critically-damped regime yields the best performance, providing universal generalization power and reliability across different tasks. We benchmarked and analyzed the \rc system using the Lorenz-63, Hénon-Heiles, Chua, Rössler, and Lorenz-96 attractors.

The optimal dynamical regime demonstrates robustness against variations in the homing force, at least until the system reaches its physical limits, such as when a collapsed droplet forms. By changing these forces when starting from the non-optimal configuration in the speed controlling forces, novel information-processing mechanisms emerge, which are, however, not optimal for performance.
In general, initially non-optimal dynamical regimes exhibit much greater sensitivity to such changes. For example, strong alignment forces between particles can enhance performance by increasing surface tension and enabling the formation of a coherent droplet -- mimicking those physical mechanisms of information processing that result in optimal performance in the near-critically-damped dynamical regime. This optimal regime enables high performance without alignment forces in place.

\subsection*{Emergent mechanisms of physical computation}

Our results suggest the emergence of interfaces may be fundamental to the functionality of computing systems in soft matter (of analog or neuromorphic kinds):
The interface that forms around the driver appears to be a crucial feature for the computational properties of the \rc. It might seem counterintuitive, as a slowly moving interface exhibits little variation, meaning there are few significant fluctuations informative for the prediction task. Indeed, \refref{\cite{Lymburn2021}} comments on ``adiabatic'' driving and the resulting difficulty for making a prediction. However, a coherent state enables sharp detections:  Specifically, transitions and fluctuations to and from the highly nonlinearly responding state reported in \sectref{subsec:rationale_RC_critically damped} can be read from signals in the observation layer. From another view, surface phenomena are known to be rich \cite{binder2008confinement}, particularly under nonequilibrium conditions \cite{Empting2021,Michely-book,Klopotek2021}, which may suggest a kind of capacity for ``physical expressivity'' -- to amplify or express subtle changes in the physical system at lower scales \cite{Flack2013}. For the Lorenz-63 driving, near the quasi-stationary state, the radial propagation of velocity fluctuations originating at the interface is clearly visible, as is the propagation of fluctuations during shear thinning when the driver moves through the medium during the highly nonlinear response. The integration of the driver within the soft matter system facilitates a continuous transfer of information from any changes in the driver's state into the system.

The distinction between quasi-stationary and highly nonlinear responses becomes less pronounced in nearby dynamical regimes (towards underdamping, near the ``gas-liquid'' transition). There, the interfaces become unstable. 

Empirically, certain tasks appear simpler due to higher performance (Rössler, Chua, Hénon-Heiles prediction), and, simultaneously, the strongly alternating response dynamics diminish. Interfaces do form near the driver, but can become deformed or channel-like, as the driving pattern has longer periods of regular motion. The chaotic Lorenz-96 driver should embody a more difficult task due to a shorter Lyapunov time. It shows that the phenomenology and optimal state are subtly different from those of the Lorenz-63 case.

We speculate that in the cases of simpler tasks, more of the information processed (as defined via the Gaussian kernels that measure local densities in structure and momentum) expresses structural changes, which are inherently slower variables and thus have more internal coherence.  This is subject to future investigation. For these simpler tasks, the dynamic quantifiers of the responding system seem to be less clear indicators of information processing abilities (in the more regularly changing environments) -- at least in the way they were defined in this study. Quantifiers taking into consideration different dynamical symmetries (i.e., macroscopic rotations) would clarify this outstanding issue.

\subsection*{Relating physical to computing properties}

\textit{Robustness:} 
Notably, it is thought that sensitivity can come with a tradeoff in robustness \cite{Daniels2017}. Yet, a system that is adaptive is also robust: For example, a biological cell can continue to render useful behavior despite major intrinsic or external changes \cite{Aldana2007,Kitano2004}.
Within the machine learning literature, the concept of robustness is defined in various ways \cite{Freiesleben2023}. As discussed in the reference, it may be crucial to test a system across different models, rather than focusing solely on datasets. 
In contrast, how robustness is expressed in physical systems is still in the early stages of research \cite{Jaeger2021,Saunders2019}. It may be associated with metastability \cite{Rabinovich2008} or other nonequilibrium features expressing self-restoration. Tests we considered in our study varied the intrinsic physical parameters and inference tasks, which represent changes in static confinement, a new dynamical environment, the occlusion of particular internal interactions, and the introduction of random noise to the system. \\

\textit{Collectivity, analog computing}: 
Outside of computing, it has been observed that collectivity can lead to intelligent and adaptive behavior \cite{Daniels2016,Flack2017,Couzin2007,Couzin2009}.
The collective mechanisms during computation in this active matter system exhibited a complex mixture of elastic-like and viscous properties. There may be an intriguing bridge to computation: Nesting irreversible components of information processing into reversible ones is part of the design of analog computers \cite{Bennett1982}.
\\

\textit{Computing near ``criticality'':}
A broadly cited hypothesis is that systems poised near criticality generally exhibit adaptive behavior when faced with changing environments \cite{Mora2011-biological_criticality,Kinouchi2006,Sune2019,Chialvo2010,Beggs2012}. In such contexts, ``criticality'' often refers to a system-wide nonequilibrium phase transition. In our work, the concept of ``critical damping'' refers to the absence of oscillations in microscopic dynamics, which alters the overall dynamical behavior of the system. Though one can analyze the scaling behavior in greater detail, a further differentiation of the term ``edge of criticality'' may be helpful \cite{Carroll2020_edge}. \\

\textit{Interpretability:} The inherent interpretability offered by information processing machines made out of matter is interesting for stimulating scientific inquiry. Standard employed artificial neural networks present significant challenges to understanding the network's overall functional operation and internal mechanisms \cite{Yan2024-opportunitiesRC,wetzel2025interpretable}. 
For physical computing systems, dynamical processes take precedence over algorithmic descriptions \cite{jaeger2023toward,Horsman2014-when_phys_compute},  and physics provides intrinsic frameworks for understanding such dynamics.

\subsection*{Real-world experimental analogies}
\new{There are a number of real-world experimental systems for which this model (from Lymburn \etal) provides and serves to be a basic analogy.  The key is to control the microscopic dynamics, in analogy to the non-Hamiltonian velocity-control forces of this model. 
Dynamics based in this non-conventional setting can be found in active colloids \cite{Bechinger2016,teVrugt2025}, where self-propulsion is inducible through a hydrodynamic coupling with the embedding solvent. A characteristic parameter is the Reynolds number, a systemic property that can be adjusted by varying the physical/chemical properties of the swimmers and the solvent. (The P\'eclet number characterizes the relative importance of directed motion of an active particle versus diffusive motion.) Tuning such chemical-reaction-driven propulsion was illustrated in Ref.~\cite{Ismagilov2002}; see also \cite{Howse2007}. In magnetic systems (Janus particles), magnetic fields can be used for control of the particles \cite{Kaiser2015}; see also \cite{Baraban2013,Baraban2013b}. Catalytically-driven nanorods \cite{Paxton2004}, and spinning magnetic microdisks \cite{Wang2022-microdisks} are further examples. (In the latter, tuning local pairwise interactions leads to different global patterning: the magnetic dipole-dipole force, the capillary force, and a hydrodynamic lift force.)}

\subsection*{Outlook}

From our above discussion on the order-of-importance of types of forces in the model, we envision seeking the truly global optimum for performance. Our next step will be to fine-tune the interparticle repulsion forces (which need not necessarily be long-ranged) and the driver-agent interaction. We aim to find further deep connections between the physical properties of the reservoir substrate and its computational properties. We have preliminarily investigated much larger systems ($N=500 \dots 10^5$) and foresee the different scaling of agent number (and observation kernel) densities to predictive performance in different dynamical regimes. In a further step, other types of higher-order velocity-dependent interparticle forces could potentially enhance performance. Finally, we are studying the effects of additional ``artificial'' or extrinsically-controllable interactions, such as those for the phase coupling of mobile oscillators (swarmalators), of interest for robotic or AI-agent systems \cite{hamann2018swarm,okeeffe:2017:oscillators,Kruijssen2024}.

Capturing the right, relevant collective variables in the observation layer poses another outstanding question. 
The kernels' distribution in space and scale is one thing to consider \cite{Lymburn2021}, the way they compute an average quantity internally is another.  

A real-world implementation of a similar \rc system could be inspired by a basic analogy of this \rc setup to a rheology experiment \cite{Puertas2014,Sagis2011}, or using spinning magnetic microdisks \cite{Wang2022b}, for instance. Many soft matter systems may serve as suitable \rc substrates. 
Exploring mixtures of active systems represents a natural next step to more expressive reservoirs. Viscoelasticity may be an important material feature to consider, just as the hierarchical structuring capabilities found in many functional materials \cite{Ikkala2002} and chemical reactions \cite{Baltussen2024,Gorecki2015}. Predicting across several time scales has been achieved for Echo State Networks through using, \eg, hierarchical (\textit{non-shallow}) reservoir architectures \cite{Manneschi2021} or by introducing band-pass neurons \cite{Wyffels2008}. Predicting classes of time series like spiking attractors, such as the Hindmarsh-Rose system \cite{Hindmarsh1984} that display dynamics on vastly different timescales, is an interesting prospect, relating to more recent developments in the \rc literature \cite{Jaurigue2024_timescale}. Active matter reservoir computing integrating multiple modes of function across physical scales may inspire new forms of unconventional computing \cite{Jaeger2021-unconventional_general}.\\

%% file: contributions.tex

\section*{Author Contributions}

Miriam Klopotek: Conceptualization, Formal analysis, Funding acquisition, Methodology, Visualization, Project administration, Supervision, Writing – original draft, Writing – review \& editing.

Mario U. Gaimann: Data curation, Formal analysis, Investigation, Methodology, Software, Validation, Visualization, Writing – original draft, Writing – review \& editing.

%% file: acknowledgements.tex

\begin{acknowledgments}
Funded by Deutsche Forschungsgemeinschaft (DFG, German Research Foundation) under Germany's Excellence Strategy -- EXC 2075 -- 390740016. We acknowledge the support of the Stuttgart Center for Simulation Science (SimTech), the Interchange Forum for Reflecting on Intelligent Systems (IRIS) at the University of Stuttgart, the International Max Planck Research School for Intelligent Systems (IMPRS-IS), and the WIN-Kolleg of the Heidelberg Academy of Science and the Humanities.
We thank Christian Holm, Mathias Niepert, Anna Levina, Francesco Ginelli, Shankar P.\ Das, Max Weinmann, Patrick Egenlauf, Gaurav Gardi, and Lasse J.\ Schulz for fruitful discussions.
\end{acknowledgments}

%% file: supp_benchmarks_details.tex

\section{\label{subsec:prediction_tasks}Chaotic Time-Series Benchmarking Details}

In this paper, we use some common chaotic attractors as benchmark systems for the time series prediction task of our active matter reservoir computer (see \sectref{sec:mm-interactions-driven}). 
In our default implementation, we integrate the coupled differential equations of the Lorenz-63 system \cite{Lorenz1963} \new{
\begin{align}
  \dot{x} &= a (y - x), \\
  \dot{y} &= x (b - z) - y, \\
  \dot{z} &= x y - c z,
\end{align}
}
with standard parameters ($a = 10; b = 28; c = 8/3$) using Euler integration with an integration time step $\Delta t = 0.02$ and initial conditions $(x_0,y_0,z_0) = (0.0, 1.0, 1.05)$ for training and $(0.0, 1.1, -1.5)$ for testing. 
For the other chaotic attractors (Hénon-Heiles, Rössler, Chua, Lorenz-96), we generate trajectories using the \texttt{dysts} library (version 0.6) \cite{dysts} (corresponding publications: Refs.\ \cite{Gilpin2021, Gilpin2023}). 
\new{The corresponding differential equations are for the N-dimensional Lorenz-96 system (we use $N=4$ here)}
\begin{align}
  \dot{x}_1 &= (x_2 - x_{N-1}) x_N - x_1 + F, \\
  \dot{x}_2 &= (x_3 - x_N) x_1 - x_2 + F, \\
  \dot{x}_i &= (x_{i+1} - x_{i-2}) x_{i-1} - x_i + F, \quad i = 3,\dots,N-1, \\
  \dot{x}_N &= (x_1 - x_{N-2}) x_{N-1} - x_N + F\,,
\end{align}

\new{For the Chua system}
\begin{align}
  \dot{x} &= \alpha \bigl( y - x - f(x) \bigr), \\
  \dot{y} &= x - y + z, \\
  \dot{z} &= -\beta\, y,
\end{align}

\new{where}
\begin{align}
  f(x) &= m_1 x + \tfrac{1}{2} (m_0 - m_1)\bigl( |x+1| - |x-1| \bigr)\,,
\end{align}

\new{For the Hénon--Heiles system}
\begin{align}
  \dot{x}   &= p_x, \\
  \dot{y}   &= p_y, \\
  \dot{p}_x &= -x - 2\lambda x y, \\
  \dot{p}_y &= -y - \lambda x^2 + \lambda y^2\,,
\end{align}

\new{and for the Rössler system}
\begin{align}
  \dot{x} &= -y - z, \\
  \dot{y} &= x + a y, \\
  \dot{z} &= b + x z - c z.
\end{align}

\begin{table*}[t]
    \centering
    \caption{\new{\textbf{Specifications of chaotic attractors used for benchmarking} described in Appendix A.}}
    \label{tab:attractors-params}
    \begin{tabular}{|l|p{0.25\textwidth}|p{0.25\textwidth}|l|}
        \hline
        \textbf{Attractor} & \textbf{Initial conditions} & \textbf{Parameters} & $\mathbf{dt}$ \\ \hline\hline

        Lorenz-63
        & \raggedright
          (0.0, 1.0, 1.05) (train)\par
          (0.0, 1.1, -1.5) (test)
        & \raggedright $a = 10$, $b = 28$, $c = \tfrac{8}{3}$
        & 0.02 \\ \hline

        Henon--Heiles 
        & \raggedright (-0.29707259, -0.0031367513, 0.42548865, 0.071905482)
        & \raggedright $\lambda = 1$
        & 0.002548 \\ \hline

        Lorenz--96 
        & \raggedright (-2.46820633, 0.09570264, 1.59270902, 10.21372147)
        & \raggedright $F = 20.0$, $N_{\text{dim}} = 4$
        & 0.0001428 \\ \hline

        R\"ossler 
        & \raggedright (6.5134412, 0.4772013, 0.34164294)
        & \raggedright $a = 0.2$, $b = 0.2$, $c = 5.7$
        & 0.0007563 \\ \hline

        Chua 
        & \raggedright (1.1262351, -0.11386474, -0.78324421)
        & \raggedright $\alpha = 15.6$, $\beta = 28.0$, $m_0 = -1.142857$, $m_1 = -0.71429$
        & 0.006605 \\ \hline
    \end{tabular}
\end{table*}

Refer to \tabref{tab:attractors-params} for a summary of employed parameters and initial conditions. \new{All generating differential equations and the employed parameters are retrieved from the files \href{https://github.com/GilpinLab/dysts/blob/4c01e6d5b752a599e398a655b56581d9fd9f7b29/dysts/flows.py}{\texttt{flows.py}} and \href{https://github.com/GilpinLab/dysts/blob/4c01e6d5b752a599e398a655b56581d9fd9f7b29/dysts/data/chaotic_attractors.json}{\texttt{chaotic\_attractors.json}} in the \texttt{dysts} library \cite{dysts}.}
To compare predictive performance, we choose to normalize these time series to the mean effective driver speed
\begin{equation}
    \bar{v}_d = \frac{1}{\Delta t (T-1)} \sum^{T-1}_{i=1} \sqrt{(x^{i+1}_d - x^i_d)^2 + (y^{i+1}_d - y^i_d)^2}
\end{equation}
for $T = 5 \cdot 10^4$ integration time steps in a single run and driver coordinates $(x^d_i, y^d_i)$ at a time point $i$. For the Lorenz-63 system we measure $\bar{v}_d = 9.91 \pm 10.02$.

Then each \texttt{dysts} trajectory is generated using the following procedure: We first generate a fine-grained trajectory with 10\textsuperscript{5} points per period and 2 $\cdot$ 10\textsuperscript{8} steps in total.
The trajectory is then fitted to a square box (the driver box) of size 8.0 and centered in a square box (the simulation box) of size 16.0. This ensures sufficient driver interaction with typical swarm sizes of $N=200$ particles.
We then determine the best sampling rate to sample the fine-grained trajectory using a simple bisection algorithm. This yields a driver trajectory with the above-determined desired mean speed $\bar{v}_d$, with a precision of 10\textsuperscript{-2}.
The final trajectories correspond to the sampled fine-grained trajectories with the following allocation: 10\textsuperscript{3} (burn-in steps training) + 5 $\cdot$ 10\textsuperscript{4} (simulation steps training) + 10\textsuperscript{3} (burn-in steps testing) + 5 $\cdot$ 10\textsuperscript{4} (simulation steps testing) steps. 
We plot the first 10\textsuperscript{3} simulation data points (excluding the burn-in period) of these systems as insets in \figref{fig:benchmarks} and provide videos of the raw trajectories in \tabref{tab:supplementary_videos_benchmarks}. We confirm that for our test trajectories of 5 $\cdot$ 10\textsuperscript{4} data points each, the mean driver speeds for the chosen subset of the total time series generated are reasonably close to those of the (non-\texttt{dysts}-generated) Lorenz-63 trajectory: 10.20 $\pm$ 2.78 (Hénon–Heiles), 10.30 $\pm$ 3.84 (Rössler), 9.91 $\pm$ 4.57 (Chua), and 10.31 $\pm$ 6.05 (Lorenz-96).
We quantify the chaoticity of the (non-rescaled) attractors using the highest estimated Lyapunov exponents (\texttt{maximum\_lyapunov\_estimated}) recorded in the provided database \href{https://github.com/williamgilpin/dysts/blob/4c01e6d5b752a599e398a655b56581d9fd9f7b29/dysts/data/chaotic_attractors.json}{\texttt{chaotic\_attractors.json}} of the \texttt{dysts} package (version 0.6) \cite{dysts}: 
$\lambda_{max}^{\text{Hénon–Heiles}} = 0.03551120315239528$,
$\lambda_{max}^{\text{Rössler}} = 0.15059688939547888$, 
$\lambda_{max}^{\text{Chua}} = 1.2301163591725595$, and
$\lambda_{max}^{\text{Lorenz-96}} = 1.3361667787286362$.

In \figref{fig:benchmarks} we show predictive performances for predicting $\approx 0.45283$ Lyapunov times ahead. In real-time, this corresponds to 
\begin{alignat*}{2}
    \Delta t_{\text{pred}} 
    & \approx 0.50\ (\text{L63}) 
    & & \\
    & \approx 12.76 \ (\text{Hénon–Heiles}) 
    & & \approx 3.00 \ (\text{Rössler}) \\
    & \approx 0.36 \ (\text{Chua}) 
    & & \approx 0.34\ (\text{L96})\,.
\end{alignat*}

%% file: supp_noise.tex

\section{\label{subsec:role_noise}Robustness against added noise to agent dynamics}

\begin{figure}[htbp]
    \centering
    \subfloat[Default speed-controller setting of Lymburn \etal (2021)]{ \includegraphics{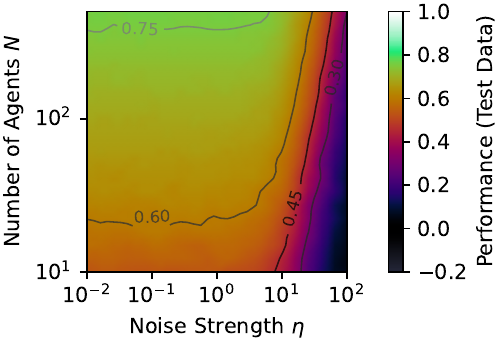}}\hfill
    \subfloat[Near-critically damped speed-controller setting]{\includegraphics{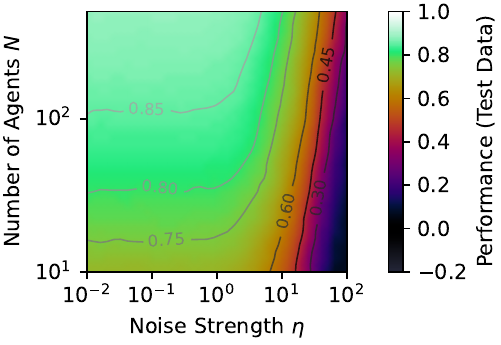}}\\

    \caption{\textbf{Predictive performances for different numbers of agents $\mathbf{N_a}$ and Gaussian white noise strengths $\mathbf{\eta}$.} The parameter scans vary $N_a \in [10^1; 5 \cdot10^2]$ and $\eta \in [10^{-2}; 10^2]$ for (a) the default speed-controller setting of Lymburn \etal (2021) ($K_{sc} = 2.0$, $s = 10.0$) \cite{Lymburn2021} and (b) a near-critically damped active matter system ($K_{sc} = 0.01$, $s = 0.1$).}
    \label{fig:noise}
\end{figure}

Here we briefly explore the role of Gaussian white noise added to the agents' motion on predictive performance. We define a random force
\begin{equation}
    \mathbf{F}_{\text{noise}} = \eta \cdot \boldsymbol{\mathcal{N}}(\mu, \sigma)
\end{equation}
that acts on each agent in the system. For each dimension, a random number is drawn from a normal distribution $\mathcal{N}$ with mean $\mu = 0.0$ and standard deviation $\sigma = 1.0$ and scaled with the force strength $\eta$.
In \figref{fig:noise} we observe that performance scales with the number of agents. This can be rationalized by the notion that the active matter system becomes more expressive by generating more diverse patterns and that the system can localize more information about the driving input signal. It also provides better statistics for the observation kernels which are a proxy for swarm dynamics by capturing agent count and velocity densities.
Higher noise strengths lead to a performance decrease. This is also expected as the Gaussian observation kernels are misled by noisy particles. In case of high noise, the soft matter system cannot resolve and process small, quasi-stationary driver motions anymore, because they occur at the same length scale as the added noise. We note that the onset of this performance decrease is later when the number of particles in the system is higher. This suggests that a higher number of particles protects the predictive capabilities of active matter reservoirs and provides robustness. When comparing the two sub-figures, the performance deterioration is qualitatively similar for the underdamped and the near-critically damped regimes.
 This analysis underlines that agent response to the external driving signal must be as consistent and predictable as possible -- ideally free from noise. Swarms operating in the previously reported ``critical'' regime at a droplet-to-gas phase transition \cite{Lymburn2021}, where agents obtain high and persistent momenta, could be understood as carrying \textit{intrinsic} noise originating from underdamped particle dynamics.
 

%% file: supp_memory_capacity.tex

\section{Short-Term Memory Capacity}\label{subsec:memcap}

\new{We measure the memory capacity, a key measure of reservoir computers, according to the definition provided by Jaeger (2001) \cite{Jaeger2001-short_term_memory}.
We first define the \textit{k-delay Short Term Memory Capacity} (STMC)
\begin{equation}
    MC_{k,d} = \frac{\operatorname{cov}^2\!\bigl(y_{\text{target,d}}(t-k\Delta t),\, y^{k}_{\text{pred,d}}(t)\bigr)}
{\sigma^2\!\bigl(y_{\text{target,d}}(t-k\Delta t)\bigr)\,\sigma^2\!\bigl(y^k_{\text{pred,d}}(t)\bigr)}\,,
\end{equation}
where $y^k_{\text{pred,d}}(t)$ is the predicted time series (for dimension $d$) using the trained readout layer for the delay $k$, $y_{\text{target,d}}(t)$ is the actual time series for dimension $d$, and $\Delta t$ is the integration time step. If the signal at a delay of $k$ steps can be reconstructed perfectly $MC_k$ becomes 1.0, and if there is no memory at all $MC_k$ becomes 0.0.
The total STMC is then given as the sum over all $k$-delays and dimensions $x$ and $y$
\begin{equation}
    MC = \sum_{d=1}^2  \sum_{k=1}^{k_{max}} MC_{k,d}
\end{equation}
and indicates how much memory about past states can be retained by the reservoir, where the maximum delay is set to $k_{max} = 100$. The maximum obtainable total memory capacity is hence 200.0 here. \\
Typically, a uniformly at random time series is used to compute the memory capacity. For our reservoir computing with active matter approach, this procedure comes with some challenges. Firstly, a driver trajectory that is sampled uniformly at random across the simulation box can result in situations where the driver is placed in the corner of the simulation box. When this happens, a repulsive driver-agent interaction with the swarm is highly unlikely. To this end, we sample the trajectory uniformly at random from a disk of radius $R$ through
\begin{equation}
    U \sim \mathrm{Uniform}(0,1), \quad r = R \sqrt{U}, \quad \theta \sim \mathrm{Uniform}(0, 2\pi),
\end{equation}
transforming to Cartesian coordinates via
\begin{equation}
    x_d = r \cos(\theta), \quad y_d = r \sin(\theta),
\end{equation}
and shifting the $(x, y)$ pairs by a constant offset so that the origin $(0.0,0.0)$ was mapped to the center of the simulation box with length $\ell_{box} = 16.0$, i.e., to $(8.0, 8.0)$ (see \figref{fig:uniform_random_time_series_circle}). We chose the radius $R = 4.0$ such that it matches the maximum radial displacement $R_{max} \approx 3.94$ with respect to the simulation box center of an undriven swarm in the near-critically damped regime (pyramid symbol), ensuring sufficient interaction with the active matter system. We note that the swarm extent for other dynamic regimes is comparable (see \tabref{tab:supplementary_videos_no_driver}).
The second issue is that changing the current driver position to a random position at each time step does not allow sufficient time for the active matter system to react to build memory. To this end, we try out multiple frequencies of driver changes (driver position changes each $\Delta s \in \{1, 2, 5, 10\}$ integration steps) and show the resulting trajectories in \figref{fig:uniform_random_time_series_circle}. The corresponding videos for the active matter simulations with random driver placements of different change frequencies can be found in \tabref{tab:supplementary_videos_random_memory_capacity_interval_1} -- \tabref{tab:supplementary_videos_random_memory_capacity_interval_10}.
We report the total memory capacity in \figref{fig:speed-controller-memcap-total} and the $k$-delay memory capacity in \figref{fig:speed-controller-memcap-k-delay} for each driver position change frequency. We observe in \figref{fig:speed-controller-memcap-total} that the total memory capacity remains quite low, with peak values ranging from 2.9 for $\Delta s = 1$ to 30.0 for $\Delta s = 10$. The memory capacity seems to be highest between the parameter combinations indicated by the square and the diamond symbol. For the near-critically damped regime (pyramid symbol), it remains higher than in the underdamped regime. In \figref{fig:speed-controller-memcap-k-delay}, it becomes clear that the buildup of memory requires a speed-controller setting in the transition regime between the underdamped and the near-critically damped regime (diamond symbol), together with sufficiently long waiting times $\Delta s$ allowing a sufficient active matter response.
In \figref{fig:speed-controller-memory-capacity}, we tested feeding a continuous input signal -- the chaotic Lorenz-63 trajectory that we use by default -- into our active matter reservoir to determine a modified version of the memory capacity. This modified method confirms the optimality of the transition regime (diamond symbol) to obtain the highest short-term memory capacity, here with a much higher value of 112.0.
These findings suggest that the near-critically damped regime possesses a reasonable amount of memory -- albeit not the highest found among different speed-controller settings -- and likely possesses other favourable properties such as non-linearity or consistency to obtain optimal performance in this regime. Future studies could further enhance the connection between physical observables and these generic reservoir computing properties.
}

%% file: supp_esn_comparison.tex

\section{\label{subsec:esn_comparison}Echo State Network Comparison}

\new{To obtain a rough estimate of the performance that we obtained using our active matter reservoir computing studies, we perform Echo State Network (ESN) simulations using standard tools (the \texttt{ReservoirPy} framework for simple ESN architectures \cite{Trouvain2020} and the \texttt{Hyperopt} framework for hyperparameter optimization \cite{Bergstra2015}), following a standard procedure \cite{Hinaut2021}. We use the same Lorenz-63 input time series for training and testing as for the active matter system, and apply z-score standardization (scaling to zero mean and unit variance) per dimension by obtaining the scaler from the training data and applying it to both train and test data. We use an ESN with the hyperparameters leaking rate $\text{lr}$, spectral radius $\text{sr}$, input scaling $\text{is}$, and Ridge parameter $\text{ridge}$. 
We first perform a coarse scan with 125 trials of hyperparameter optimization with 25 instances per trial using the default method (random search) with reservoir size fixed to $N = 100$, spectral radius $\texttt{sr} \sim \mathrm{LogUniform}(10^{-2}, 10^{1})$, leaking rate
$\texttt{lr} \sim \mathrm{LogUniform}(10^{-3}, 1)$, input scaling $\texttt{is} = 1.0$, and ridge
regularization parameter $\texttt{ridge} \sim \mathrm{LogUniform}(10^{-9}, 10^{-3})$.
We then reduce our search intervals to $\texttt{sr} \sim \mathrm{LogUniform}(10^{-2}, 3)$, leaking rate $\texttt{lr} \sim \mathrm{LogUniform}(10^{-1}, 1)$, and ridge regularization parameter $\texttt{ridge} \sim \mathrm{LogUniform}(10^{-9}, 10^{-3})$ and use 10,000 trials with 25 instances per trial.
The optimal values found are $\texttt{sr} = 1.584012$ and $\texttt{lr} = 0.6585810$ (see also \figref{fig:hyperopt-trials-summary}).
We then check for the optimal input scaling and vary $\texttt{is} \sim \mathrm{LogUniform}(10^{-2}, 10^{2})$ and $\texttt{ridge} \sim \mathrm{LogUniform}(10^{-9}, 10^{-3})$ using 1,000 trials with 25 instances per trial, and find the optimal value $\texttt{is} = 1.051989$.
We then choose $N=600$ neurons to match the number of observation kernel outputs that define our active matter reservoir state for a reasonable comparison, noting that we did not optimize hyperparameters for our active matter reservoir computer. 
We find the best regularization parameter by searching for the optimal Ridge parameter $\texttt{ridge} \sim \mathrm{LogUniform}(10^{-9}, 10^{-3})$ using 1,000 trials with 30 instances per trial. We found an optimal value of $\texttt{ridge} = 9.000225 \cdot 10^{-4}$. The final predictive performances for this hyperparameter combination ($N= 600$, $\texttt{sr} = 1.584012$, $\texttt{lr} = 0.6585810$, $\texttt{is} = 1.051989$, $\texttt{ridge} = 9.000225 \cdot 10^{-4}$) as as follows: Pearson correlation coefficient for the $x$ dimension $P= 0.833$, for the $y$ dimension $P_y = 0.746$,  $\mathrm{NRMSE}=0.619$, $\mathrm{NMSE}=0.384$, $\mathrm{sMAPE}=7.23\,\%$. 
In the active matter speed-controller scan (\figref[a]{fig:Lymburn_critical_speed_controller_scan_larger}), we obtain in the near-critically damped regime (pyramid symbol) the scores $P = 0.879$, 
$P_y = 0.786$, 
$\mathrm{NRMSE}= 0.542$, 
$\mathrm{NMSE}=0.294$, 
$\mathrm{sMAPE}= 6.42\,\%$. 
This means that the performance of a near-critically damped active matter reservoir computer for the given task is -- despite not having optimized its Ridge hyperparameter and further physical parameters -- about 5.5\,\% higher than the performance of a comparable Echo State Network. 
}

%% file: supp_data.tex

\section{Data Availability}

Data generated in this study can be accessed in the Data Repository of the University of Stuttgart (DaRUS), in the dataverse Stuttgart Center for Simulation Science EXC 2075 / Project 6-15:\\
\href{https://doi.org/10.18419/darus-4620}{\texttt{https://doi.org/10.18419/darus-4620}} \cite{DARUS-4620_2025}\\
All data in this manuscript were generated using the ResoBee software for active matter reservoir computing. It is currently being prepared for publication in an open-source software journal.\\

%% file: supp_figs.tex

\section{Supplementary Figures}

\subsection{Methods}

\begin{figure}[H]
    \centering
    \includegraphics[width=5cm]{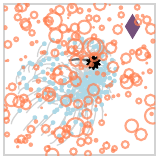}
    \caption{
        \textbf{Random placement of the Gaussian kernel observers} (orange circles) for extracting coarse-grained information from the soft matter system. The circle center points correspond to kernel positions $\mathbf{c}_m$ and circle widths correspond to the kernel widths $w_m$ as defined in \eqqref{eq:gaussian-kernel}.
    }
\label{fig:gaussian_kernels_placement}
\end{figure}

\vspace{3cm}

\subsection{Reproductions of Lymburn \etal (2021)}

\begin{figure}[H]
    \centering 
    \includegraphics{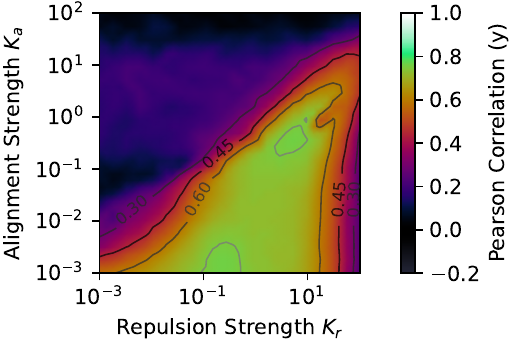}
    \caption{
        \textbf{Predictive performance for different active matter regimes in a repulsion/alignment strength parameter scan} (reproduction of Fig.\ 8a in \refref{\cite{Lymburn2021}}). Predictive performance is quantified by the Pearson correlation coefficient between the actual and the predicted time series using the active matter reservoir computer with a given repulsion/alignment force strength parameter combination. 
    }
    \label{fig:reproduction_lymburn_alignment_repulsion_performance}
\end{figure}

\begin{figure}[H]
    \centering
    \subfloat[Undriven System, Polarity]{\includegraphics{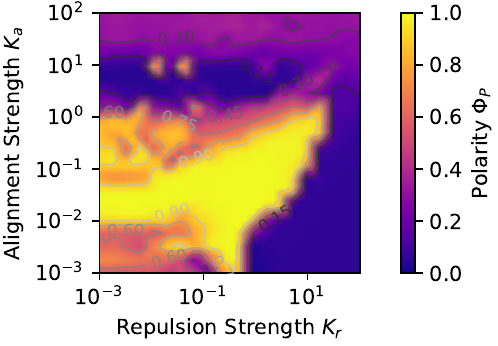}}\hfill
    \subfloat[Driven System, Polarity]{\includegraphics{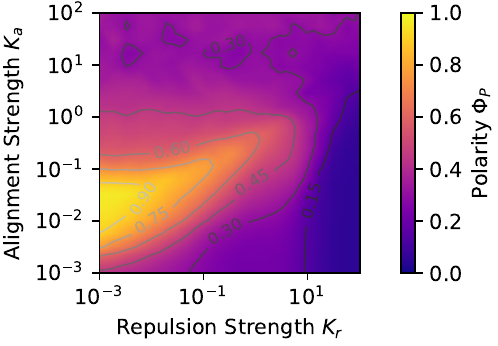}}\\
    \vspace{1em}
    \subfloat[Undriven System, Rotation]{\includegraphics{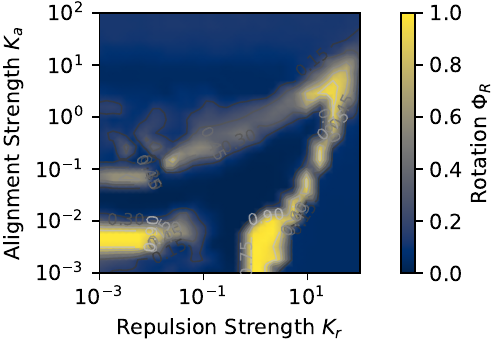}}\hfill
    \subfloat[Driven System, Rotation]{\includegraphics{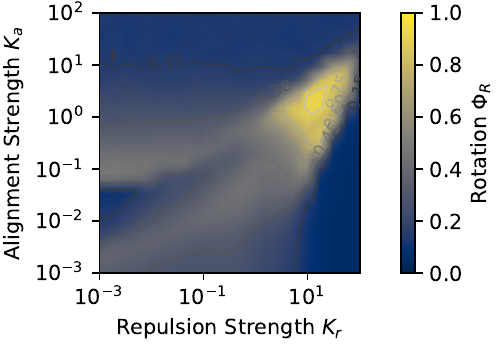}}

    \caption{\textbf{Polarity and rotation order parameters in repulsion/alignment force strength parameter scans} -- with and without an external driving force, characterizing different active matter regimes (reproduction of Fig.\ 6 in \refref{\cite{Lymburn2021}}).}
    \label{fig:reproduction_lymburn_alignment_repulsion_order_parameters}
\end{figure}

\subsection{Speed-controller Parameter Scan}

\begin{figure}[H]
    \centering
    \includegraphics[scale=.8]{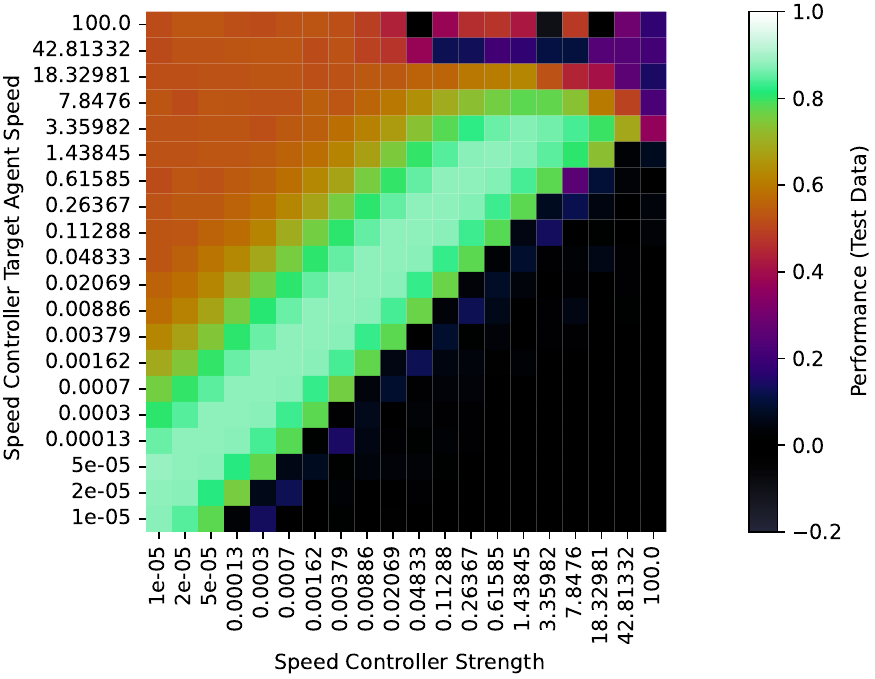}
  \caption{\new{\textbf{Non-interpolated predictive performances for the speed-controller parameter scan} shown in \figref[a]{fig:Lymburn_critical_speed_controller_scan_larger}.}}
    \label{fig:speed-controller-not-interpolated}
\end{figure}

\begin{figure}[H]
    \centering
    \subfloat[MSDs per agent]{\includegraphics{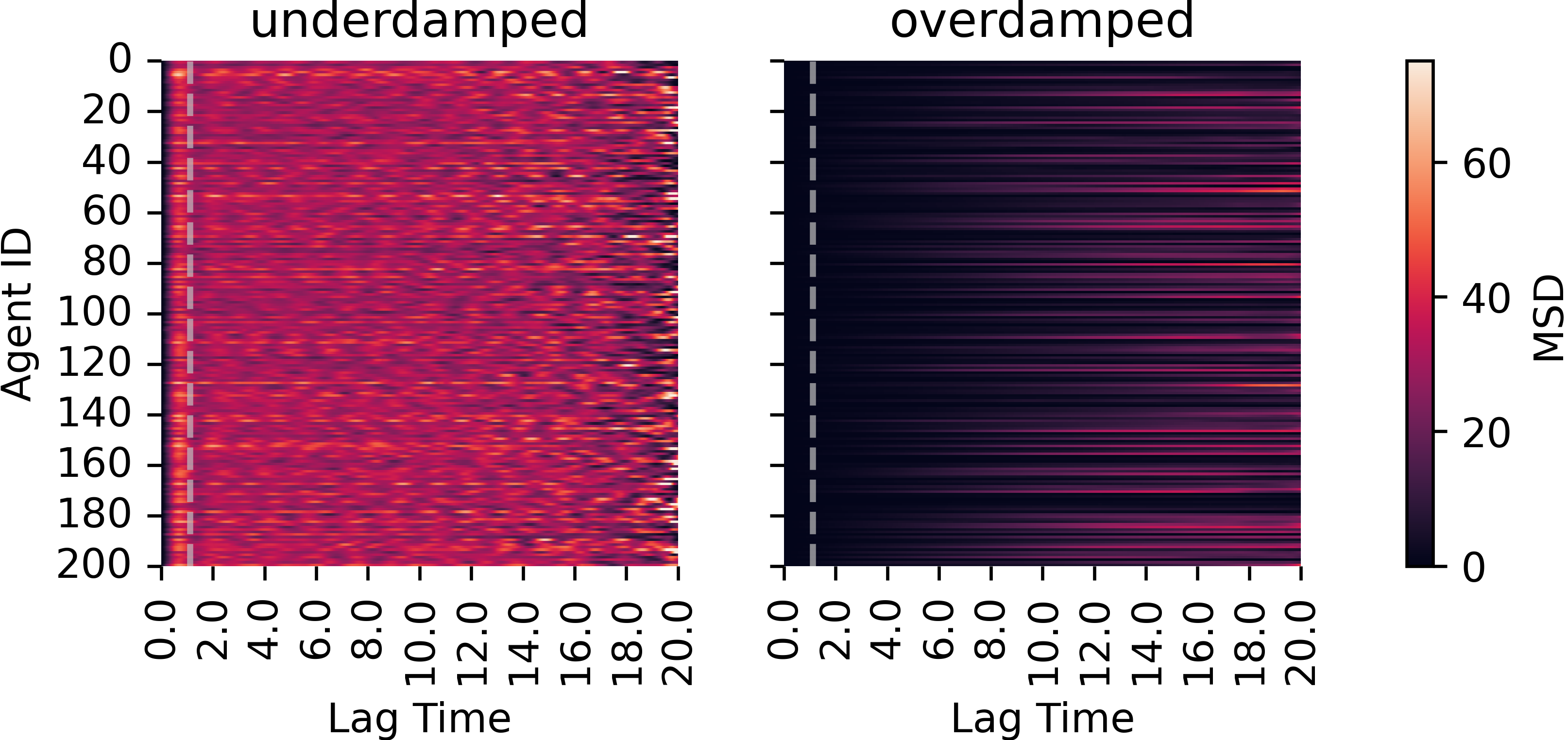}}\\
    \vspace{1em}
    \subfloat[Agent-averaged MSDs]{\includegraphics{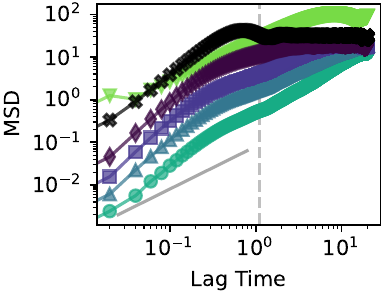}}

 \caption{
        \textbf{Mean squared displacements (MSDs) for 200 agents in the speed-controller parameter scan.} Grey dashed lines indicate the Lyapunov time $t^{L63}_{\text{lyap}} \approx (0.90566)^{-1} \approx 1.1$ \cite{Viswanath1998} of the studied Lorenz-63 external driving signal. The agent-averaged MSD values at this time are taken for the $\text{MSD}(t^{\text{L63}}_{\text{lyap}})$ plot in the main text in \figref{fig:Lymburn_critical_speed_controller_scan_larger}.
        (a) MSDs \textit{per agent} in the underdamped (\figref{fig:Lymburn_critical_speed_controller_scan_larger} cross symbol) and the overdamped damped (circle symbol) regimes. In the underdamped regime agents experience high mean squared displacements with rare returns to their original positions. On the contrary, in the overdamped regime, the displacement over the Lyapunov time is much lower than in the underdamped regime. In the overdamped regime, agents return to their undriven ground-state positions after external driving, rendering a well-predictable agent response. For some agents the external driving leads to agents leaving their undriven ground-state positions, which eventually results in higher MSDs after several Lyapunov times.
        (b) \textit{Agent-averaged} mean squared displacements (MSD) for different parameter combinations (symbols) in the speed-controller parameter scan in \figref{fig:Lymburn_critical_speed_controller_scan_larger}. The solid grey line indicates normal diffusion (Brownian motion) with $\text{MSD}(\tau) \sim \tau$.
    }
    \label{fig:msds}
\end{figure}

\begin{figure}[H]
    \centering
    \subfloat[Polarity]{\includegraphics{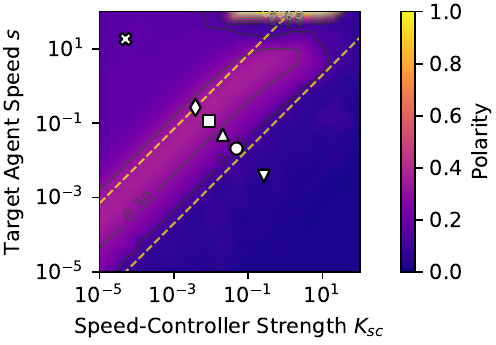}}\hfill
    \subfloat[Rotation]{\includegraphics{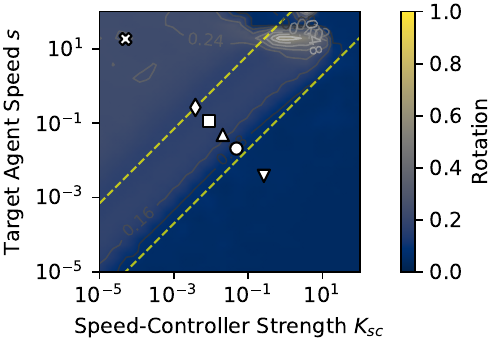}}

     \caption{\textbf{Time-averaged polarity $\Phi_P$ and rotation $\Phi_R$ order parameters for varied speed-controller parameters.} 
    The diagonal high $\Phi_P$ region (diamond, square symbols) correlates with high predictive performance regions (near-critical damping) shown in \figref[a]{fig:Lymburn_critical_speed_controller_scan_larger}, but is slightly shifted. When moving towards higher ratios $r = K_{sc}/s$ (overdamped regime, circle symbol), $\Phi_P$ decreases since agents mostly remain at their undriven ground state positions with $|\textbf{v}_i| \approx 0$. Collectivity decreases for smaller ratios $r$ (cross symbol). }
    \label{fig:Lymburn_critical_speed_controller_scan_heatmap_canoncial_order_parameters}
\end{figure}

\begin{figure}[H]
    \centering
    \includegraphics{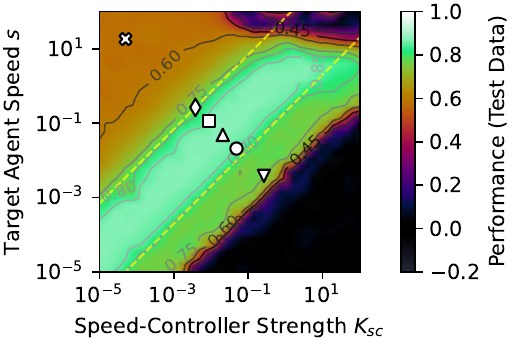}
    \caption{
        \textbf{Predictive performance for a speed-controller parameter scan with a smaller integration time step of $\Delta t = 0.002$} instead of the default integration time step $\Delta t = 0.02$ used in \figref[a]{fig:Lymburn_critical_speed_controller_scan_larger}. Compared to \figref[a]{fig:Lymburn_critical_speed_controller_scan_larger}, there is no sharp drop in predictive performance at the lower dashed yellow line. The arrested regime is shifted outwards towards the lower right-hand side of the diagram, expanding the range for predictive performances $> 0.75$ for the Lorenz-63 prediction. Here, the superposition of oscillations (alternating forward-backward motion) observed for the original timestep of $\Delta t = 0.02$, disappears for the circle and nabla symbols. We also observe a higher performance beyond the upper yellow dashed line in the upper left-hand side of the diagram. This could be the effect of reduced variations within the Lorenz-63 trajectory integrated with the smaller integration time step, which yields a slightly simpler prediction task.
    } 
    \label{fig:Lymburn_critical_scans_speed_controller_smaller_integration_time_step}
\end{figure}

\begin{figure}[H]
    \centering
    \subfloat[12 integration time steps ahead ($\Delta t = 0.24$)]{\includegraphics{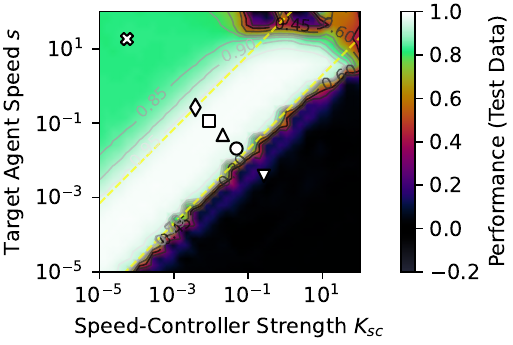}}\hfill
    \subfloat[50 integration time steps ahead ($\Delta t = 1.0$)]{\includegraphics{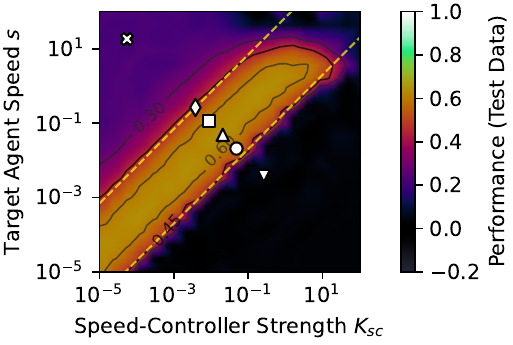}}\\
   \vspace{1em}
    \subfloat[100 integration time steps ahead ($\Delta t = 2.0$)]{\includegraphics{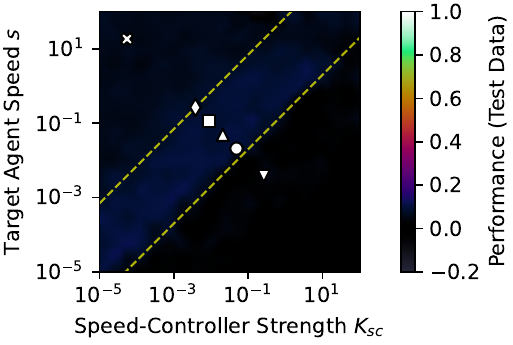}}\hfill
    \subfloat[Scaling]{\includegraphics{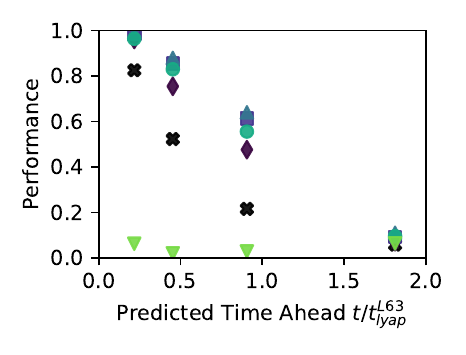}}

    \caption{\textbf{Predictive performances for speed-controller parameter scans with different prediction times ahead} analogous to \figref[a]{fig:Lymburn_critical_speed_controller_scan_larger}. The near-critically damped regime remains the optimal dynamical regime for making predictions with the employed active matter reservoir computing setup for different prediction times ahead.}
    \label{fig:speed-controller-varied-prediction-ahead}
\end{figure}

\begin{figure}[H]
    \centering
     \includegraphics{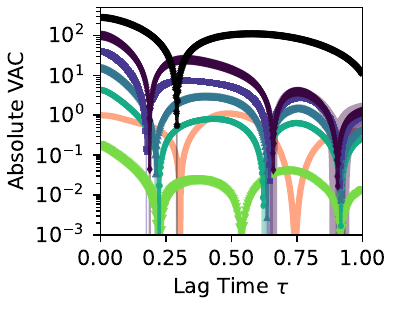}
    
    \caption{
       \textbf{Velocity auto-correlation (VAC) for the speed-controller parameter scan with a smaller integration time step of $\Delta t = 0.002$} shown in \figref{fig:Lymburn_critical_scans_speed_controller_smaller_integration_time_step}, in analogy to \figref[c]{fig:damping}. The smaller integration time step resolves the oscillations observed in the overdamped (circle symbol) and arrested (nabla symbol) regimes observed for the default integration time step of  $\Delta t = 0.02$. Better statistics and less variation in the Lorenz-63 trajectory integrated with a smaller integration time step allow a better resolution of the absolute VAC curves (total time simulated: $T = 1,000.0$; with $\Delta t = 0.002$: $N = 5 \cdot 10^{5}$ simulation steps; with $\Delta t = 0.02$: $5 \cdot 10^{4}$ simulation steps).
    }
    \label{fig:vacf_0.002}
\end{figure}

\begin{figure}[H]
    \centering
    \includegraphics{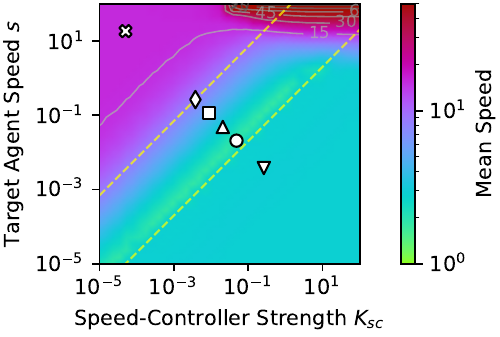}
    \caption{\textbf{Mean particle speeds for the speed-controller parameter scan} displayed in \figref[a]{fig:Lymburn_critical_speed_controller_scan_larger}. Lower mean speeds are observed as the speed-controlling force becomes stronger with respect to the total force acting on an agent, which causes stronger damping (moving from the cross symbol toward the circle symbol).  } \label{fig:Lymburn_critical_scans_speed_controller_observable_heatmaps_mean_speed}
\end{figure}

\begin{figure}[H]
    \centering
    \includegraphics[scale=1.0]{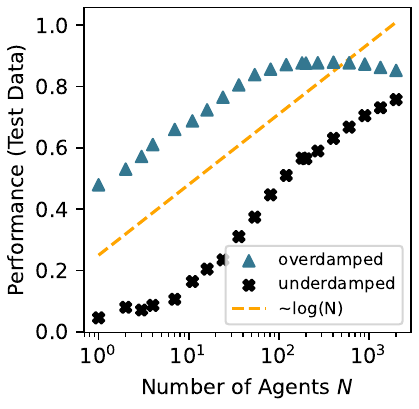}
  \caption{\new{\textbf{Scaling behavior of predictive performance with the number of agents in a reservoir for different dynamical regimes:} overdamped (target agent speed $s = 0.0016237767$) and underdamped ($s = 1.4384498883$); speed controller strength $K_{sc} = 0.001$. All other parameters are fixed as specified in \sectref{sec:mm-simulation-details}. }}
    \label{fig:n-agents-scaling}
\end{figure}

\subsection{Intrinsic Agent Relaxation Dynamics}

\begin{figure}[H]
    \centering
    \subfloat[\centering Structural excitation, \textit{without} agent-agent interactions ($\Delta t = 0.002$)]{\includegraphics{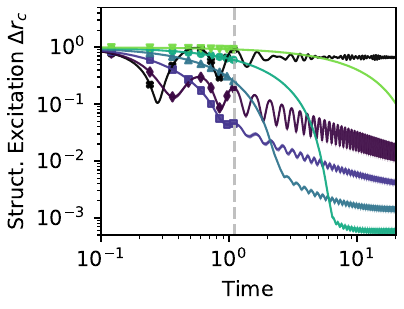}}\hfill
    \subfloat[\centering Dynamical excitation, \textit{without} agent-agent interactions ($\Delta t = 0.002$)]{\includegraphics{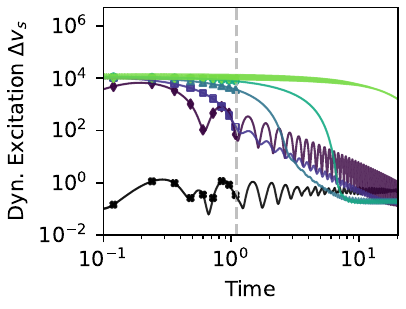}}\\
       \vspace{0.5em}
    \subfloat[\centering Structural excitation, with full interactions ($\Delta t = 0.02$)]{\includegraphics{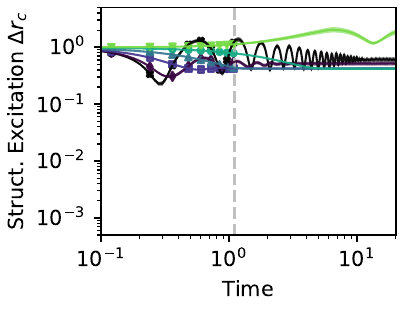}}\hfill
    \subfloat[\centering Dynamical excitation, with full interactions ($\Delta t = 0.02$)]{\includegraphics{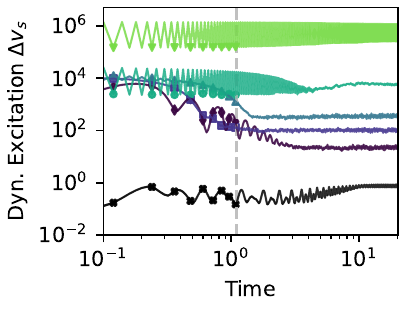}}\\
       \vspace{0.5em}
    \subfloat[\centering Structural excitation, with full interactions ($\Delta t = 0.002$)]{\includegraphics{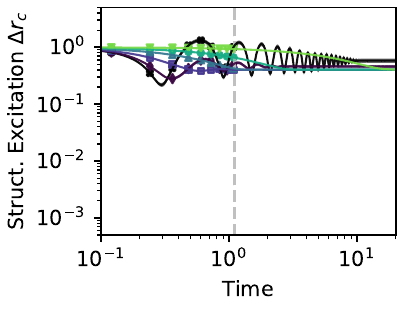}}\hfill
    \subfloat[\centering Dynamical excitation, with full interactions ($\Delta t = 0.002$)]{\includegraphics{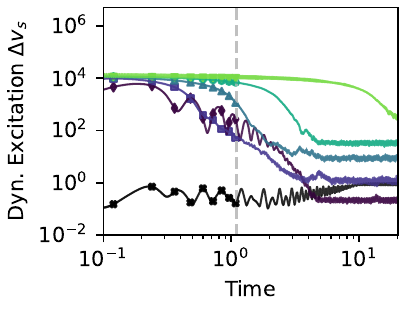}}

    \caption{
       \textbf{Intrinsic agent relaxation dynamics for simulations \textit{without} (a,b) and \emph{with} (c-f) agent-agent interactions in place and with different integration time steps} analogously to \figref[a,b]{fig:damping}. The smaller integration time step of $\Delta t = 0.002$ allows resolving relaxation dynamics in the far overdamped regime (nabla symbol) that is otherwise rendered as an arrested regime for $\Delta t = 0.02$.  The presence of agent-agent interactions (d,f) (which also applies to (c,e)) prominently hinders the oscillations in the relaxation dynamics above critical damping (square, diamond, cross), and subtly shifts the time-points of relaxation for the near-critical and overdamped regimes (pyramid, circle).
    }
    \label{fig:damping_variations}
\end{figure}

\subsection{Speed-controller Parameter Scans for Different Chaotic Time Series Benchmarks}

\begin{figure}[H]
    \centering
    \subfloat[Hénon–Heiles]{\includegraphics{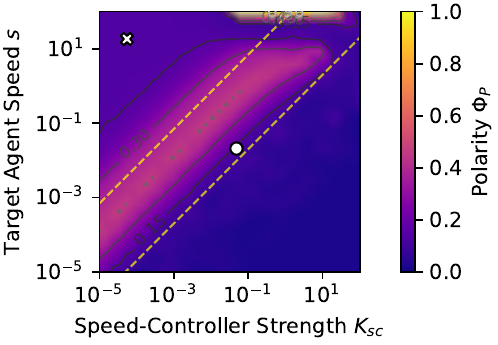}}\hfill
    \subfloat[Rössler]{\includegraphics{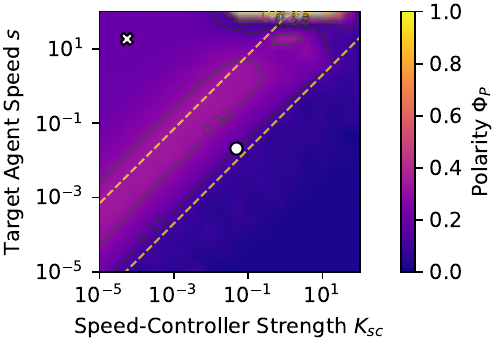}}\\
          \vspace{1em}
    \subfloat[Chua]{\includegraphics{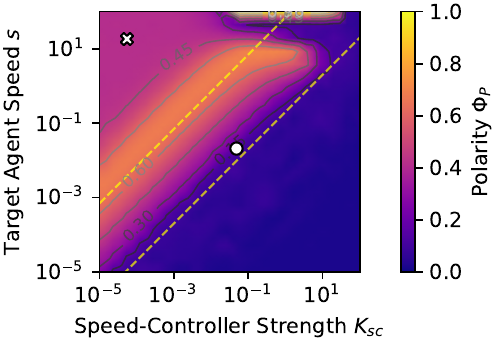}}\hfill
    \subfloat[Lorenz-96]{\includegraphics{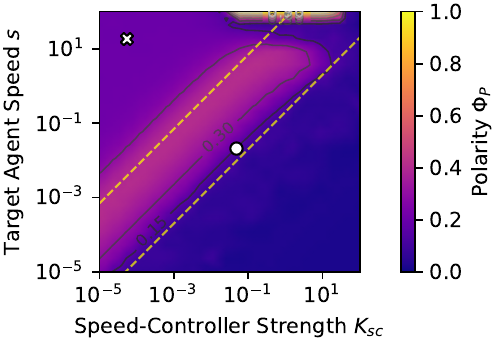}}

    \caption{\textbf{Polarity order parameter for various chaotic time-series benchmarks} presented in \figref{fig:benchmarks} and analogous to \figref[a]{fig:Lymburn_critical_speed_controller_scan_heatmap_canoncial_order_parameters} for the Lorenz-63 attractor. }
    \label{fig:benchmarks-scalar-polarity}
\end{figure}

\begin{figure}[H]
    \centering
    \subfloat[Hénon–Heiles]{\includegraphics{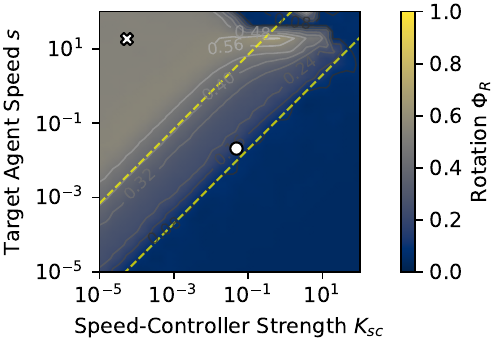}}\hfill
    \subfloat[Rössler]{\includegraphics{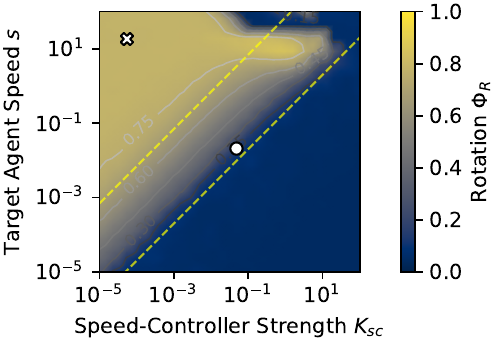}}\\
          \vspace{1em}
    \subfloat[Chua]{\includegraphics{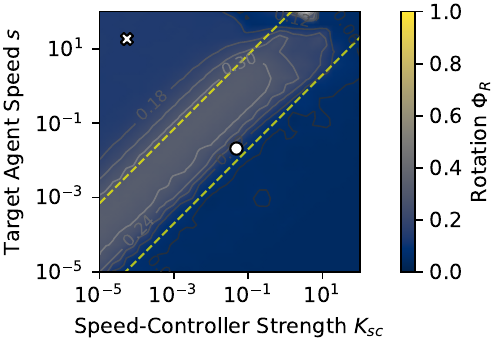}}\hfill
    \subfloat[Lorenz-96]{\includegraphics{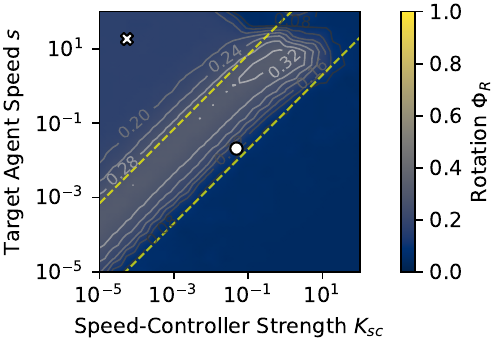}}

    \caption{\textbf{Rotation order parameter for various chaotic time-series benchmarks} presented in \figref{fig:benchmarks} and analogous to \figref[b]{fig:Lymburn_critical_speed_controller_scan_heatmap_canoncial_order_parameters} for the Lorenz-63 attractor.}
    \label{fig:benchmarks-scalar-rotation}
\end{figure}

\begin{figure}[H]
    \centering
    \subfloat[Hénon–Heiles]{\includegraphics{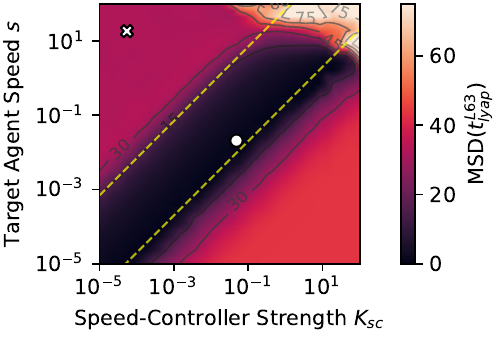}}\hfill
    \subfloat[Rössler]{\includegraphics{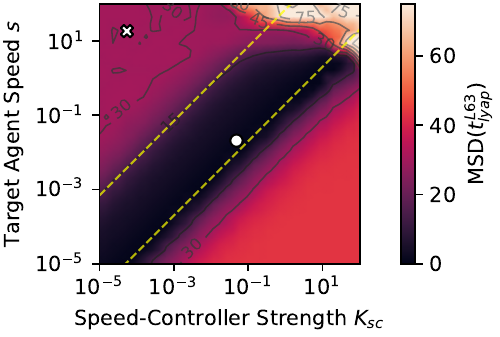}}\\
          \vspace{1em}
    \subfloat[Chua]{\includegraphics{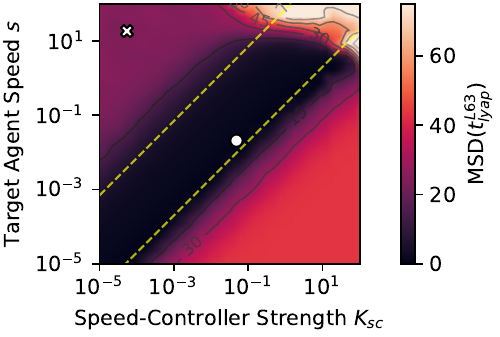}}\hfill
    \subfloat[Lorenz-96]{\includegraphics{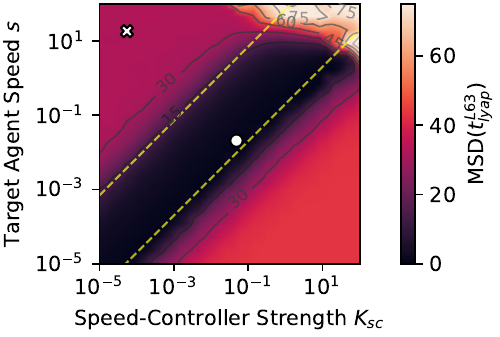}}
    
    \caption{\textbf{Agent-averaged mean squared displacements (MSDs) after the Lorenz-63 Lyapunov time} $\mathbf{t^{L63}_{\text{\textbf{lyap}}}}$ \textbf{for various chaotic time-series benchmarks} presented in \figref{fig:benchmarks}. The MSD is studied at the Lyapunov time of the default Lorenz-63 system -- $t^{\text{L63}}_{\text{lyap}} \approx (0.90566)^{-1} \approx 1.1$ \cite{Viswanath1998} -- and is considered here also for the other attractors for simplicity. As for the Lorenz-63 system in \figref[a]{fig:Lymburn_critical_speed_controller_scan_larger}, low MSDs characterize the near-critically damped high-performance regimes in \figref{fig:benchmarks}.}
    \label{fig:benchmarks-agent-avg-msds}
\end{figure}

\begin{figure}[H]
    \centering
    \subfloat[Hénon–Heiles]{\includegraphics{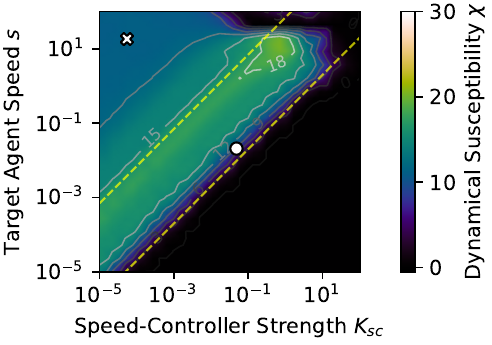}}\hfill
    \subfloat[Rössler]{\includegraphics{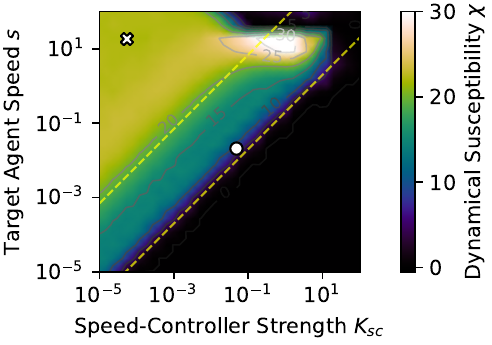}}\\
          \vspace{1em}
    \subfloat[Chua]{\includegraphics{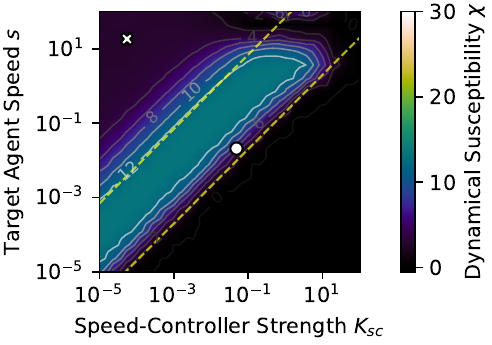}}\hfill
    \subfloat[Lorenz-96]{\includegraphics{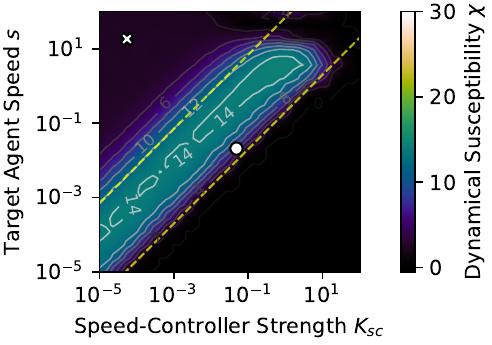}}
    
    \caption{\textbf{Dynamical susceptibility for various chaotic time-series benchmarks} presented in \figref{fig:benchmarks} and analogous to \figref[a]{fig:speed_controller_fluctuations} for the Lorenz-63 attractor in the main text. High dynamical susceptibility correlates with the near-critically damped high-performance region in \figref{fig:benchmarks} for swarms with low rotation, which is displayed in \figref{fig:benchmarks-scalar-rotation}. Note that rotations and dilatations were not taken into account when computing the velocity fluctuations as a prerequisite for the dynamical susceptibility (see \eqqref{eq:velocity-fluctuation} and its discussion). Because of this, we measure high dynamical susceptibilities $\chi > 15.0$ in the underdamped regimes (upper left-hand pyramid) of the Hénon–Heiles and Rössler attractors, which feature high rotation as shown in \figref[a,b]{fig:benchmarks-scalar-rotation}.}
    \label{fig:benchmarks-attanasi-susceptibility}
\end{figure}

\begin{figure}[H]
    \centering
    \subfloat[Hénon–Heiles]{\includegraphics{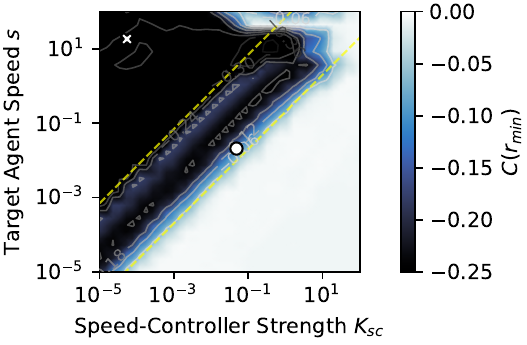}}\hfill
    \subfloat[Rössler]{\includegraphics{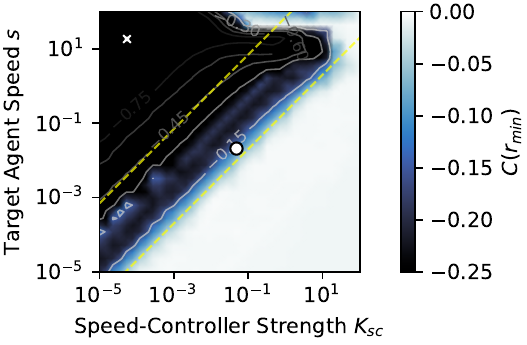}}\\
          \vspace{1em}
    \subfloat[Chua]{\includegraphics{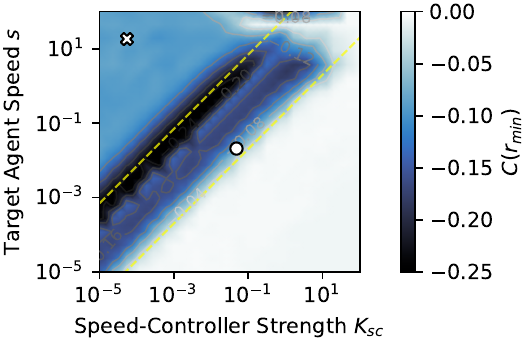}}\hfill
    \subfloat[Lorenz-96]{\includegraphics{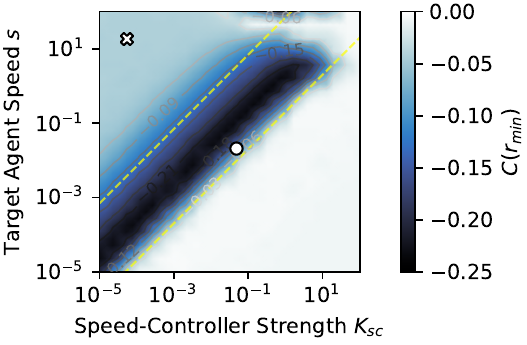}}
    
    \caption{\textbf{First local minimum of the connected velocity correlation function $\textbf{CVC}(\mathbf{r}_{\text{min}})$ (see \eqqref{eq:connected_velocity_correlation}) for various chaotic time-series benchmarks} presented in \figref{fig:benchmarks}. The color bar is the same as used in the main text for the Lorenz-63 attractor in \figref[b]{fig:speed_controller_fluctuations}; values below -0.25 are indicated here by contour lines.}
    \label{fig:benchmarks-minimum-conn-vel-corr}
\end{figure}

\begin{figure}[H]
    \centering
     \includegraphics{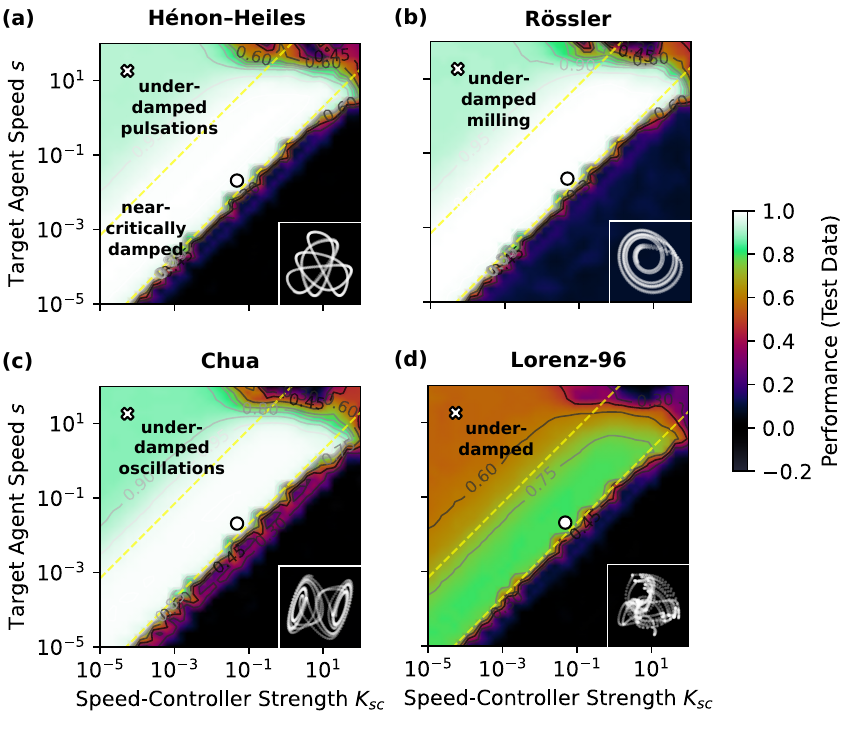}
    \caption{
        \textbf{Predictive performance for the speed-controller parameter scans for different chaotic attractors with the same \textit{real time} (not Lyapunov time) predicted ahead for all attractors} (25 integration time steps of $\Delta t = 0.02$, $\Delta t_{\text{pred}} = 0.50$). In contrast, the main text Lyapunov time-adjusted prediction times ahead are presented in \figref{fig:benchmarks}. 
        In attractor Lyapunov times this corresponds to $\Delta t_{\text{pred}} \equiv 0.45283\ t_{\text{lyap}}^{\text{L63}} \approx 0.0753\ t_{\text{lyap}}^{\text{Rössler}} \approx 0.0178\ t_{\text{lyap}} ^{\text{Hénon–Heiles}} \approx 0.615\ t_{\text{lyap}}^{\text{Chua}} \approx 0.668\ t_{\text{lyap}}^{\text{L96}}$. The Rössler attractor can be predicted with a performance of up to 0.9991, Chua with up to 0.9905, Hénon–Heiles with up to 0.9965, and Lorenz-96 with up to 0.7983.
    }
    \label{fig:benchmarks-same-real-time-predicted-ahead}
\end{figure}

\begin{figure}[H]
    \centering
    \subfloat[Hénon–Heiles (cross symbol)]{\includegraphics{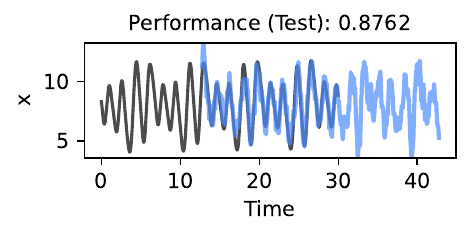}}\hfill
    \subfloat[Hénon–Heiles (circle symbol)]{\includegraphics{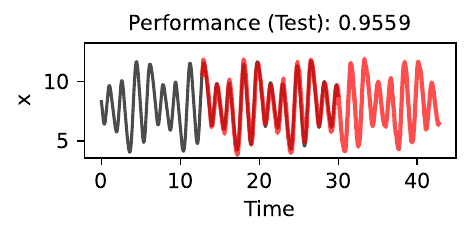}}\\
          \vspace{1em}
    \subfloat[Chua (cross symbol)]{\includegraphics{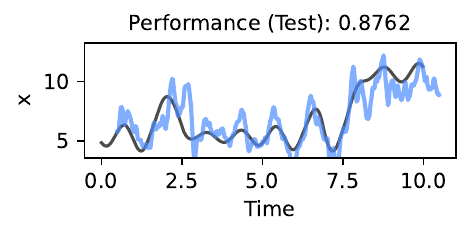}}\hfill
    \subfloat[Chua (circle symbol)]{\includegraphics{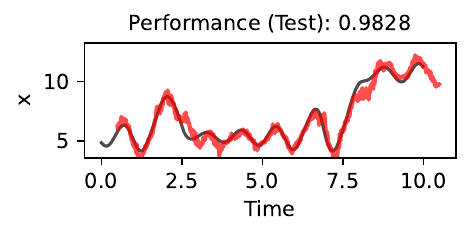}}\\
          \vspace{1em}
    \subfloat[Rössler (cross symbol)]{\includegraphics{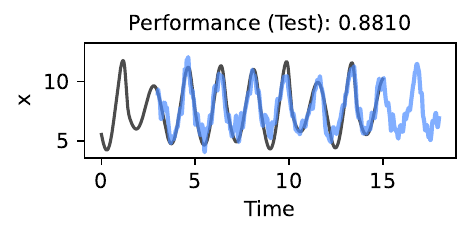}}\hfill
    \subfloat[Rössler (circle symbol)]{\includegraphics{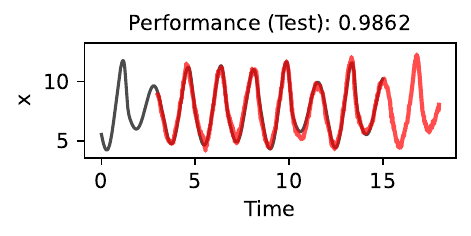}}\\
          \vspace{1em}
    \subfloat[Lorenz-96 (cross symbol)]{\includegraphics{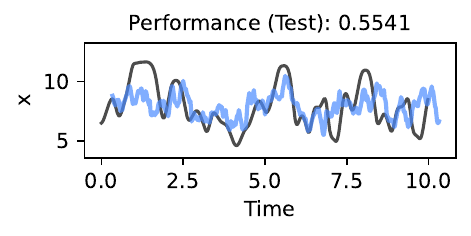}}\hfill
    \subfloat[Lorenz-96 (circle symbol)]{\includegraphics{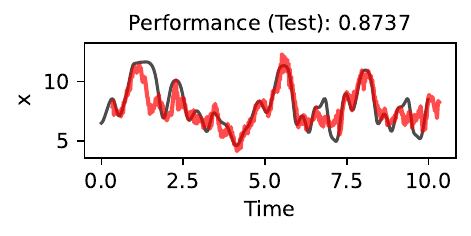}}

    \caption{\textbf{Actual (black line) and \textit{0.45283 Lyapunov times ahead predicted} time series for different chaotic attractors using underdamped (cross symbol, blue line) and overdamped (circle symbol, red line) active matter reservoir computers}. Displayed are only the first 500 integration time steps.  Shifts in start and end times between the actual and the predicted time series indicate the predicted time ahead. Dynamical regimes correspond to systems with $N=200$ agents and parameter combinations displayed in \figref{fig:benchmarks}.}
    \label{fig:benchmarks-actual-vs-predicted-timeseries}
\end{figure}

\begin{figure}[H]
    \centering
    \subfloat[Hénon–Heiles (cross symbol)]{\includegraphics{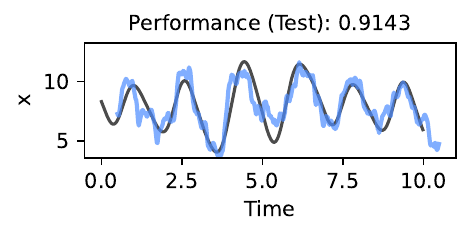}}\hfill
    \subfloat[Hénon–Heiles (circle symbol)]{\includegraphics{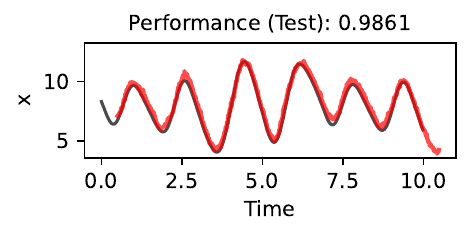}}\\
          \vspace{1em}
    \subfloat[Chua (cross symbol)]{\includegraphics{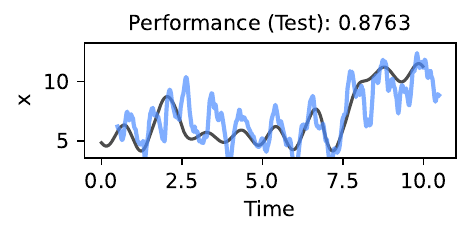}}\hfill
    \subfloat[Chua (circle symbol)]{\includegraphics{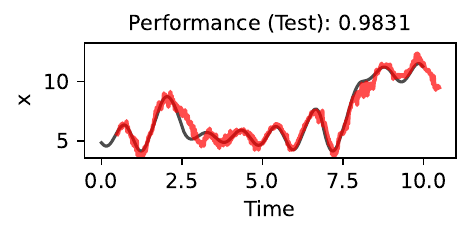}}\\
          \vspace{1em}
    \subfloat[Rössler (cross symbol)]{\includegraphics{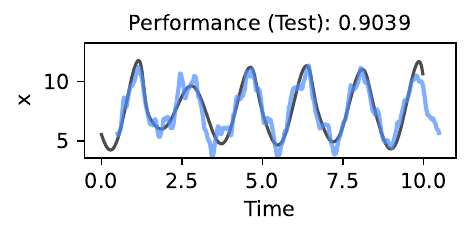}}\hfill
    \subfloat[Rössler (circle symbol)]{\includegraphics{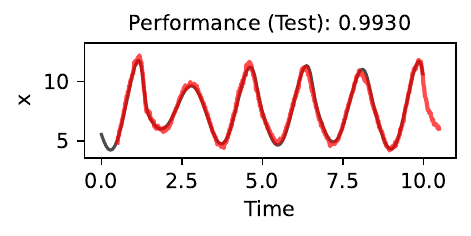}}\\
          \vspace{1em}
    \subfloat[Lorenz-96 (cross symbol)]{\includegraphics{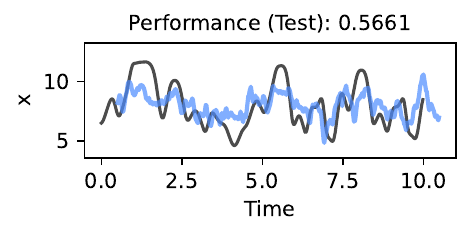}}\hfill
    \subfloat[Lorenz-96 (circle symbol)]{\includegraphics{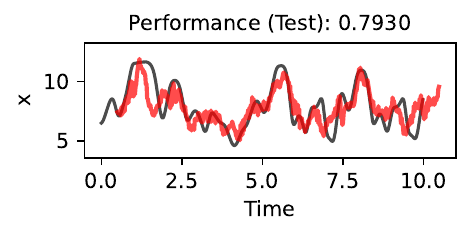}}

    \caption{\textbf{Actual (black line) and \textit{25 integration time steps ahead predicted} time series for different chaotic attractors predicted using underdamped (cross symbol, blue line) and overdamped (circle symbol, red line) active matter reservoir computers.} Displayed are only the first 500 integration time steps for Chua and Lorenz-96, the first 750 for Rössler, and the first 1500 for Hénon–Heiles. Shifts in start and end times between the actual and the predicted time series indicate the predicted time ahead. Parameter combinations (symbols) correspond to systems with $N=200$ agents and parameter combinations displayed in \figref{fig:benchmarks-same-real-time-predicted-ahead}.}
    \label{fig:benchmarks-actual-vs-predicted-timeseries-25-integration-steps-ahead}
\end{figure}

\subsection{Alignment Force Parameter Scan}

\begin{figure}[H]
    \centering
    \subfloat[Polarity]{\includegraphics{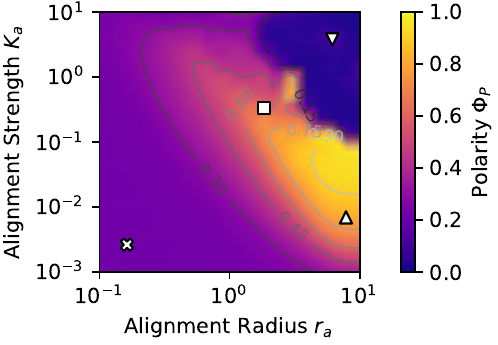}}\hfill
    \subfloat[Rotation]{\includegraphics{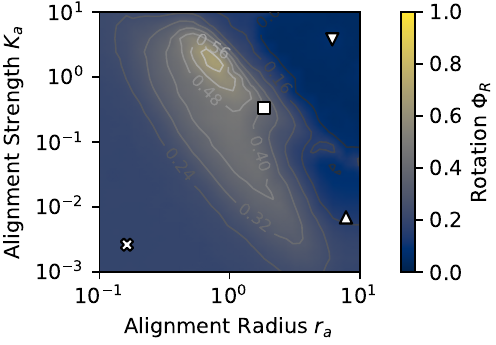}}\\
       \vspace{1em}
    \subfloat[Mean Squared Displacements]{\includegraphics{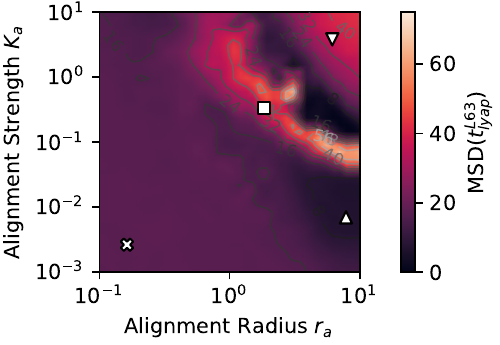}}\\
       \vspace{1em}
    \subfloat[Dynamical Susceptibility\newline]{\includegraphics{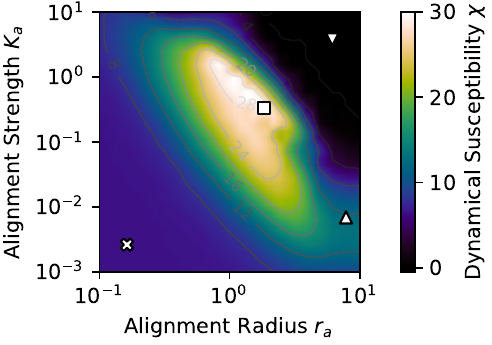}}\hfill
    \subfloat[Connected Velocity Correlation Function at First Local Minimum]{\includegraphics{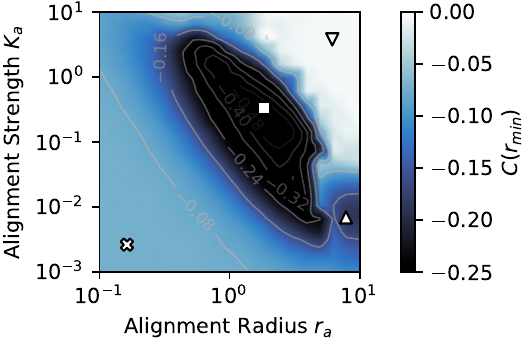}}

    \caption{
        \textbf{Various observables for the alignment force parameter scan} displayed in \figref[a]{fig:alignment}. The polarity heatmap in sub-figure (a) captures the globally frustrated regime well (nabla symbol).
    }
    \label{fig:alignment-observables}
\end{figure}

\subsection{Homing Force Strength Parameter Scan}

\begin{figure}[H]
    \centering
    \includegraphics{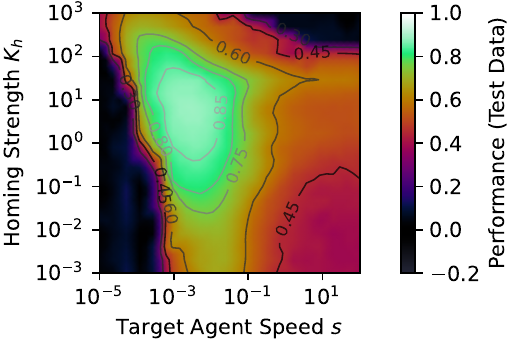}
    \caption{
        \textbf{Predictive performance for the homing force strength $K_h$ / target agent speed $s$ parameter scan} for a fixed speed-controller strength of $K_{sc} = 10^{-3}$. The active matter dynamics for a standard homing force strength of $K_h = 2.0$ are near-critically damped for $2.0 \cdot 10^{-4} \lessapprox s \lessapprox 10^{-1}$ and underdamped for $s \gtrapprox 10^{-1}$ (see \figref[a]{fig:Lymburn_critical_speed_controller_scan_larger}). To achieve maximum performances above 0.85 in the near-critically damped regime, the homing force strength must be tuned as shown here. This parameter scan is complementary to the homing force strength -- speed-controller strength scan shown in \figref[a]{fig:homing}.
    }
    \label{fig:homing_target_agent_speed_scan}
\end{figure}

\begin{figure}[htbp]
    \centering
    \includegraphics[scale=1.0]{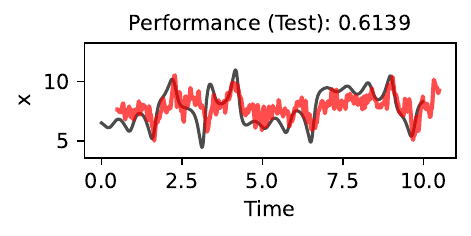}
  \caption{\new{\textbf{Actual versus predicted time series for a near-critically damped active matter system subject to a high homing force strength}, corresponding to the parameter combination indicated by the square symbol in \figref{fig:homing}.}}
    \label{fig:strong-homing-actual-vs-predicted}
\end{figure}

\begin{figure}[htbp]
    \centering
    \includegraphics[scale=1.0]{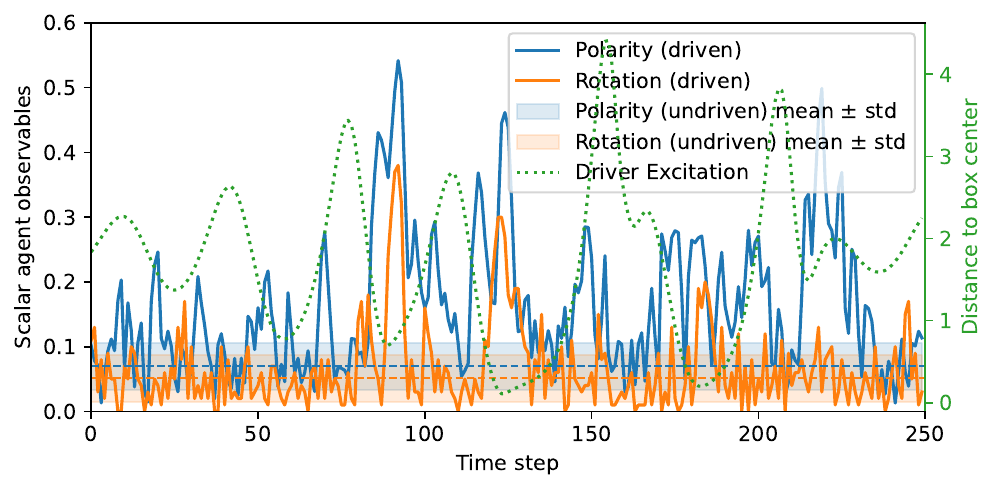}
  \caption{\new{\textbf{Canonical scalar agent observables polarity (blue solid line) and rotation (orange solid line)} for the first 250 integration time steps, found at the parameter combination indicated by the square symbol in \figref{fig:homing}. The blue and orange dashed lines correspond to the mean values of the same simulation without an external driving force, and the bands correspond to the standard deviation. The green dotted line shows the distance of the driver to the center of the simulation box. }}
    \label{fig:strong-homing-actual-observables}
\end{figure}

\subsection{Additional Performance Measures}
\label{sec:additional-performance-measures}

\begin{figure}[H]
    \centering
    \subfloat[Pearson Correlation (y)]{\includegraphics[scale=1.0]{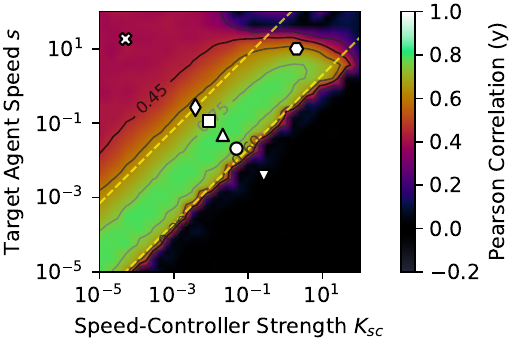}}\hfill
    \subfloat[NRMSE]{\includegraphics[scale=1.0]{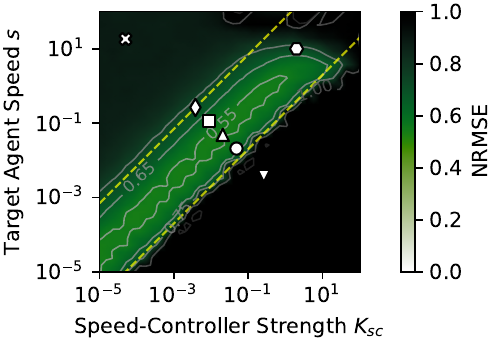}}\\
   \vspace{1em}
    \subfloat[NMSE]{\includegraphics[scale=1.0]{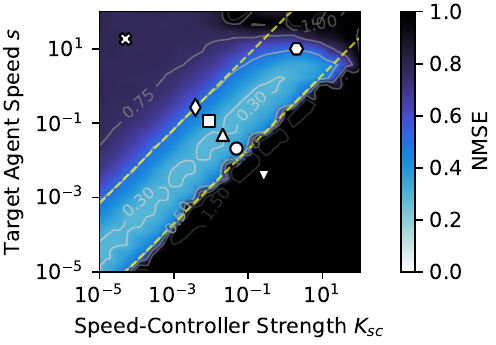}}\hfill
    \subfloat[sMAPE]{\includegraphics[scale=1.0]{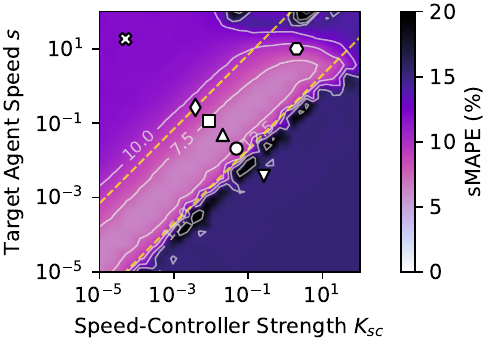}}

    \caption{\new{\textbf{Different performance measures for the speed-controller parameter scan} shown in \figref[a]{fig:Lymburn_critical_speed_controller_scan_larger} in the main text: (a) Pearson correlation coefficient for the second dimension ($y$), (b) Normalized Root Mean Squared Error by standard deviation (NRMSE), (c) Normalized Mean Squared Error (NMSE) by variance, and (d) Symmetric Mean Absolute Percentage Error (sMAPE).}}
    \label{fig:speed-controller-more-metrics}
\end{figure}

\begin{figure}[H]
    \centering
    \subfloat[Pearson Correlation (y)]{\includegraphics[scale=1.0]{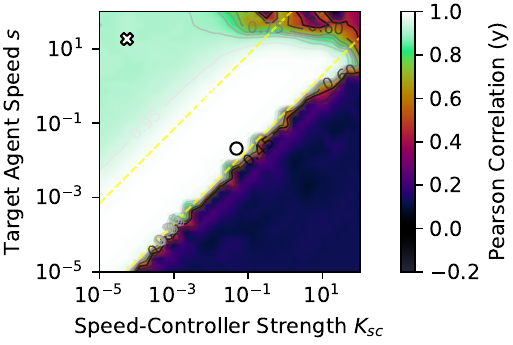}}\hfill
    \subfloat[NRMSE]{\includegraphics[scale=1.0]{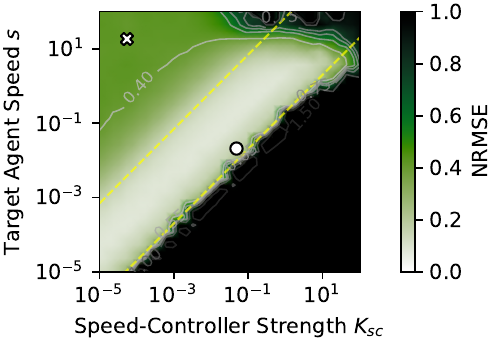}}\\
    \vspace{1em}
    \subfloat[NMSE]{\includegraphics[scale=1.0]{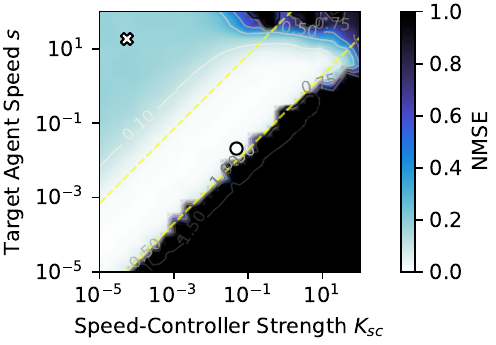}}\hfill
    \subfloat[sMAPE]{\includegraphics[scale=1.0]{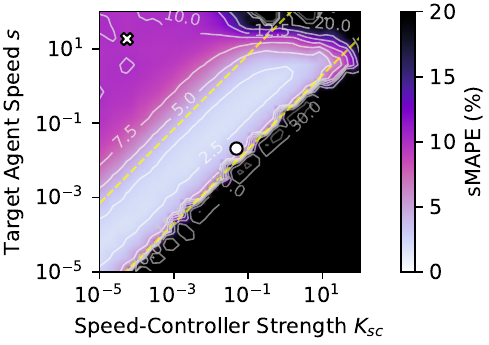}}
    \caption{\new{\textbf{Different predictive performance measures for a speed-controller parameter scan using the  Hénon–Heiles driving protocol}, corresponding to the \figref[a]{fig:benchmarks} in the main text.}}
    \label{fig:2024-11-10-attractors_varied_friction_HenonHeiles}
\end{figure}

\begin{figure}[H]
    \centering
    \subfloat[Pearson Correlation (y)]{\includegraphics[scale=1.0]{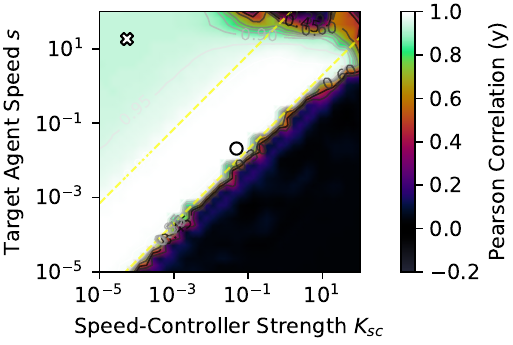}}\hfill
    \subfloat[NRMSE]{\includegraphics[scale=1.0]{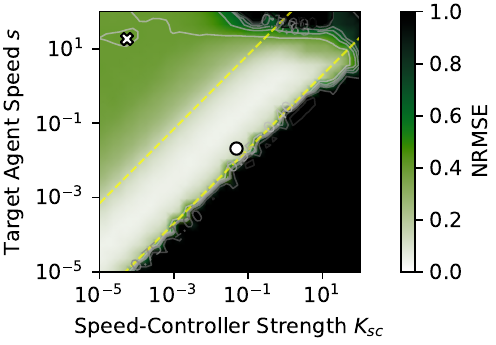}}\\
    \vspace{1em}
    \subfloat[NMSE]{\includegraphics[scale=1.0]{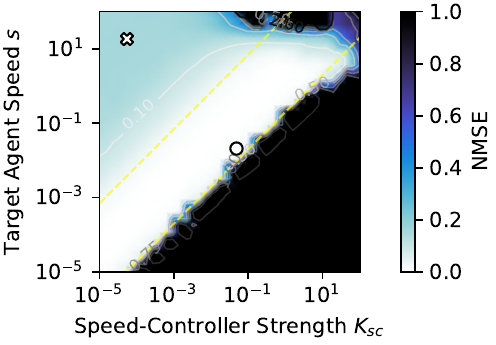}}\hfill
    \subfloat[sMAPE]{\includegraphics[scale=1.0]{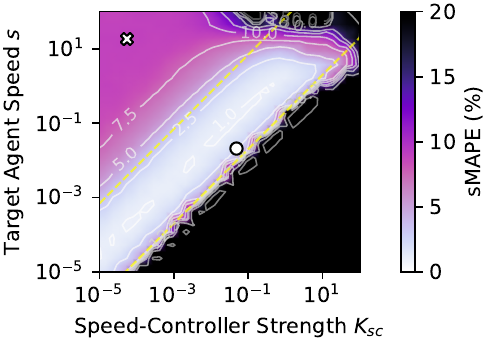}}
    \caption{\new{\textbf{Different predictive performance measures for a speed-controller parameter scan using the Rössler driving protocol}, corresponding to the \figref[b]{fig:benchmarks} in the main text.}}
    \label{fig:2024-10-14-attractors_varied_friction_Rossler}
\end{figure}

\begin{figure}[H]
    \centering
    \subfloat[Pearson Correlation (y)]{\includegraphics[scale=1.0]{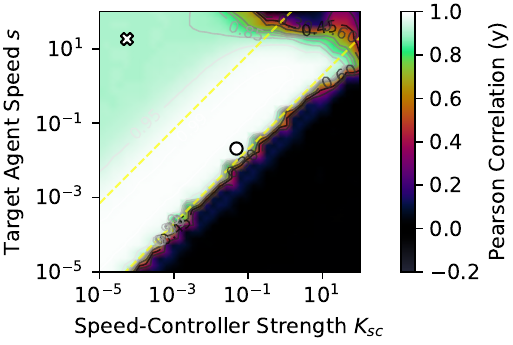}}\hfill
    \subfloat[NRMSE]{\includegraphics[scale=1.0]{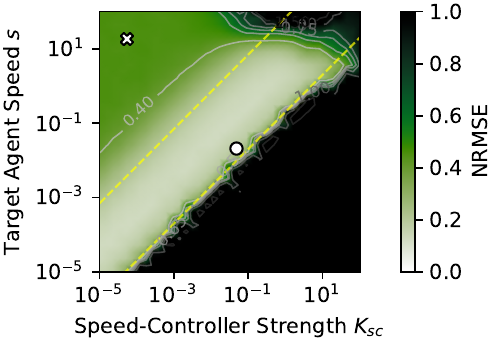}}\\
    \vspace{1em}
    \subfloat[NMSE]{\includegraphics[scale=1.0]{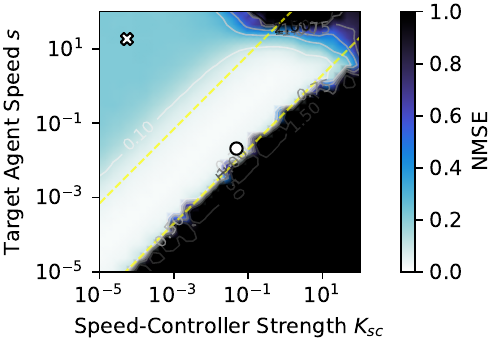}}\hfill
    \subfloat[sMAPE]{\includegraphics[scale=1.0]{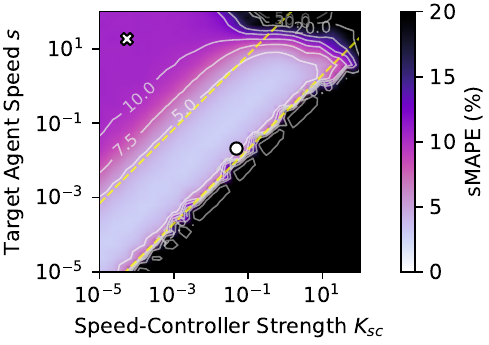}}
    \caption{\textbf{Different predictive performance measures for a speed-controller parameter scan using the Chua driving protocol}, corresponding to the \figref[c]{fig:benchmarks} in the main text.}
    \label{fig:2024-10-14-attractors_varied_friction_Chua}
\end{figure}

\begin{figure}[H]
    \centering
    \subfloat[Pearson Correlation (y)]{\includegraphics[scale=1.0]{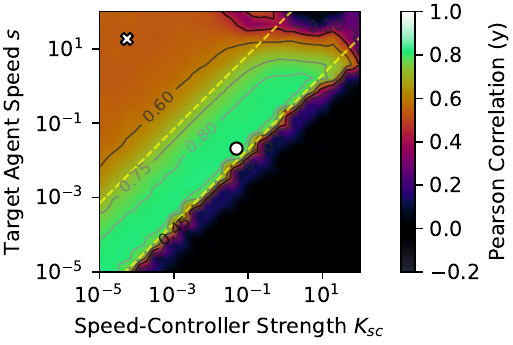}}\hfill
    \subfloat[NRMSE]{\includegraphics[scale=1.0]{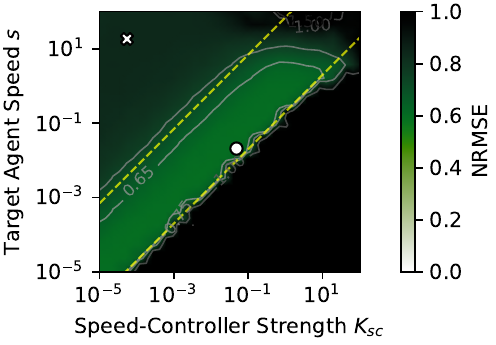}}\\
    \vspace{1em}
    \subfloat[NMSE]{\includegraphics[scale=1.0]{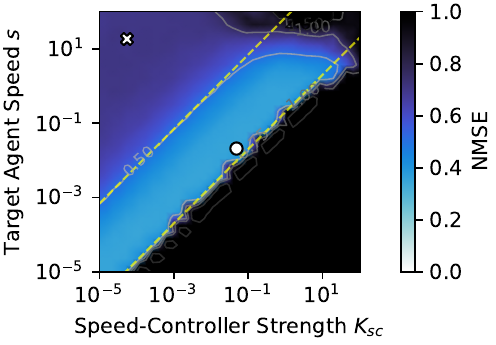}}\hfill
    \subfloat[sMAPE]{\includegraphics[scale=1.0]{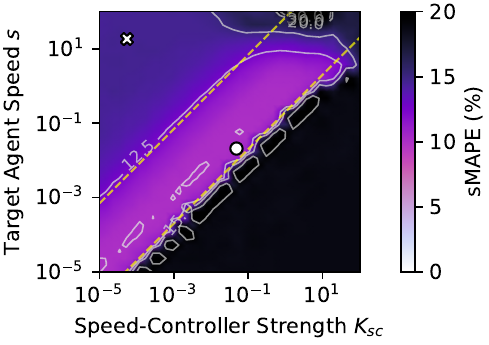}}
    \caption{\new{\textbf{Different predictive performance measures for a speed-controller parameter scan using the Lorenz-96 driving protocol}, corresponding to the \figref[d]{fig:benchmarks} in the main text.}}
    \label{fig:2024-10-14-attractors_varied_friction_Lorenz96}
\end{figure}

\begin{figure}[H]
    \centering
    \subfloat[Pearson Correlation (y)]{\includegraphics[scale=1.0]{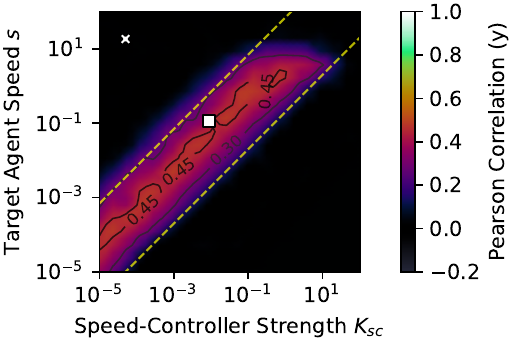}}\hfill
    \subfloat[NRMSE]{\includegraphics[scale=1.0]{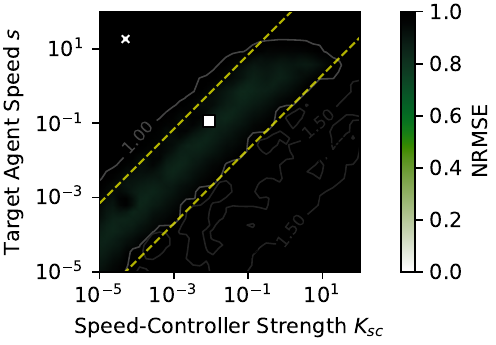}}\\
    \vspace{1em}
    \subfloat[NMSE]{\includegraphics[scale=1.0]{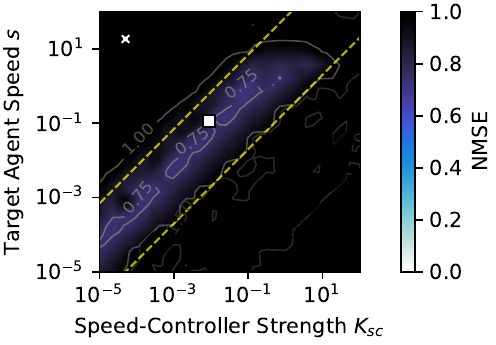}}\hfill
    \subfloat[sMAPE]{\includegraphics[scale=1.0]{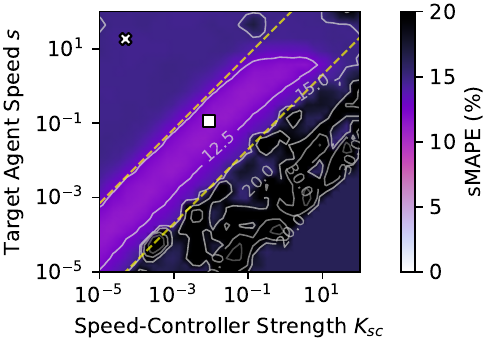}}
    \caption{\new{\textbf{Different predictive performance measures for the speed-controller parameter scan using a single agent}, corresponding to \figref[a]{fig:few_particles} in the main text.}}
    \label{fig:2024-04-22-Lymburn_critical_varied_friction_1_agent}
\end{figure}

\begin{figure}[H]
    \centering
    \subfloat[Pearson Correlation (y)]{\includegraphics[scale=1.0]{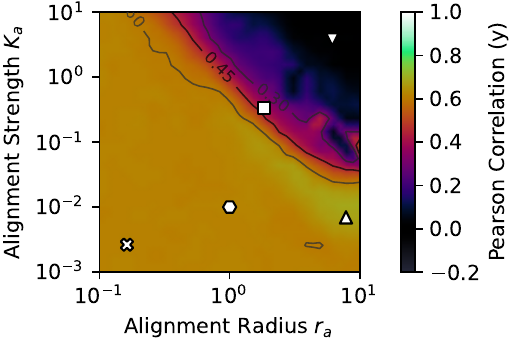}}\hfill
    \subfloat[NRMSE]{\includegraphics[scale=1.0]{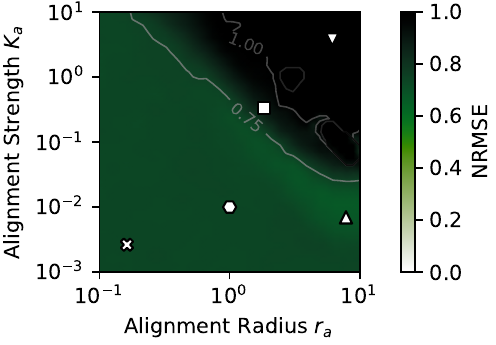}}\\
    \vspace{1em}
    \subfloat[NMSE]{\includegraphics[scale=1.0]{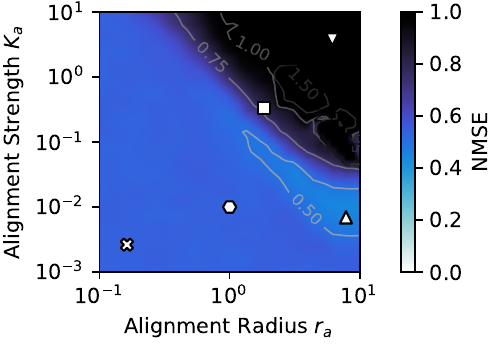}}\hfill
    \subfloat[sMAPE]{\includegraphics[scale=1.0]{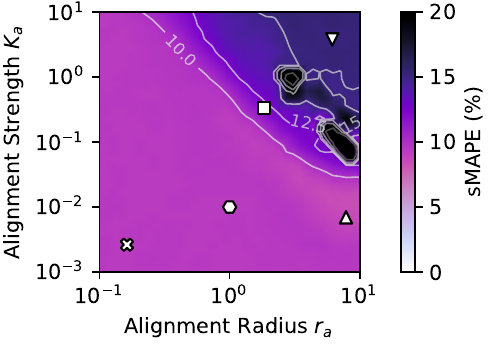}}
    \caption{\new{\textbf{Different predictive performance measures for varying the parameters of the agent-agent alignment interaction using a speed-controller setting of Lymburn \etal (2021)}, corresponding to \figref[a]{fig:alignment} in the main text.}}
    \label{fig:2024-10-18-Lymburn_critical_varied_alignment}
\end{figure}

\begin{figure}[H]
    \centering
    \subfloat[Pearson Correlation (y)]{\includegraphics[scale=1.0]{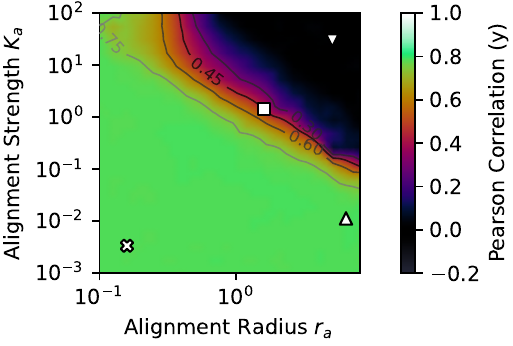}}\hfill
    \subfloat[NRMSE]{\includegraphics[scale=1.0]{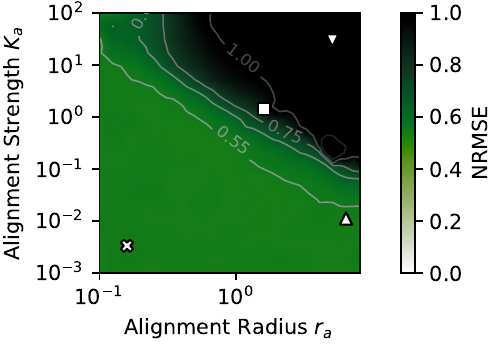}}\\
    \vspace{1em}
    \subfloat[NMSE]{\includegraphics[scale=1.0]{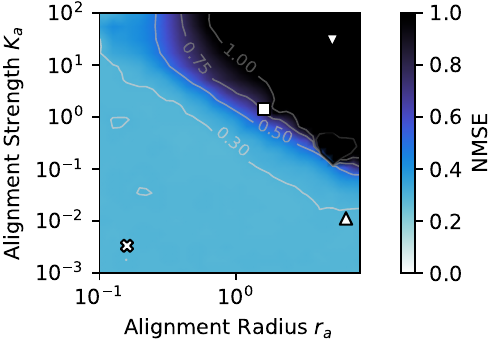}}\hfill
    \subfloat[sMAPE]{\includegraphics[scale=1.0]{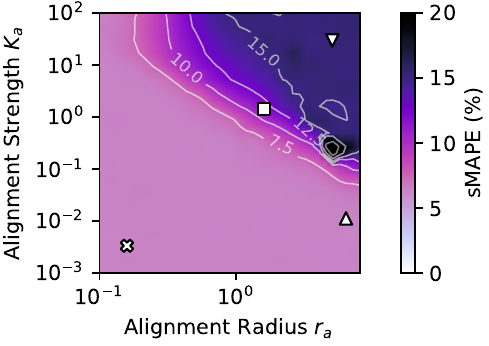}}
    \caption{\new{\textbf{Different predictive performance measures for varying the parameters of the agent-agent alignment interaction using a near-critically damped speed-controller setting}, corresponding to \figref[b]{fig:alignment} in the main text.}}
    \label{fig:2024-10-21-overdamped_varied_alignment}
\end{figure}

\begin{figure}[H]
    \centering
    \subfloat[Pearson Correlation (y)]{\includegraphics[scale=1.0]{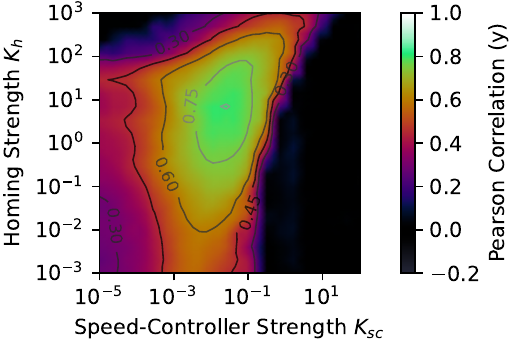}}\hfill
    \subfloat[NRMSE]{\includegraphics[scale=1.0]{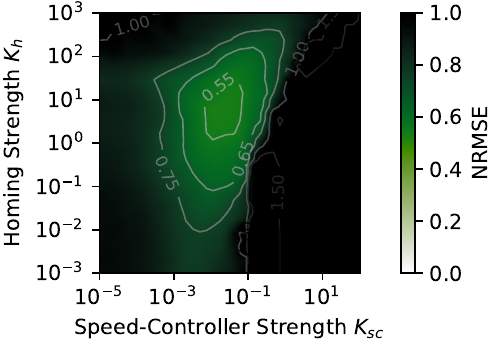}}\\
    \vspace{1em}
    \subfloat[NMSE]{\includegraphics[scale=1.0]{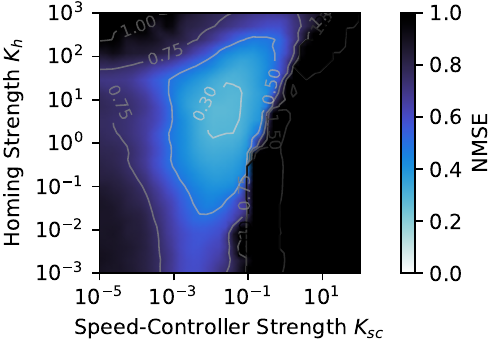}}\hfill
    \subfloat[sMAPE]{\includegraphics[scale=1.0]{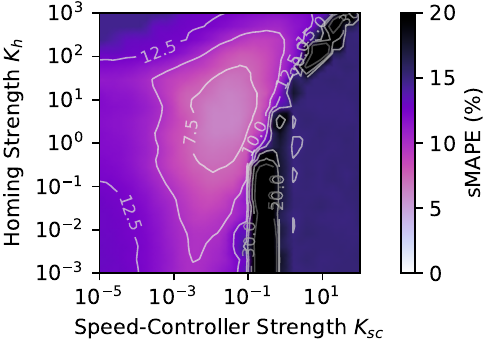}}
    \caption{\new{\textbf{Different predictive performance measures for varied homing force strength and speed-controlling force strength}, corresponding to \figref[a]{fig:homing} in the main text.}}
    \label{fig:2024-10-22-overdamped_varied_homing_strength_and_speed_controller_strength}
\end{figure}

\begin{figure}[H]
    \centering
    \subfloat[Pearson Correlation (y)]{\includegraphics[scale=1.0]{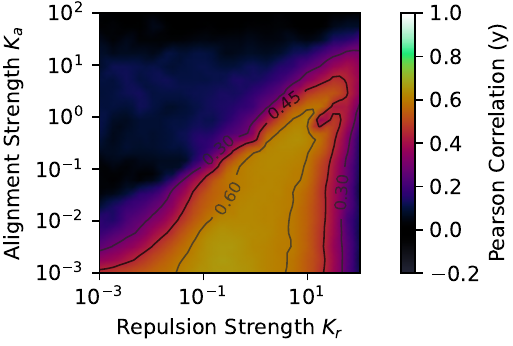}}\hfill
    \subfloat[NRMSE]{\includegraphics[scale=1.0]{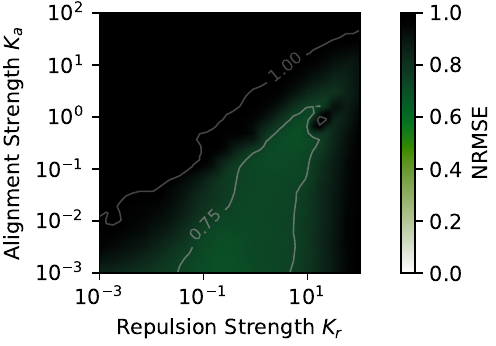}}\\
    \vspace{1em}
    \subfloat[NMSE]{\includegraphics[scale=1.0]{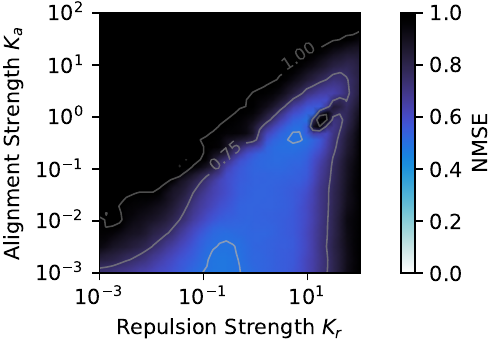}}\hfill
    \subfloat[sMAPE]{\includegraphics[scale=1.0]{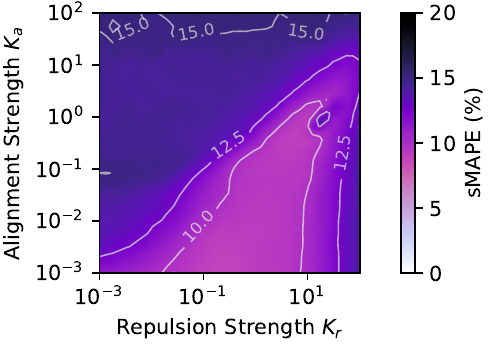}}
    \caption{\new{\textbf{Different predictive performance measures for different active matter regimes in a repulsion/alignment strength parameter scan (reproduction of Fig.\ 8a in \refref{\cite{Lymburn2021}})}, corresponding to \figref{fig:reproduction_lymburn_alignment_repulsion_performance}.}}
    \label{fig:2024-11-21-reproduction_Lymburn_alignment_repulsion_scan_Fig6B}
\end{figure}

\newpage
\subsection{Short-term memory capacity and Echo State Network comparison}
\label{sec:additional-appendix}

\begin{figure*}[htbp]
    \centering
    \subfloat[$\Delta s = 1$]{\includegraphics[scale=1.0]{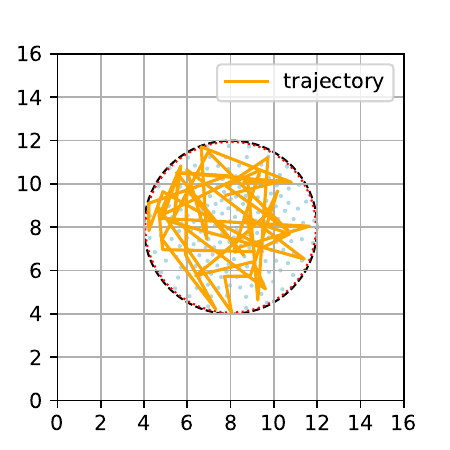}}\hfill
    \subfloat[$\Delta s = 2$]{\includegraphics[scale=1.0]{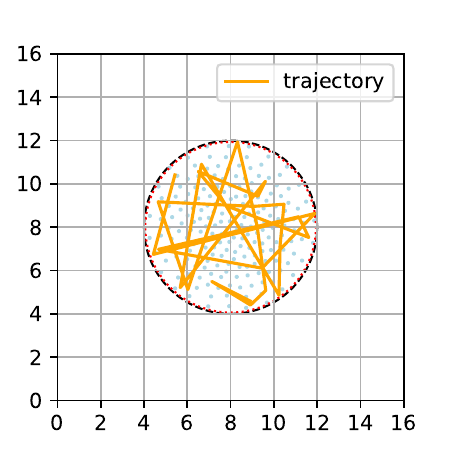}}\\
   \vspace{1em}
    \subfloat[$\Delta s = 5$]{\includegraphics[scale=1.0]{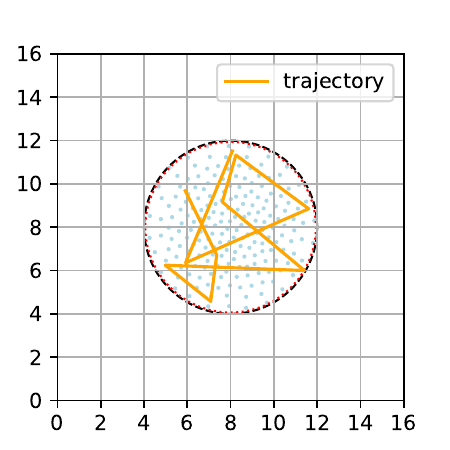}}\hfill
    \subfloat[$\Delta s = 10$]{\includegraphics[scale=1.0]{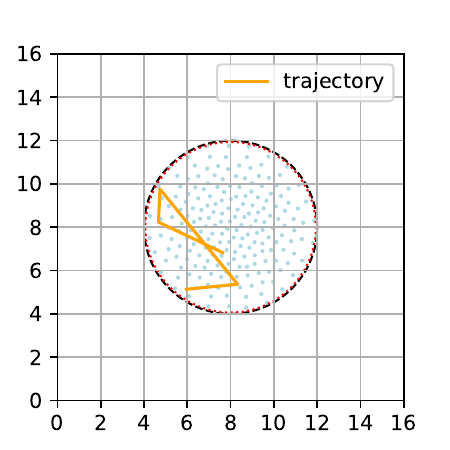}}

     \caption{\new{\textbf{Uniformly at random time series confined to a circle within the simulation box.} The time series (orange solid line) changes each $\Delta s$ steps and is sampled in a circle with radius $R = 4.0$ (black dashed line). The light blue dots correspond to agent positions of an undriven swarm in the near-critically damped regime (pyramid symbol) at $t = 0.0$. The red dotted line corresponds to the maximum radial displacement $R_{max} \approx 3.94$ of a swarm agent with respect to the simulation box center computed using the first 1,000 time steps. Visualized are the first 50 time steps of each time series.}
    }
    \label{fig:uniform_random_time_series_circle}
\end{figure*}

\begin{figure*}[htbp]
    \centering
    \subfloat[$\Delta s = 1$]{\includegraphics[scale=1.0]{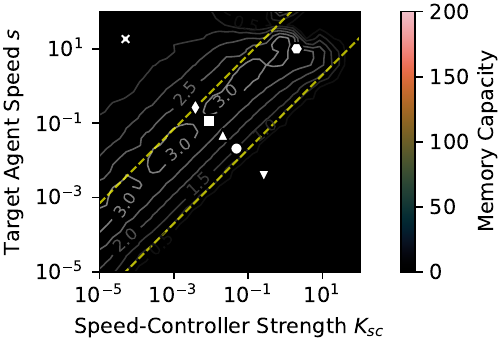}}\hfill
    \subfloat[$\Delta s = 2$]{\includegraphics[scale=1.0]{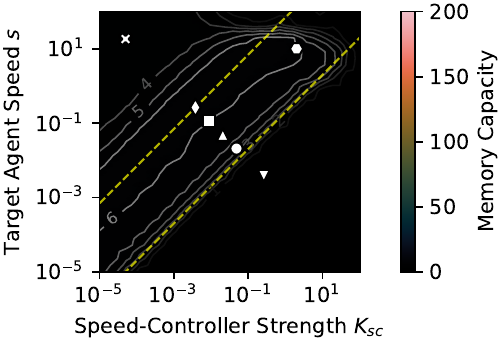}}\\
   \vspace{1em}
    \subfloat[$\Delta s = 5$]{\includegraphics[scale=1.0]{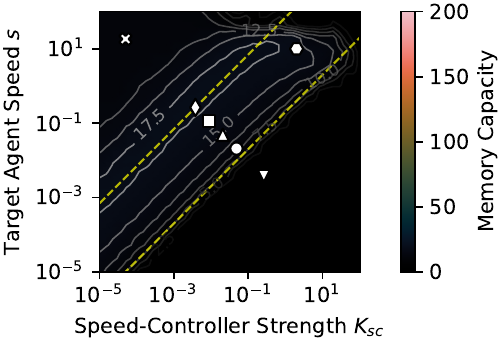}}\hfill
    \subfloat[$\Delta s = 10$]{\includegraphics[scale=1.0]{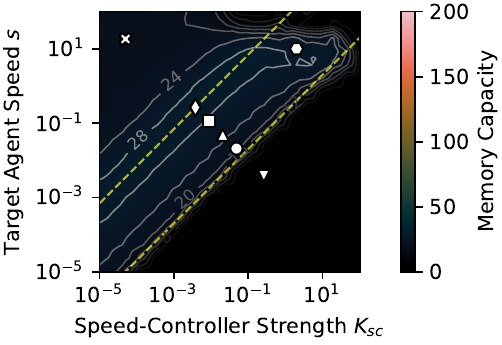}}

  \caption{\new{\textbf{Short-term memory capacity measurements for varying speed controller parameters and different driver change intervals $\Delta s$ for randomly at uniform driver positions sampled form a circle with radius $R = 4$ around the center of the simulation box} (see also \figref{fig:uniform_random_time_series_circle}). }}
    \label{fig:speed-controller-memcap-total}
\end{figure*}

\begin{figure*}[htbp]
    \centering
    \subfloat[$\Delta s = 1$]{\includegraphics[scale=1.0]{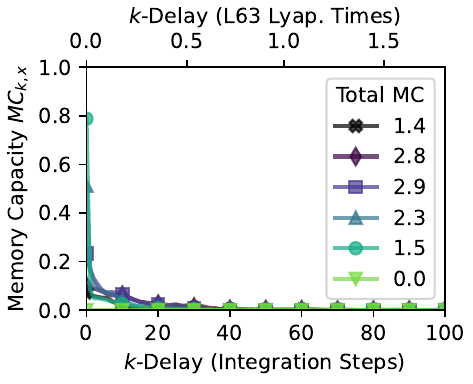}}\hfill
    \subfloat[$\Delta s = 2$]{\includegraphics[scale=1.0]{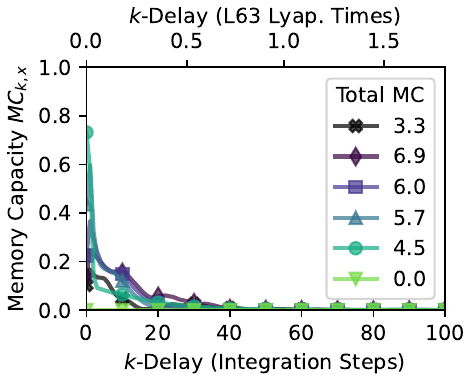}}\\
   \vspace{1em}
    \subfloat[$\Delta s = 5$]{\includegraphics[scale=1.0]{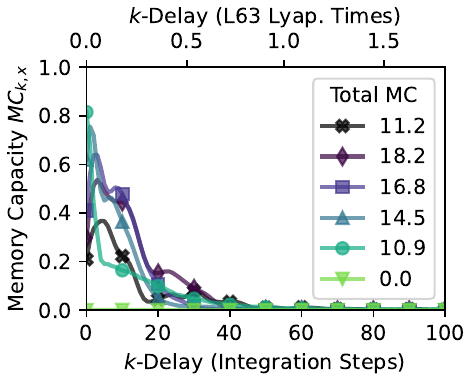}}\hfill
    \subfloat[$\Delta s = 10$]{\includegraphics[scale=1.0]{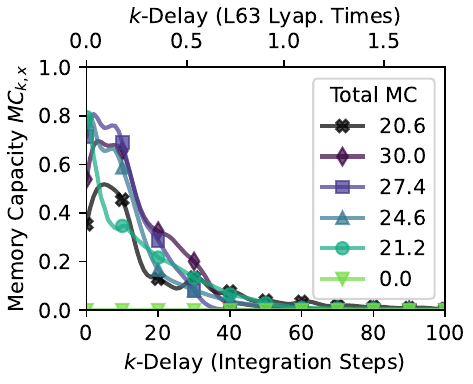}}

  \caption{\new{\textbf{$k$-delay memory capacity measurements for varying speed controller parameters and different uniformly at random driver change intervals $\Delta s$}, matching the total memory capacities reported in \figref{fig:speed-controller-memcap-total}. }}
    \label{fig:speed-controller-memcap-k-delay}
\end{figure*}

\begin{figure*}[htbp]
    \centering
    \subfloat[Total Memory Capacity]{\includegraphics[scale=1.0]{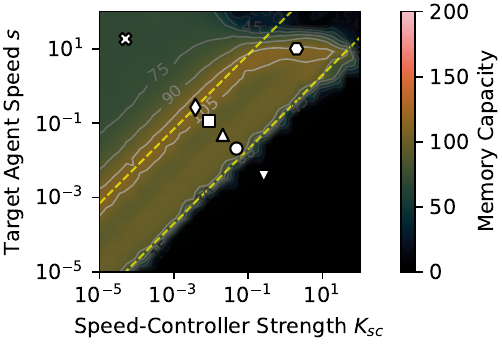}}\hfill
    \subfloat[$k$-delay Memory Capacity, $x$ dimension]{\includegraphics[scale=0.9]{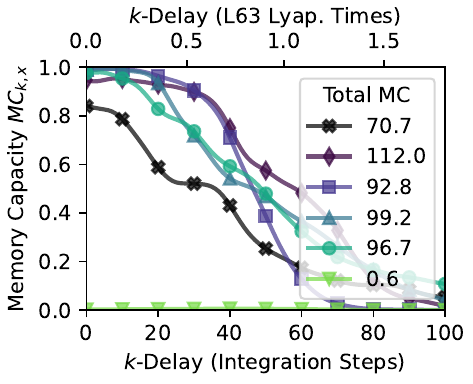}}
  \caption{\new{\textbf{Short-term memory capacity for varying speed-controller parameters measured using the default Lorenz-63 trajectory as input instead signal of the random uniform signal} used in \figref{fig:speed-controller-memcap-total}, corresponding to the predictive performance plot shown in \figref[a]{fig:Lymburn_critical_speed_controller_scan_larger}.}}
    \label{fig:speed-controller-memory-capacity}
\end{figure*}

\begin{figure*}[htbp]
    \centering
    \includegraphics{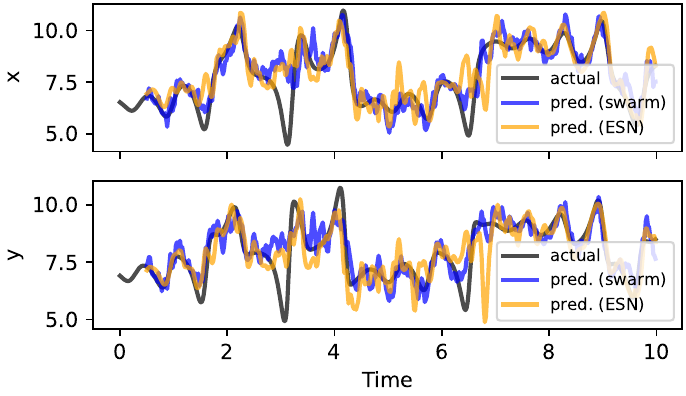}
    \caption{\new{\textbf{Predicted versus actual time series} for a near-critically damped active matter system (swarm, blue) and an Echo State Network with $N=600$ neurons and properties described in \axref{subsec:esn_comparison} (orange). }}
    \label{fig:time-series-comparison}
\end{figure*}

\begin{figure*}[htbp]
    \centering
    \includegraphics[width=0.92\textwidth]{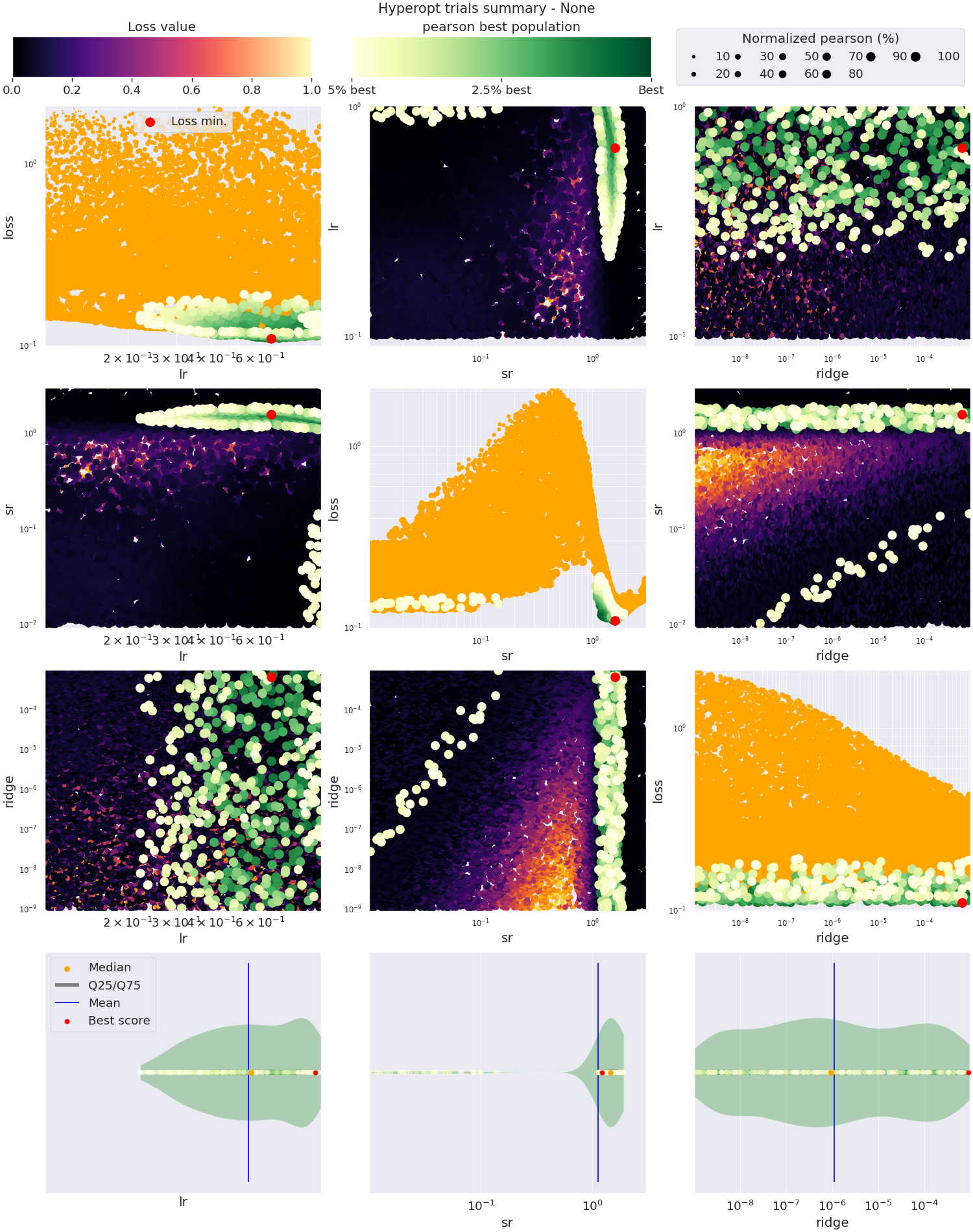}
    \caption{\new{\textbf{Reservoirpy hyperopt trials summary} for the optimization of the hyperparameters leaking rate $\text{lr}$, spectral radius $\text{sr}$, and Ridge parameter $\text{ridge}$, with constant input scaling $\text{is} = 1.0$ and number of neurons $N = 100$ (step B1 of the general method described in Ref.~\cite{Hinaut2021}). The red point marks the configuration with the lowest loss, and the green points mark the best configurations according to the chosen metric (here: Pearson correlation coefficient between the actual and predicted time series for the $x$ dimension).}}
    \label{fig:hyperopt-trials-summary}
\end{figure*}

%% file: supp_vids.tex

\section{\label{ax:supplementary_videos} Supplementary Videos}

All supplementary videos can be accessed in the Data Repository of the University of Stuttgart (DaRUS), in the dataverse Stuttgart Center for Simulation Science EXC 2075 / Project 6-15:\\
\href{https://doi.org/10.18419/darus-4619}{\texttt{https://doi.org/10.18419/darus-4619}} \cite{DARUS-4619_2025} \\
Snapshots show the simulated soft matter systems at $t = 3.3$ (after 165 integration time steps of $\Delta t = 0.02$) if not mentioned otherwise.\\

\begin{table}[H]
    \small
    \begin{center}
    \begin{tabular}{|>{\centering\arraybackslash}p{\vidTableIdColumnWidth}|>{\centering\arraybackslash}p{\vidTableSnapshotColumnWidth}|>{\centering\arraybackslash}p{\vidTableDescriptionColumnWidth}|}
    \hline
    \textbf{ID} & \textbf{Snapshot} & \textbf{Description} \\ 
    \hline
    \hline
    
    {1} & \includegraphics{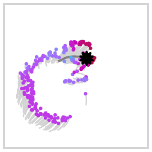} & Droplet phase ($K_a = 0.01, K_r = 0.01$) ($P = 0.409$)\\ 
    \hline
    {2} & \includegraphics{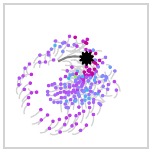} & "Critical" phase ($K_a = 0.01, K_r = 2.0$) ($P = 0.719$)\\ 
    \hline
    {3} & \includegraphics{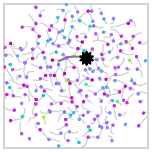} & Gas-like phase ($K_a = 0.01, K_r = 50$) ($P = 0.397$)\\ 
    \hline
    {4} & \includegraphics{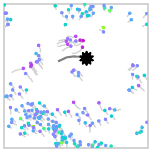} & Arrested phase ($K_a = 10.0, K_r = 1.0$) ($P = 0.210$)\\ 
    \hline
    \end{tabular}
    \end{center}
    
    \caption{ 
        \textbf{Supplementary snapshots and corresponding videos for different active matter regimes using four exemplary repulsion/alignment force strength parameter combinations,} reproducing Fig.\ 7 in \refref{\cite{Lymburn2021}}. The measured predictive performances $P$ using a similar active matter reservoir computing procedure as \refref{\cite{Lymburn2021}} (specified in \sectref{sec:materials-methods}) are indicated in parentheses. Video 2 corresponds to the parameter combination marked with a hexagon/``L'' symbol in \figref[a,d]{fig:Lymburn_critical_speed_controller_scan_larger}, \figref[a,b]{fig:speed_controller_fluctuations}, and \figref[a]{fig:alignment}.\newline 
        Label in repository: \texttt{reproduction-Lymburn2021}.
    }
    \label{tab:supplementary_videos_lymburn_reproductions}
\end{table}

\begin{table}[H]
    \small
    \begin{center}
    \begin{tabular}{|>{\centering\arraybackslash}p{\vidTableIdColumnWidth}|>{\centering\arraybackslash}p{\vidTableSnapshotColumnWidth}|>{\centering\arraybackslash}p{\vidTableDescriptionColumnWidth}|}
    \hline
    \textbf{ID} & \textbf{Snapshot} & \textbf{Description} \\ 
    \hline
    \hline
    
    {1} & \includegraphics{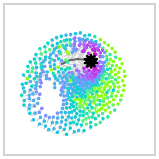} & Near-critically damped system, $N = 500$ agents\\ 
    \hline
    \end{tabular}
    \end{center}
    
    \caption{\textbf{Supplementary video for the overdamped phenomenology study with $N=500$ agents} in \figref{fig:phenomenology}. \newline Label in repository: \texttt{overdamped-phenomenology}.}
    \label{tab:supplementary_video_phenomenology}
\end{table}

\begin{table}[H]
    \small
    \begin{center}
    \begin{tabular}{|>{\centering\arraybackslash}p{\vidTableIdColumnWidth}|>{\centering\arraybackslash}p{\vidTableSnapshotColumnWidth}|>{\centering\arraybackslash}p{\vidTableDescriptionColumnWidth}|}
    \hline
    \textbf{ID} & \textbf{Snapshot} & \textbf{Description} \\ 
    \hline
    \hline
    
    {1} & \includegraphics{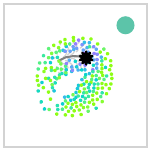} & Speed controller scan (circle symbol) \\ 
    \hline
    {2} & \includegraphics{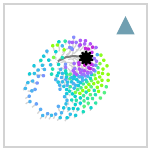} & Speed controller scan (pyramid symbol) \\ 
    \hline
    {3} & \includegraphics{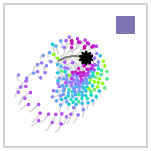} & Speed controller scan (square symbol) \\ 
    \hline
    {4} & \includegraphics{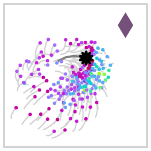} & Speed controller scan (diamond symbol) \\ 
    \hline
    {5} & \includegraphics{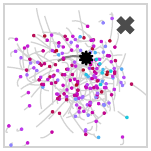} & Speed controller scan (cross symbol)\\ 
    \hline
    {6} & \includegraphics{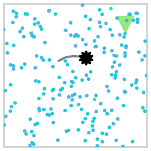} & Speed controller scan (nabla symbol)\\ 
    \hline
    {7} & \includegraphics{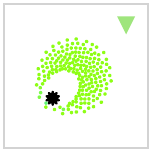} & Speed controller scan ($\Delta t = 0.002$, nabla symbol). \\ 
    \hline
    \end{tabular}
    \end{center}
    
    \caption{\textbf{Supplementary videos for default speed-controller parameter scans} in \figref{fig:Lymburn_critical_speed_controller_scan_larger} (and \figref{fig:Lymburn_critical_scans_speed_controller_smaller_integration_time_step} for the scan with the smaller integration time step of $\Delta t =0.002$ instead of $0.02$ by default; Video 7). \newline Label in repository: \texttt{speed-controller}.}
    \label{tab:supplementary_videos_speed_controller}
\end{table}

\begin{table}[H]
    \small
    \begin{center}
    \begin{tabular}{|>{\centering\arraybackslash}p{\vidTableIdColumnWidth}|>{\centering\arraybackslash}p{\vidTableSnapshotColumnWidth}|>{\centering\arraybackslash}p{\vidTableDescriptionColumnWidth}|}
    \hline
    \textbf{ID} & \textbf{Snapshot} & \textbf{Description} \\ 
    \hline
    \hline
    
    {1} & \includegraphics{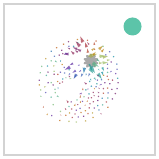} & Speed controller scan (circle symbol) \\ 
    \hline
    {2} & \includegraphics{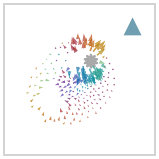} & Speed controller scan (pyramid symbol) \\ 
    \hline
    {3} & \includegraphics{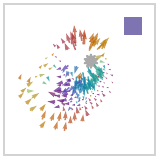} & Speed controller scan (square symbol) \\ 
    \hline
    {4} & \includegraphics{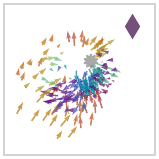} & Speed controller scan (diamond symbol) \\ 
    \hline
    {5} & \includegraphics{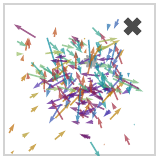} & Speed controller scan (cross symbol)\\ 
    \hline
    {6} & \includegraphics{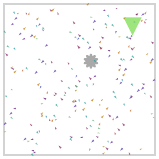} & Speed controller scan (nabla symbol)\\ 
    \hline
    \end{tabular}
    \end{center}
    
    \caption{\textbf{Supplementary videos to visualize velocity fluctuations for the default speed-controller parameter scan} in \figref{fig:Lymburn_critical_speed_controller_scan_larger}, as displayed in \figref{fig:speed_controller_fluctuations}). \newline Label in repository: \texttt{speed-controller-velocity-fluctuations}.}
    \label{tab:supplementary_videos_speed_controller_velocity_fluctuations}
\end{table}

\begin{table}[H]
    \small
    \begin{center}
    \begin{tabular}{|>{\centering\arraybackslash}p{\vidTableIdColumnWidth}|>{\centering\arraybackslash}p{\vidTableSnapshotColumnWidth}|>{\centering\arraybackslash}p{\vidTableDescriptionColumnWidth}|}
    \hline
    \textbf{ID} & \textbf{Snapshot} & \textbf{Description} \\ 
    \hline
    \hline
    
    {1} & \includegraphics{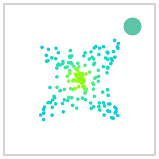} & Damping analysis, non-interacting (circle symbol) \\ 
    \hline
    {2} & \includegraphics{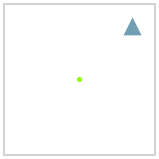} & Damping analysis, non-interacting (pyramid symbol) \\ 
    \hline
    {3} & \includegraphics{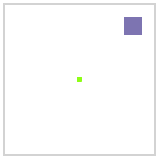} & Damping analysis, non-interacting (square symbol) \\ 
    \hline
    {4} & \includegraphics{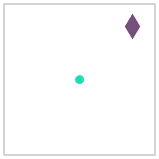} & Damping analysis, non-interacting (diamond symbol) \\ 
    \hline
    {5} & \includegraphics{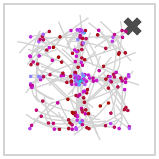} & Damping analysis, non-interacting (cross symbol)\\ 
    \hline
    {6} & \includegraphics{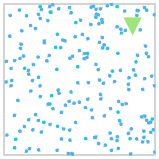} & Damping analysis, non-interacting (nabla symbol)\\ 
    \hline
    \end{tabular}
    \end{center}
    
    \caption{\textbf{Supplementary videos for the intrinsic relaxation dynamics studies for different speed-controller settings with \textit{non-interacting} agents and an integration time step of $\Delta t =0.02$} in \figref[a]{fig:damping}. \newline
    Label in repository: \texttt{damping-analysis-non-interacting}.}
    \label{tab:supplementary_videos_damping_non_interacting}
\end{table}

\begin{table}[H]
    \small
    \begin{center}
    \begin{tabular}{|>{\centering\arraybackslash}p{\vidTableIdColumnWidth}|>{\centering\arraybackslash}p{\vidTableSnapshotColumnWidth}|>{\centering\arraybackslash}p{\vidTableDescriptionColumnWidth}|}
    \hline
    \textbf{ID} & \textbf{Snapshot} & \textbf{Description} \\ 
    \hline
    \hline
    
    {1} & \includegraphics{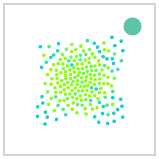} & Damping analysis, interacting (circle symbol) \\ 
    \hline
    {2} & \includegraphics{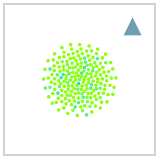} & Damping analysis, interacting (pyramid symbol) \\ 
    \hline
    {3} & \includegraphics{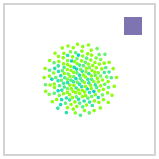} & Damping analysis, interacting (square symbol) \\ 
    \hline
    {4} & \includegraphics{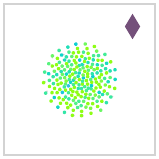} & Damping analysis, interacting (diamond symbol) \\ 
    \hline
    {5} & \includegraphics{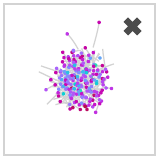} & Damping analysis, interacting (cross symbol)\\ 
    \hline
    {6} & \includegraphics{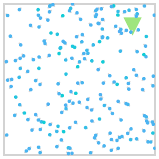} & Damping analysis, interacting (nabla symbol)\\ 
    \hline
    \end{tabular}
    \end{center}
    
    \caption{\textbf{Supplementary videos for the intrinsic relaxation dynamics studies for different speed-controller settings with \textit{interacting} agents and an integration time step of $\Delta t =0.02$} in \figref[b]{fig:damping} (all interactions present described in \sectref{sec:mm-interactions}). \newline Label in repository: \texttt{damping-analysis-interacting}.}
    \label{tab:supplementary_videos_damping_interacting}
\end{table}

\begin{table}[H]
    \small
    \begin{center}
    \begin{tabular}{|>{\centering\arraybackslash}p{\vidTableIdColumnWidth}|>{\centering\arraybackslash}p{\vidTableSnapshotColumnWidth}|>{\centering\arraybackslash}p{\vidTableDescriptionColumnWidth}|}
    \hline
    \textbf{ID} & \textbf{Snapshot} & \textbf{Description} \\ 
    \hline
    \hline
    
    {1} & \includegraphics{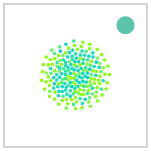} & Undriven speed controller scan (circle symbol) \\ 
    \hline
    {2} & \includegraphics{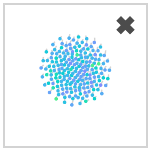} & Undriven state speed controller scan (cross symbol) \\ 
    \hline
    \end{tabular}
    \end{center}
    
    \caption{\textbf{Supplementary videos for speed-controller parameter scans without external driving force (steady-state simulations, no reservoir computing).} 
    In the undriven underdamped regime (\figref[a]{fig:Lymburn_critical_speed_controller_scan_larger}, cross symbol) agents do not rest, causing loose (low $K_{sc}$) or strict (higher $K_{sc}$) collective motion in a preserved ordered structure. We observe vertically oscillating active crystals (sustained polarity, video 2) and milling states (sustained rotation). On the contrary, the overdamped regime (video 1) features an almost static spatial distribution with zero center of mass motion and only minimal deviations around resting positions. Hence, to identify the regimes the out-of-steady-state properties play a significant role. One useful metric is to measure the initial transient time it takes to settle a randomly initialized system into its steady state (see \tabref{tab:supplementary_videos_initial_transients}). \newline Label in repository: \texttt{speed-controller-undriven}.}
    \label{tab:supplementary_videos_no_driver}
\end{table}

\begin{table}[H]
    \small
    \begin{center}
    \begin{tabular}{|>{\centering\arraybackslash}p{\vidTableIdColumnWidth}|>{\centering\arraybackslash}p{\vidTableSnapshotColumnWidth}|>{\centering\arraybackslash}p{\vidTableDescriptionColumnWidth}|}
    \hline
    \textbf{ID} & \textbf{Snapshot} & \textbf{Description} \\ 
    \hline
    \hline
    {1} & \includegraphics{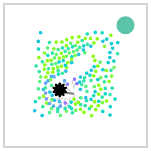} & Initial transient, speed controller scan (circle symbol) \\ 
    \hline
    {2} & \includegraphics{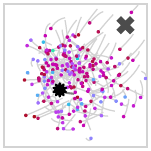} & Initial transient, speed controller scan (cross symbol) \\ 
    \hline
    \end{tabular}
    \end{center}
    
    \caption{\textbf{Supplementary videos for visualizing transients from an initial random system configuration to a driven out-of-steady-state configuration.} \newline Label in repository: \texttt{speed-controller-initial-transient}.}
    \label{tab:supplementary_videos_initial_transients}
\end{table}

\begin{table}[H]
    \small
    \begin{center}
    \begin{tabular}{|>{\centering\arraybackslash}p{\vidTableIdColumnWidth}|>{\centering\arraybackslash}p{\vidTableSnapshotColumnWidth}|>{\centering\arraybackslash}p{\vidTableDescriptionColumnWidth}|}
    \hline
    \textbf{ID} & \textbf{Snapshot} & \textbf{Description} \\ 
    \hline
    \hline
    {1} & \includegraphics[width=\vidTableSnapshotColumnWidth]{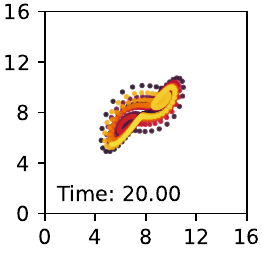} & Lorenz-63 driver trajectory \\ 
    \hline
    {2} & \includegraphics[width=\vidTableSnapshotColumnWidth]{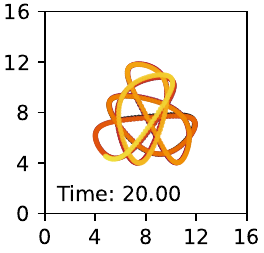} & Hénon–Heiles driver trajectory \\ 
    \hline
    {3} & \includegraphics[width=\vidTableSnapshotColumnWidth]{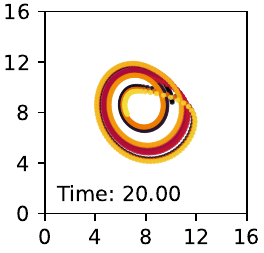} & Rössler driver trajectory \\ 
    \hline
    {4} & \includegraphics[width=\vidTableSnapshotColumnWidth]{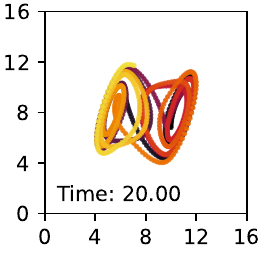}  & Chua driver trajectory \\ 
    \hline
    {5} & \includegraphics[width=\vidTableSnapshotColumnWidth]{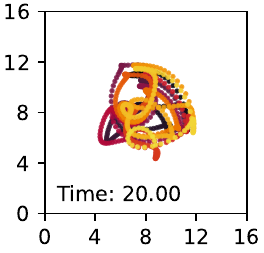} & Lorenz-96 driver trajectory \\ 
    \hline
    \end{tabular}
    \end{center}
    
    \caption{\textbf{Supplementary videos for visualizing driver trajectories for different chaotic attractor benchmarks} in \figref{fig:Lymburn_critical_speed_controller_scan_larger} (Video 1) and \figref{fig:benchmarks} (Videos 2-5). The first 1,000 integration time steps are displayed in the actual simulation box. Color changes as time progresses to aid inspections of the trajectory evolution. \newline Label in repository: \texttt{driver-trajectory}.}
    \label{tab:supplementary_videos_driver_trajectories}
\end{table}

\begin{table}[H]
    \small
    \begin{center}
    \begin{tabular}{|>{\centering\arraybackslash}p{\vidTableIdColumnWidth}|>{\centering\arraybackslash}p{\vidTableSnapshotColumnWidth}|>{\centering\arraybackslash}p{\vidTableDescriptionColumnWidth}|}
    \hline
    \textbf{ID} & \textbf{Snapshot} & \textbf{Description} \\ 
    \hline
    \hline
    
    {1} & \includegraphics{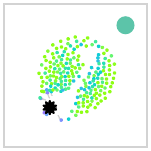} & Hénon–Heiles speed-controller scan (circle symbol) \\
    \hline
    {2} & \includegraphics{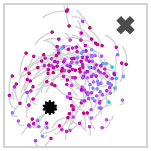} & Hénon–Heiles speed-controller scan (cross symbol) \\
    \hline
    {3} & \includegraphics{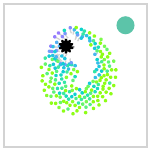} & Rössler speed-controller scan (circle symbol) \\
    \hline
    {4} & \includegraphics{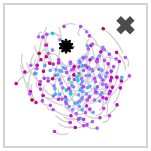} & Rössler speed-controller scan (cross symbol) \\
    \hline
    {5} & \includegraphics{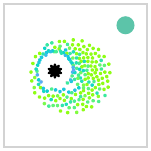} & Chua speed-controller scan (circle symbol) \\
    \hline
    {6} & \includegraphics{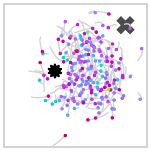} & Chua speed-controller scan (cross symbol) \\
    \hline
    {7} & \includegraphics{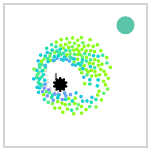} & Lorenz-96 speed-controller scan (circle symbol) \\
    \hline
    {8} & \includegraphics{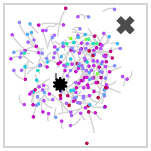} & Lorenz-96 speed-controller scan (cross symbol) \\
    
    \hline
    \end{tabular}
    \end{center}
    
    \caption{\textbf{Supplementary videos for the speed-controller scans for different chaotic time series driving protocols} in \figref{fig:benchmarks}. \newline Label in repository: \texttt{speed-controller-benchmarks}.}
    \label{tab:supplementary_videos_benchmarks}
\end{table}

\begin{table}[H]
    \small
    \begin{center}
    \begin{tabular}{|>{\centering\arraybackslash}p{\vidTableIdColumnWidth}|>{\centering\arraybackslash}p{\vidTableSnapshotColumnWidth}|>{\centering\arraybackslash}p{\vidTableDescriptionColumnWidth}|}
    \hline
    \textbf{ID} & \textbf{Snapshot} & \textbf{Description} \\ 
    \hline
    \hline

    {1} & \includegraphics{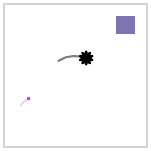} & Speed-controller scan, single agent (square symbol) \\
    \hline
    {2} & \includegraphics{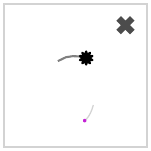} & Speed-controller scan, single agent (cross symbol) \\
    \hline
    
    \end{tabular}
    \end{center}
    \caption{
        \textbf{Supplementary videos for the speed-controller scan with a single agent} in \figref[a]{fig:few_particles}.\newline Label in repository: \texttt{speed-controller-single-agent}.
    }
    \label{tab:supplementary_videos_few_agents}
\end{table}

\begin{table}[H]
    \small
    \begin{center}
    \begin{tabular}{|>{\centering\arraybackslash}p{\vidTableIdColumnWidth}|>{\centering\arraybackslash}p{\vidTableSnapshotColumnWidth}|>{\centering\arraybackslash}p{\vidTableDescriptionColumnWidth}|}
    \hline
    \textbf{ID} & \textbf{Snapshot} & \textbf{Description} \\ 
    \hline
    \hline
     
    {1} & \includegraphics{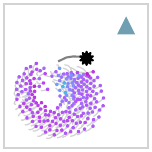} & Alignment force scan (pyramid symbol) \\
    \hline
    {2} & \includegraphics{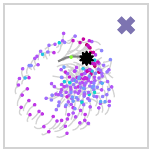} & 
    Alignment force scan (cross symbol) \\
    \hline
    {3} & \includegraphics{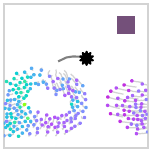} & Alignment force scan (square symbol) \\
    \hline
    {4} & \includegraphics{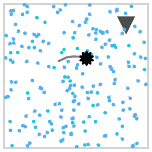} & Alignment force scan (nabla symbol) \\
    \hline

    \end{tabular}
    \end{center}
    \caption{\textbf{Supplementary videos for the alignment force scan with a speed-controller setting used in \refref{\cite{Lymburn2021}}} ($K_{sc} = 2.0$, $s = 10.0$) as displayed in  \figref[a]{fig:alignment}. \newline Label in repository: \texttt{alignment}.}
    \label{tab:supplementary_videos_alignment}
\end{table}

\begin{table}[H]
    \small
    \begin{center}
    \begin{tabular}{|>{\centering\arraybackslash}p{\vidTableIdColumnWidth}|>{\centering\arraybackslash}p{\vidTableSnapshotColumnWidth}|>{\centering\arraybackslash}p{\vidTableDescriptionColumnWidth}|}
    \hline
    \textbf{ID} & \textbf{Snapshot} & \textbf{Description} \\ 
    \hline
    \hline

    {1} & \includegraphics{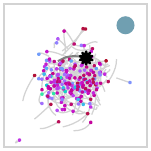} & Homing force scan (circle symbol)\\
    \hline
    {2} & \includegraphics{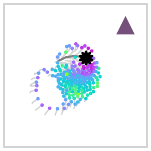} & Homing force scan (pyramid symbol)\\
    \hline
    {3} & \includegraphics{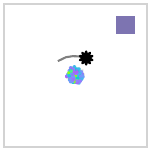} & Homing force scan (square symbol)\\
    \hline
    {4} & \includegraphics{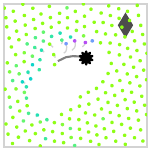} & Homing force scan (diamond symbol) \\
    \hline

    \end{tabular}
    \end{center}
    \caption{\textbf{Supplementary videos for varied homing force and speed controller strengths} in \figref{fig:homing}. \newline Label in repository: \texttt{homing}.}
    \label{tab:supplementary_videos_homing_strength_and_speed_controller_strength}
\end{table}

\begin{table}[H]
    \small
    \begin{center}
    \begin{tabular}{|>{\centering\arraybackslash}p{\vidTableIdColumnWidth}|>{\centering\arraybackslash}p{\vidTableSnapshotColumnWidth}|>{\centering\arraybackslash}p{\vidTableDescriptionColumnWidth}|}
    \hline
    \textbf{ID} & \textbf{Snapshot} & \textbf{Description} \\ 
    \hline
    \hline
    
    {1} & \includegraphics{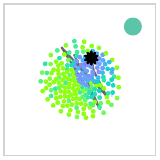} & Speed controller scan, random uniform driver ($\Delta s = 1$) (circle symbol) \\ 
    \hline
    {2} & \includegraphics{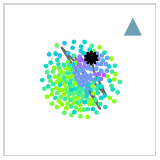} & Speed controller scan, random uniform driver ($\Delta s = 1$) (pyramid symbol) \\ 
    \hline
    {3} & \includegraphics{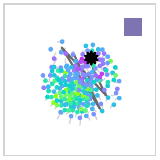} & Speed controller scan, random uniform driver ($\Delta s = 1$) (square symbol) \\ 
    \hline
    {4} & \includegraphics{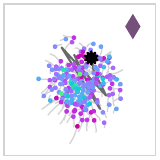} & Speed controller scan, random uniform driver ($\Delta s = 1$) (diamond symbol) \\ 
    \hline
    {5} & \includegraphics{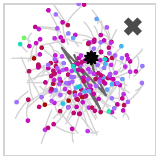} & Speed controller scan, random uniform driver ($\Delta s = 1$) (cross symbol)\\ 
    \hline
    {6} & \includegraphics{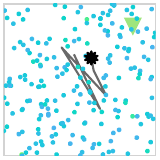} & Speed controller scan, random uniform driver ($\Delta s = 1$) (nabla symbol)\\
    \hline
    \end{tabular}
    \end{center}
    
    \caption{\new{\textbf{Supplementary videos for the speed-controller parameter scan with a driver sampled uniformly at random within a circle of radius $R=4$ around the center of the simulation box and a change interval of $\Delta s = 1$ step for the determination of the short-term memory capacity} in \figref[a]{fig:speed-controller-memcap-total} and \figref[a]{fig:speed-controller-memcap-k-delay}. \newline Label in repository: \texttt{speed-controller-random-uniform-driver-interval-1}. }}
    \label{tab:supplementary_videos_random_memory_capacity_interval_1}
\end{table}

\begin{table}[H]
    \small
    \begin{center}
    \begin{tabular}{|>{\centering\arraybackslash}p{\vidTableIdColumnWidth}|>{\centering\arraybackslash}p{\vidTableSnapshotColumnWidth}|>{\centering\arraybackslash}p{\vidTableDescriptionColumnWidth}|}
    \hline
    \textbf{ID} & \textbf{Snapshot} & \textbf{Description} \\ 
    \hline
    \hline
    
    {1} & \includegraphics{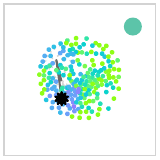} & Speed controller scan, random uniform driver ($\Delta s = 2$) (circle symbol) \\ 
    \hline
    {2} & \includegraphics{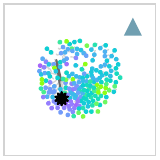} & Speed controller scan, random uniform driver ($\Delta s = 2$) (pyramid symbol) \\ 
    \hline
    {3} & \includegraphics{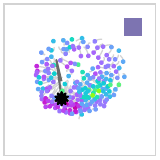} & Speed controller scan, random uniform driver ($\Delta s = 2$) (square symbol) \\ 
    \hline
    {4} & \includegraphics{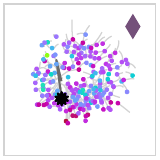} & Speed controller scan, random uniform driver ($\Delta s = 2$) (diamond symbol) \\ 
    \hline
    {5} & \includegraphics{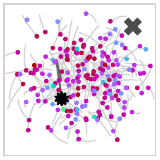} & Speed controller scan, random uniform driver ($\Delta s = 2$) (cross symbol)\\ 
    \hline
    {6} & \includegraphics{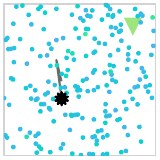} & Speed controller scan, random uniform driver ($\Delta s = 2$) (nabla symbol)\\
    \hline
    \end{tabular}
    \end{center}
    
    \caption{\new{\textbf{Supplementary videos for the speed-controller parameter scan with a driver sampled uniformly at random within a circle of radius $R=4$ around the center of the simulation box and a change interval of $\Delta s = 2$ steps for the determination of the short-term memory capacity} in \figref[b]{fig:speed-controller-memcap-total} and \figref[b]{fig:speed-controller-memcap-k-delay}.. \newline Label in repository: \texttt{speed-controller-random-uniform-driver-interval-2}. }}
    \label{tab:supplementary_videos_random_memory_capacity_interval_2}
\end{table}

\begin{table}[H]
    \small
    \begin{center}
    \begin{tabular}{|>{\centering\arraybackslash}p{\vidTableIdColumnWidth}|>{\centering\arraybackslash}p{\vidTableSnapshotColumnWidth}|>{\centering\arraybackslash}p{\vidTableDescriptionColumnWidth}|}
    \hline
    \textbf{ID} & \textbf{Snapshot} & \textbf{Description} \\ 
    \hline
    \hline
    
    {1} & \includegraphics{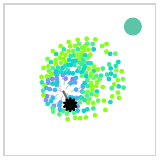} & Speed controller scan, random uniform driver ($\Delta s = 5$) (circle symbol) \\ 
    \hline
    {2} & \includegraphics{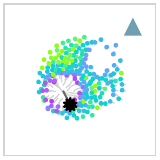} & Speed controller scan, random uniform driver ($\Delta s = 5$) (pyramid symbol) \\ 
    \hline
    {3} & \includegraphics{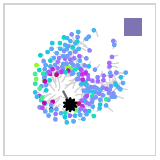} & Speed controller scan, random uniform driver ($\Delta s = 5$) (square symbol) \\ 
    \hline
    {4} & \includegraphics{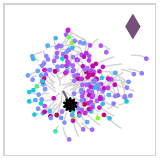} & Speed controller scan, random uniform driver ($\Delta s = 5$) (diamond symbol) \\ 
    \hline
    {5} & \includegraphics{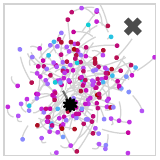} & Speed controller scan, random uniform driver ($\Delta s = 5$) (cross symbol)\\ 
    \hline
    {6} & \includegraphics{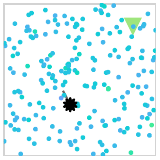} & Speed controller scan, random uniform driver ($\Delta s = 5$) (nabla symbol)\\
    \hline
    \end{tabular}
    \end{center}
    
    \caption{\new{\textbf{Supplementary videos for the speed-controller parameter scan with a driver sampled uniformly at random within a circle of radius $R=4$ around the center of the simulation box and a change interval of $\Delta s = 5$ steps for the determination of the short-term memory capacity} in \figref[c]{fig:speed-controller-memcap-total} and \figref[c]{fig:speed-controller-memcap-k-delay}. \newline Label in repository: \texttt{speed-controller-random-uniform-driver-interval-5}. }}
    \label{tab:supplementary_videos_random_memory_capacity_interval_5}
\end{table}

\begin{table}[H]
    \small
    \begin{center}
    \begin{tabular}{|>{\centering\arraybackslash}p{\vidTableIdColumnWidth}|>{\centering\arraybackslash}p{\vidTableSnapshotColumnWidth}|>{\centering\arraybackslash}p{\vidTableDescriptionColumnWidth}|}
    \hline
    \textbf{ID} & \textbf{Snapshot} & \textbf{Description} \\ 
    \hline
    \hline
    
    {1} & \includegraphics{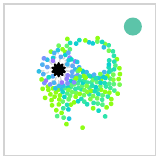} & Speed controller scan, random uniform driver ($\Delta s = 10$) (circle symbol) \\ 
    \hline
    {2} & \includegraphics{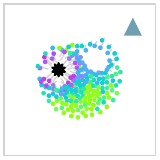} & Speed controller scan, random uniform driver ($\Delta s = 10$) (pyramid symbol) \\ 
    \hline
    {3} & \includegraphics{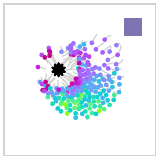} & Speed controller scan, random uniform driver ($\Delta s = 10$) (square symbol) \\ 
    \hline
    {4} & \includegraphics{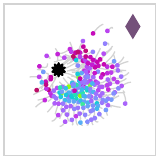} & Speed controller scan, random uniform driver ($\Delta s = 10$) (diamond symbol) \\ 
    \hline
    {5} & \includegraphics{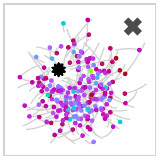} & Speed controller scan, random uniform driver ($\Delta s = 10$) (cross symbol)\\ 
    \hline
    {6} & \includegraphics{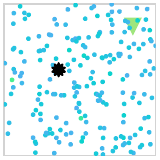} & Speed controller scan, random uniform driver ($\Delta s = 10$) (nabla symbol)\\ 
    \hline
    \end{tabular}
    \end{center}
    
    \caption{\new{\textbf{Supplementary videos for the speed-controller parameter scan with a driver sampled uniformly at random within a circle of radius $R=4$ around the center of the simulation box and a change interval of $\Delta s = 10$ steps for the determination of the short-term memory capacity} in \figref[d]{fig:speed-controller-memcap-total} and \figref[d]{fig:speed-controller-memcap-k-delay}. \newline Label in repository: \texttt{speed-controller-random-uniform-driver-interval-10}. }}
    \label{tab:supplementary_videos_random_memory_capacity_interval_10}
\end{table}